\documentclass[11pt]{article}

\usepackage[margin=2cm]{geometry}
\usepackage{amsmath,amssymb,amsthm,array,booktabs,cite,dsfont,graphicx,mathtools,multirow,rotating,placeins,subcaption, tensor,verbatim}
\usepackage[bookmarksnumbered,linktocpage,pdfstartview=FitH]{hyperref}
\usepackage[all]{hypcap}
\usepackage[all]{xy}
\usepackage[margin=2cm]{geometry}

\graphicspath{{figures/}}

\newcolumntype{M}[1]{>{$}{#1}<{$}}
\newcolumntype{C}[1]{>{\centering}m{#1}}

\newcommand{\rep}[1]{\ensuremath{\mathbf{#1}}}

\newcolumntype{B}[1]{>{\mathbf\bgroup}{#1}<{\egroup}}
\newcolumntype{K}{>{\lvert}{c}<{\rangle}}

\newtheorem*{theorem*}{Theorem}

\numberwithin{equation}{section}

\DeclareMathOperator{\Aut}{Aut}
\DeclareMathOperator{\Str}{Str}
\DeclareMathOperator{\tr}{tr}

\DeclareMathOperator{\Hom}{Hom}
\DeclareMathOperator{\SO}{SO}

\DeclareMathOperator{\SL}{SL}
\DeclareMathOperator{\USp}{USp}
\DeclareMathOperator{\SU}{SU}

\DeclareMathOperator{\Spin}{Spin}

\newcommand{\be}{\begin{equation}}
\newcommand{\ee}{\end{equation}}
\newcommand{\bea}{\begin{eqnarray}}
\newcommand{\eea}{\end{eqnarray}}

\newcommand{\half}{\tfrac{1}{2}}

\newcommand{\J}{\mathfrak{J}}

\newcommand{\alg}{\mathds{A}}

\newcommand{\F}{\mathds{F}}
\newcommand{\R}{\mathds{R}}
\newcommand{\C}{\mathds{C}}
\newcommand{\Q}{\mathds{H}}
\newcommand{\s}{\mathds{S}}
\newcommand{\Ts}{\mathds{T}_s}
\newcommand{\Cs}{\mathds{C}_s}

\newcommand{\T}{\mathds{T}}
\newcommand{\Z}{\mathds{Z}}
\newcommand{\Oct}{\mathds{O}}

\newcommand{\N}{\mathcal{N}}
\newcommand{\bn}{\mathbf{n}}
\newcommand{\blf}[2]{\langle#1 , #2\rangle}
\newcommand{\f}[1]{\mathfrak{#1}}
\newcommand{\pd}{\,\dot{+}\,}

\newcommand{\sst}[1]{{\scriptscriptstyle #1}}
\newcommand\nn{\nonumber}
\newcommand{\ft}[2]{\tfrac{#1}{#2}}
\newcommand{\del}{\partial}
\def\sz{{\sst{0}}}
\def\0{{\sst{(0)}}}
\def\1{{\sst{(1)}}}
\def\2{{\sst{(2)}}}
\def\3{{\sst{(3)}}}
\def\4{{\sst{(4)}}}
\def\5{{\sst{(5)}}}
\def\6{{\sst{(6)}}}
\def\7{{\sst{(7)}}}

\def\mi{{\sst{(-1)}}}
\def\mii{{\sst{(-2)}}}

\begin{document}

\begin{titlepage}
\begin{center}
\hfill DIAS-STP-17-05\\
\hfill DFPD/2015/TH-23\\

\vskip 1.5cm

{\Huge \bf A Kind of Magic}

\vskip 1.5cm

{\bf Leron Borsten${}^{1}$ and
Alessio Marrani${}^{2,3}$}

\vskip 20pt
 {\it ${}^1$School of Theoretical Physics, Dublin Institute for Advanced Studies,\\
10 Burlington Road, Dublin 4, Ireland}\\\vskip 5pt
 {\it ${}^2$Museo Storico della Fisica e Centro Studi e Ricerche ``Enrico Fermi'',\\
Via Panisperna 89A, I-00184, Roma, Italy}\\\vskip 5pt
 {\it ${}^3$Dipartimento di Fisica e Astronomia ``Galileo Galilei'', Univ. di Padova,\\
and INFN, Sez. di Padova, Via Marzolo 8, I-35131 Padova, Italy}\\\vskip 5pt

\texttt{leron@stp.dias.ie}\\
\texttt{alessio.marrani@pd.infn.it}

\end{center}

\vskip 2cm

\begin{center} {\bf ABSTRACT}\\[3ex]\end{center}
We introduce the extended Freudenthal-Rosenfeld-Tits magic square based on six   algebras: the reals $\mathds{R}$, complexes $\mathds{C}$, ternions $\T$, quaternions $\mathds{H}$, sextonions $\s$ and octonions $\mathds{O}$.  The sextonionic row/column of the magic square appeared previously and was shown to yield the non-reductive Lie algebras, $\mathfrak{sp}_{6\scriptscriptstyle{\frac{1}{2}}}$, $\mathfrak{sl}_{6\scriptscriptstyle{\frac{1}{2}}}$,
$\mathfrak{so}_{12\scriptscriptstyle{\frac{1}{2}}}$, $\mathfrak{so}_{12\scriptscriptstyle{\frac{3}{4}}}$ and $\mathfrak{e}_{7\scriptscriptstyle{\frac{1}{2}}}$, for $\mathds{R}, \mathds{C}, \mathds{H}, \s$ and $\mathds{O}$ respectively. The fractional ranks are used to denote the semi-direct extension of the simple Lie algebra in question by a unique (up to equivalence) Heisenberg algebra. The  ternionic row/column yields the non-reductive Lie algebras, $\mathfrak{sl}_{3\scriptscriptstyle{\frac{1}{4}}}$, $[\mathfrak{sl}_{3}\oplus\mathfrak{sl}_3]_{\scriptscriptstyle{\frac{1}{4}}}$, $[\mathfrak{sl}_{3}\oplus\mathfrak{sl}_3]_{\scriptscriptstyle{\frac{1}{2}}}$
$\mathfrak{sl}_{6\scriptscriptstyle{\frac{1}{4}}}$, $\mathfrak{sl}_{6\scriptscriptstyle{\frac{3}{4}}}$ and $\mathfrak{e}_{6\scriptscriptstyle{\frac{1}{4}}}$, for $\mathds{R}, \mathds{C}, \T, \mathds{H}, \s$ and $\mathds{O}$ respectively. The fractional ranks  here are used to denote the semi-direct extension of the semi-simple Lie algebra in question by a unique (up to equivalence) nilpotent ``Jordan'' algebra.  We present all possible real forms of the extended magic square.
It is demonstrated that the algebras of the extended magic square appear quite naturally as the symmetries of supergravity Lagrangians. The sextonionic row (for appropriate choices of real forms) gives the non-compact global symmetries of the Lagrangian for the $D=3$ maximal $\mathcal{N}=16$, magic $\mathcal{N}=4$ and magic non-supersymmetric theories, obtained by dimensionally reducing the $D=4$ parent theories on a circle, with the graviphoton  left undualised. In particular, the extremal intermediate non-reductive Lie algebra  $\tilde{\mathfrak{e}}_{7(7)\scriptscriptstyle{\frac{1}{2}}}$ (which is not a subalgebra of $\mathfrak{e}_{8(8)}$) is  the non-compact global symmetry algebra  of $D=3$, $\mathcal{N}=16$ supergravity as obtained by dimensionally reducing $D=4$, $\mathcal{N}=8$ supergravity with $\mathfrak{e}_{7(7)}$  symmetry on a circle.
On the other hand, the ternionic row (for appropriate choices of real forms) gives the non-compact global symmetries  of the Lagrangian for the $D=4$ maximal $\mathcal{N}=8$, magic $\mathcal{N}=2$ and magic non-supersymmetric theories, as obtained by dimensionally reducing the parent $D=5$  theories on a circle. In particular, the Kantor-Koecher-Tits intermediate non-reductive Lie algebra  $\mathfrak{e}_{6(6)\scriptscriptstyle{\frac{1}{4}}}$ is  the non-compact global symmetry algebra  of $D=4$, $\mathcal{N}=8$ supergravity as obtained by dimensionally reducing $D=5$, $\mathcal{N}=8$ supergravity with $\mathfrak{e}_{6(6)}$  symmetry on a circle.




\vfill


\end{titlepage}

\newpage \setcounter{page}{1} \numberwithin{equation}{section}

\newpage

\makeatletter
\def\l@subsubsection#1#2{}
\makeatother
\tableofcontents
\newpage

\section{Introduction}

The Freudenthal-Rosenfeld-Tits magic square \cite{Freudenthal:1954,Tits:1955, Freudenthal:1959,Rosenfeld:1956, Tits:1966} is a $4\times 4$ array $\f{m}(\alg_1, \alg_2)$ of semi-simple Lie algebras given by pairs of composition algebras $\alg_1, \alg_2=\R, \C, \Q, \Oct$. Here we introduce an extended  $6\times 6$ magic square of Lie algebras, as given in   \autoref{tab:ems1}. We show that the new columns/rows appear naturally as  the global non-compact symmetries  of the bosonic sectors of supergravity theories for specific choices of dualisations.

The additional rows and columns of the  extended  $6\times 6$ magic square are derived by considering the two \emph{intermediate} composition\footnote{In an abuse of conventional terminology we will  refer to unital algebras satisfying \eqref{comp} as ``composition'' even in the case that the quadratic form is degenerate.} algebras:
\begin{enumerate}
\item The three-dimensional ternions $\T$,  sitting in-between the complexes and quaternions, $\C\subset \T\subset \Q$. \item The six-dimensional  sextonions $\s$,  sitting in-between the  quaternions and octonions, $\Q\subset \s\subset \Oct$.
\end{enumerate}
The sextonions $\s$ were introduced by Kleinfeld in \cite{kleinfeld1968extensions} and subsequently  exploited in \cite{jeurissen} to study the conjugacy classes in the smallest exceptional Lie algebra $\f{g}_{2}$ in characteristics other than 2 or 3. They were also constructed in \cite{racine1974maximal} (cf. Th. 5 therein) and proved to be a maximal subalgebra of the split octonions. Sextonions were used to construct the sextonionic row/column  of the extended magic square by Westbury \cite{westbury2006sextonions}, and Landsberg and Manivel \cite{landsberg2006sextonions}.  The $(\s, \Oct)$ entry gives a non-reductive Lie algebra sitting in between $\f{e}_{7}$ and $\f{e}_{8}$ that was named $\f{e}_{7\sst{\frac{1}{2}}}$ in \cite{landsberg2006sextonions}. A construction of the sextonions as Zorn matrices was given in \cite{Marrani:2015nta} and used to re-derive the sextonionic rows/columns using Jordan pairs. A three-dimensional analogue of $\s$ is given by the ternions $\T$, see for example \cite{dray2015geometry}. As far as we are aware, the ternionic row/column of the magic square has not been  considered previously. Its construction is closely analogous to that of the sextonionic row/column, but also differs in some important respects, as will be made clear. In both cases the corresponding Lie algebras are non-reductive, as indicated by the fractional ranks used in \autoref{tab:ems1}.

  \begin{table}[ht]
 \begin{center}
\begin{tabular}{|c|cccccccc|}
\hline
\hline
  && $\R$ &$\C$ & $\T$  & $\Q$ & $\s$ & $\Oct$ & \\
 \hline
 &&&&&&&&\\
   $\R$ && $\mathfrak{sl}_2$ & $\mathfrak{sl}_3$   &$\mathfrak{sl}_{3\frac{1}{4}}$ &$\mathfrak{sp}_6$ & $\mathfrak{sp}_{6\frac{1}{2}}$ & $\mathfrak{f}_4$  & \\[12pt]
  $\C$ && $\mathfrak{sl}_3$ & $\mathfrak{sl}_3\oplus \mathfrak{sl}_3$   &$[\mathfrak{sl}_{3}\oplus \mathfrak{sl}_{3}]_\frac{1}{4}$& $\mathfrak{sl}_6$ & $\mathfrak{sl}_{6\frac{1}{2}}$ & $\mathfrak{e}_6$   &\\[12pt]
    $\T$ && $\mathfrak{sl}_{3\frac{1}{4}}$ & $[\mathfrak{sl}_{3}\oplus \mathfrak{sl}_{3}]_\frac{1}{4}$ & $[\mathfrak{sl}_{3}\oplus \mathfrak{sl}_{3}]_{\frac{1}{4}+\frac{1}{4}}$  & $\mathfrak{sl}_{6\frac{1}{4}}$ & $\mathfrak{sl}_{6(\frac{1}{4}+\frac{1}{2})}$ & $\mathfrak{e}_{6\frac{1}{4}}$   &\\[12pt]
  $\Q$ && $\mathfrak{sp}_6$ & $\mathfrak{sl}_6$  & $\mathfrak{sl}_{6\frac{1}{4}}$& $\mathfrak{so}_{12}$ &  $\mathfrak{so}_{12\frac{1}{2}}$ & $\mathfrak{e}_7$  & \\[12pt]
  $\s$ && $\mathfrak{sp}_{6\frac{1}{2}}$ & $\mathfrak{sl}_{6\frac{1}{2}}$ & $\mathfrak{sl}_{6(\frac{1}{4}+\frac{1}{2})}$ & $\mathfrak{so}_{12\frac{1}{2}}$ & $\mathfrak{so}_{12(\frac{1}{2}+\frac{1}{2})}$ & $\mathfrak{e}_{7\frac{1}{2}}$  & \\[12pt]
   $\Oct$ && $\mathfrak{f}_4$ & $\mathfrak{e}_6$ & $\mathfrak{e}_{6\frac{1}{4}}$ & $\mathfrak{e}_7$ & $\mathfrak{e}_{7\frac{1}{2}}$ & $\mathfrak{e}_8$&   \\[12pt]
   \hline
   \hline
\end{tabular}
\caption{The   extended magic square $\f{m}(\alg_1, \alg_2)$ over complexified $\alg_1, \alg_2=\R, \C, \T, \Q, \s, \Oct$. \label{tab:ems1}}
 \end{center}
\end{table}

In a mathematical context, the magic square ties together various aspects of algebra and geometry, yielding a number of intriguing surprises. For example,  the remarkable Deligne, Cohen and de Man dimension formulae \cite{deligne1996exceptional, cohen1996computational} for the  irreducible representations appearing in  $\bigotimes^k\f{g}$, for $k\leq4$ and $\f{g}$ an exceptional  simple Lie algebra, were derived, simplified and generalised in terms of composition algebras and the magic square by Landsberg-Manivel  \cite{Landsberg200259}. Indeed,  this composition algebraic recasting suggested that a gap appearing in the dimension formulae could be filled by  a six dimensional algebra sitting in between $\Q$ and $\Oct$ that would also imply a new row/column in the magic square, as  proposed and demonstrated in  \cite{westbury2006sextonions, landsberg2006sextonions} using the sextonions. It is tempting to speculate that the ternionic entries should have a related significance, although what this might be is left for future work.

The magic square also has made a number of appearances in the context of supergravity. There are three instances that have particular significance to the present contribution. The first\footnote{As far as we are aware, this is the first such example  of magic square like structures appearing in supergravity.} is  a particular real form of the $3\times 3$ inner $\C, \Q, \Oct$  square   \cite{Julia:1980gr}. Here, Julia considered  the oxidation of $\mathcal{N}$-extended  $D=3$ dimensional supergravity theories, which yields a ``trapezoid'' of theories for $D=3,4,\ldots 11$ and $0, 2^0, 2^1,\ldots 2^7$ supercharges\footnote{It also includes the affine Kac-Moody algebras, $\f{e}_{9}=\f{e}_{8}^{+}, \f{e}_{7}^{+}, \f{e}_{6}^{+}, \f{so}_{10}^{+}$ in $D=2$.}. The subset of algebras in the trapezoid given by $D=3,4,5$ and $2^5, 2^6, 2^7$ supercharges fits into the $3\times 3$ inner $\C, \Q, \Oct$ part of the magic  square, excluding the $(\C, \C)$ entry.  Note, the exact symmetry of this square is broken by the precise set of real forms obtained, which  are not actually given by any known magic square formula\footnote{Although, they are given by the magic \emph{pyramid} formula of \cite{Anastasiou:2015vba}.} \cite{Julia:1982gx}. Dispensing with the requirement of supersymmetry, the same $3\times 3$ square but  with maximally non-compact real forms was derived and  extended to a ``magic triangle'' of theories in $3\leq D\leq 11$ spacetime dimensions \cite{Cremmer:1999du}. The entries of triangle are parametrised by the dimension $D$  of the theory and the rank $0\leq n\leq 8$ of its symmetry algebra. The complete magic triangle  displays  a remarkable symmetry\footnote{Recently, such a result was retrieved and further generalized in \cite{Marrani:2017aqc}.} on interchanging   $D$ and  $n$. From our current perspective these examples are related to the  extended magic by the following results: ($i$) the $(\T, \Oct)$ and $(\s, \Oct)$ entries of the extended magic square describe the symmetries of maximally supersymmetric theories in $D=4,3$ with a certain choices of dualisation and ($ii$) the full $5\times 5$ inner $\C, \T, \Q, \s, \Oct$ square  with maximally non-compact real forms gives the symmetry algebras of the corresponding portion of the non-supersymmetric  magic triangle.  The third particularly  relevant and well-known example is given by the G\"unaydin-Sierra-Townsend $\mathcal{N}=2$ magic supergravities \cite{Gunaydin:1983rk, Gunaydin:1983bi, Gunaydin:1984ak}; the global symmetries of these theories in $D=5,4,3$ are given by the $\C, \Q, \Oct$ rows of the magic square with an appropriate choice of real forms. Again, we consider the magic supergravities in $D=4,3$ and demonstrate how the full $\T$ and $\s$ rows of extended magic square can be realised as their symmetry algebras.

More generally, since these initial examples there have been a number of, sometimes related, applications of the magic square and associated structures in a variety of contexts, ranging from generalised spacetimes \cite{Gunaydin:1992zh, Gunaydin:2005zz, Gunaydin:2000xr, Gunaydin:2007bg} to  black holes \cite{Ferrara:1997uz, Gunaydin:2005gd, Gunaydin:2005mx, Ferrara:2006xx, Bellucci:2006xz, Gunaydin:2007bg,Borsten:2008wd, Borsten:2009zy, Borsten:2010aa,  Borsten:2011ai} to gauge theories \cite{Marrani:2012uu} to quantum entanglement \cite{Borsten:2008, levay-2008, Levay:2008mi, Levay:2009, LevayVrana} and much else besides \cite{Bars:1978yx, LevayVrana, Chiodaroli:2011pp, Borsten:2013bp,Anastasiou:2013hba,  Chiodaroli:2014xia}. One could consider the role of the extended magic square in  each of these varied topics, although its relevance is not always naively obvious. For example, the magic square was recently shown to follow from ``squaring'' super Yang-Mills theories in $D=3$ \cite{Borsten:2013bp} and it seems reasonable to expect its extension to make an appearance also in this context, but we leave such speculations for future work. 

\medskip

Let us now outline the structure of the present contribution and summarise its results. In \autoref{alg} we review the required aspects of algebras and their symmetries. We begin by introducing  a generalised notion of  an ``intermediate algebra'', which is later used to define the extended magic square. We then discuss the composition algebras, and introduce the ternions $\T$ as the split-null extension of the split complexes, in direct analogy to the construction of sextonions $\s$ as the split-null extension of the split quaternions, which is also reviewed. Alternatively, $\T$ and $\s$ can be regard as  intermediate algebras of split $\Q$ and split $\Oct$, respectively. Finally, we described the orthogonal, derivation and triality algebras defined on the composition algebras, including $\T$ and $\s$. There is a subtlety in the case of $\s$: the restriction homomorphisms  from the subalgebras preserving $\s$ in $\Oct$ to the algebras on $\s$ have a 1-dimensional kernel, which must be quotiented out, implying that the algebras on $\s$ are not subalgebras of those on $\Oct$. In contrast, the algebras over $\T$ \emph{are} subalgebras of those on $\Q$.

In \autoref{msquare} we bring these elements together to build the complexified extended magic square, reviewing in some detail the sextononic entries previously given in \cite{westbury2006sextonions, landsberg2006sextonions}, as well as  the new  ternionic entries. There are at least four possible definitions for the $\s$ row/column. If we require that the $\s$ row/column constitute  subalgebras of the $\Oct$ row/column, we must use the $\s$-preserving subalgebras of those defined on $\Oct$.  This yields two possibilities, $\f{m}(\s, \alg)$ and $\f{m}(\s, \alg)_\sz$, which are given by  extremal intermediate subalgebras of the octonionic entries,  $\f{ext}(\f{m}(\Oct, \alg))$ and  $\f{ext}_\sz(\f{m}(\Oct, \alg))$ respectively, as will be explained in \autoref{complexms}. If we instead use the algebras on $\s$ directly, we  again obtain two possibilities, denoted $\widetilde{\f{m}}(\s, \alg)$ and $\widetilde{\f{m}}(\s, \alg)_\sz$. They  are isomorphic to $\widetilde{\f{ext}}(\f{m}(\Oct, \alg))$ and  $\widetilde{\f{ext}}_\sz(\f{m}(\Oct, \alg))$, respectively, which are the extremal intermediate subalgebras quotiented by their   grade two 1-dimensional ideals; consequently, they are not subalgebras of $\f{m}(\Oct, \alg)$.  For the $\T$ row/column there are only two possibilities, $\f{m}(\T, \alg)$ and $\f{m}(\T, \alg)_\sz$, which are given by  Kantor-Koecher-Tits intermediate subalgebras of the quaternionic  entries,  $\f{ k }(\f{m}(\Q, \alg))$ and  $\f{ k }_\sz(\f{m}(\Q, \alg))$, respectively, as will be explained again in \autoref{complexms}. The non-reductive part of  $\f{ k }_{\sz}(\f{m}(\Q, \alg))$ is a nilpotent radical ideal and consequently there are  no further non-trivial possibilities; quotienting reduces  $\f{ k }(\f{m}(\Q, \alg))$ and  $\f{ k }_\sz(\f{m}(\Q, \alg))$ to $\f{gl}_1\oplus \f{m}(\C, \alg)$ and $\f{m}(\C, \alg)$, respectively.

We then present all possible real forms of the extended magic square. Over the reals, the ternions and sextonions only exist in their split forms, $\C_s\subset \T_s\subset \Q_s$ and $\Q_s\subset \s_s\subset \Oct_s$, which implies the use of both $\alg_1, \alg_2$ split\footnote{For an explicit proof for the sextonions, cfr. e.g. \cite{Marrani:2015nta}.}. This yields two possibilities, according as to whether one uses a Euclidean  or Lorentzian Jordan algebra in the Tits construction. However, they only differ in the $(\R, \R)$ slot; see \autoref{tab:emsreal1}. With respect to the global symmetries of supergravity, it is also useful to consider the semi-extended magic squares with  rows including $\T_s, \s_s$, but not the columns, which yields six  further  sets of reals forms, as given in  \autoref{tab:emsreal3} and \autoref{tab:emsreal5}.

In \autoref{sugra} it is shown that the $\T_s$ and $\s_s$ rows appear as the global symmetry algebras of various supergravity theories in $D=4,3$. In particular,
\be
\widetilde{\f{m}}(\s_s, \Oct_s)\cong \widetilde{\f{e}}_{7(7)\frac{1}{2}}
\ee
constitutes the global symmetry algebra of $D=3$, $\N=16$ supergravity obtained by dimensionally reducing $D=4$, $\N=8$ supergravity with $\f{e}_{7(7)}$ symmetry algebra on a circle.  Similarly, the $\T_s$ and $\s_s$ rows of the maximally non-compact  \autoref{tab:emsreal1} are shown to be  symmetry algebras of the purely bosonic theories of \cite{Cremmer:1999du} (recently reconsidered in \cite{Marrani:2017aqc}) with a judicious choice of dualisations.
Using the real forms of \autoref{tab:emsreal3a} the sextonionic row $(\s_s, \alg)$ gives the global symmetry algebras of the magic supergravities in $D=3$.  The ternionic row $(\T_s, \alg)$ (for appropriate choices of real forms) gives the  global symmetry algebras  of $D=4$, maximal $\N=8$ and magic $\N=2$, supergravity Lagrangians obtained by dimensionally reducing  $D=5$, maximal $\N=8$ and magic $\N=2$, theories, with $\f{e}_{6(6)}$ and $\f{e}_{6(-26)}$ invariant Lagrangians, respectively.

\subsection{Notation}

We will use $\pd$ to denote the direct sum of vector spaces and reserve $\oplus$ to denote the direct sum of algebras. Generic algebras will be denoted with calligraphic Roman capitals $\mathcal{A}$. Composition algebras will be denoted with blackboard bold Roman capitals $\alg$. This includes the normed division algebras $\R, \C, \Q, \Oct$,  their split-signature cousins $\C_s, \Q_s, \Oct_s$, and subalgebras thereof. Lie algebras will be denoted with lower-case Gothic Roman letters,   $\f{g}, \f{h}\ldots$ and so forth.

\section{Algebras and symmetries}\label{alg}

\subsection{Intermediate algebras}
For an algebra $\mathcal{A}$ we will use  ``intermediate algebra of $\mathcal{A}$'', generically denoted  $\f{int}(\mathcal{A})$, to loosely mean some positively graded subalgebra of $\mathcal{A}$ under some grading operator in $\mathcal{A}$. In this section we will present two important examples of this idea, starting with the definition of an intermediate algebra given in \cite{westbury2006sextonions}, which we shall refer to here as the extremal intermediate algebra.

\paragraph{Extremal intermediate algebras} The symmetries algebras on the sextonions (see \autoref{sexts}) and consequently the sextonionic row/column of the extended magic square, are given by  intermediate algebras following from a 5-grading \cite{westbury2006sextonions}, which we shall refer to as \emph{extremal} and denote by $\mathfrak{ext}(\mathfrak{g})$, for a
complex simple Lie algebra $\mathfrak{g}$. Let
${\mathfrak{h}}$ be the centralizer of an extremal triple
$\{h, e, f\}$ in $\mathfrak{g}$. Then $\mathfrak{g}$ admits a
5-grading in terms of the eigenspaces of $h$: \be
\mathfrak{g}=\C_{\mii}\pd  V_{\mi}\pd  \underbrace{[\mathfrak{h}\oplus\mathfrak{gl}_1(\C)]}_{\mathfrak{g}_{\0}}{}_{\0}\pd  V_{\1}\pd \C_{\2}, \ee where
$V$ is a symplectic representation of ${\mathfrak{h}}$; by symplectic,  we mean that $\wedge^2V$ contains the trivial representation of ${\mathfrak{h}}$. Note, $V$ is simple, except for $\f{g}$ of type $A$ where it is the sum of two simple complex-conjugate representations,  and minuscule.  The extremal
intermediate algebra $\mathfrak{ext}(\f{g})$ is defined as
the positively $\mathfrak{gl}_1(\C)$-graded subalgebra
\be
\mathfrak{ext}(\f{g}):=[{\mathfrak{h}}\oplus
\mathfrak{gl}_1(\C)]_{\0}\pd  V_{\1}\pd \C_{\2}.
\ee
The codimension one (except for $\f{g}$ of type $A$ for which  the centre is two dimensional) derived  extremal intermediate algebras are defined by,
\be
\mathfrak{ext}_\sz(\f{g}):=[\mathfrak{g}_{\0}, \mathfrak{g}_{\0}]\pd V\pd \C,
\ee
 where $[\mathfrak{g}_{\0},\mathfrak{g}_{\0}]$ is the semi-simple part of $\f{g}_\0$. Note that $\mathfrak{ext}_\sz(\f{g})$ is a subalgebra of $\mathfrak{ext}(\f{g})$, and thus of $\f{g}$.  
 The strictly positive graded $\f{g}_\0$-module $V\pd \C$ generates a Heisenberg algebra $H(V, \omega)$, which exponentiates to give the Heisenberg group,
\be
(a, \lambda)(b, \tau)=( a +  b, \half\omega(a,b)+\tau+\lambda),\quad a,b \in V, \quad \tau, \lambda \in \C,
\ee
where $\omega$ is the $\f{h}$-invariant symplectic form on $V$. For a Lie algebra $\f{g}$ with a sympletic representation $V$ of dimension $2n$, we will use the notation $\f{g}\ltimes H_{2n}$ to denote $\f{g}$ semi-direct sum the Heisenberg algebra $H_{2n}=H(V, \omega)$.

The grade two components of $\f{ext}$ and $\f{ext}_\sz$ are a one-dimensional ideals.  Quotienting by these ideals we have,
\be
\widetilde{\mathfrak{ext}}(\f{g}):=\mathfrak{ext}(\f{g})/\C_{\2}=\mathfrak{g}_{\0}\pd  V_{\1},\qquad \widetilde{\mathfrak{ext}}_{\sz}(\f{g}):=\mathfrak{ext}_\sz(\f{g})/\C=[\mathfrak{g}_{\0}, \mathfrak{g}_{\0}]\pd  V.
\ee
We recall here the discussion of these quotients as they arise naturally from algebras defined on the sextonions $\s$ and as the symmetries of supergravity. 

Note, there is no natural inclusion of $\widetilde{\mathfrak{ext}}$ and $\widetilde{\mathfrak{ext}}_{\sz}$ into ${\mathfrak{ext}}$ and $\mathfrak{ext}_\sz$. Rather, if $\mathfrak{h}$ itself is semi-simple ${\mathfrak{ext}}$ is obtained from $\widetilde{\mathfrak{ext}}_{\sz}$ by adding a central extension and then a grading operator.  Although $\widetilde{\mathfrak{ext}}(\f{g})$ and $\widetilde{\mathfrak{ext}}_{\sz}(\f{g})$ can be identified, as vector spaces, with the grade zero and one components of $\mathfrak{ext}(\f{g})$ and $\mathfrak{ext}_\sz(\f{g})$ respectively, their Lie algebra structures are different, due to the quotient by the 1-dimensional grade two ideal $\C_{\2}$. It should also be noted that when $\mathfrak{h}$ is semi-simple $\mathfrak{ext}_\sz(\f{g})$ and $\widetilde{\mathfrak{ext}}(\f{g})$ are the same as vector spaces, but  have distinct Lie algebraic structures; in particular, $\mathfrak{ext}_\sz(\f{g})$ is a subalgebra of $\mathfrak{ext}(\f{g})$, and thus of $\f{g}$, while - as mentioned - there is no natural inclusion of $\widetilde{\mathfrak{ext}}(\f{g})$ into $\mathfrak{ext}(\f{g})$, and thus into $\f{g}$. The situation is summarised for semi-simple $\mathfrak{h}$ by the commutative diagram \cite{westbury2006sextonions}:
\be\label{com} \xymatrix{ \mathfrak{ext}_\sz(\f{g})=\mathfrak{h}\pd V\pd \C \ar[r] \ar[d]  & \mathfrak{ext}(\f{g})=[{\mathfrak{h}}\oplus
\mathfrak{gl}_1(\C)]_{\0}\pd  V_{\1}\pd \C_{\2} \ar[d] \\ \widetilde{\mathfrak{ext}}_{\sz}(\f{g})=\mathfrak{h}\pd  V  \ar[r]  &  \widetilde{\mathfrak{ext}}(\f{g})=[{\mathfrak{h}}\oplus\mathfrak{gl}_1(\C)]_{\0}\pd  V_{\1} }
\ee
The horizontal arrows are Lie algebra   inclusions and the vertical arrows are surjective homomorphisms with 1-dimensional kernel.  

\paragraph{Kantor-Koecher-Tits (KKT) intermediate algebras} The symmetries algebras on the ternions $\T$ (see \autoref{tris}), and consequently the  ternionic  row/column of the extended magic square, are given by an intermediate algebra that follows from a 3-grading as given by the Kantor-Koecher-Tits construction, which we shall accordingly denote by $\f{ k }(\f{g})$. Let
 $\mathfrak{g}$ be a complex semi-simple Lie algebra with
3-grading, \be
\mathfrak{g}=\overline{V}_{\mii}\pd  \underbrace{[\mathfrak{h}\oplus\mathfrak{gl}_1(\C)]}_{\mathfrak{g}_{\0}}{}_{\0}\pd  V_{\2}, \ee where
$\overline{V}_{\mii}, V_{\2}$ constitute a Jordan pair.   The Kantor-Koecher-Tits
intermediate algebra  is, as for the 5-grading example, defined as
the positively $\mathfrak{gl}_1(\C)$-graded subalgebra
\be\label{3grade k }
\mathfrak{k}(\f{g}):=[{\mathfrak{h}}\oplus
\mathfrak{gl}_1(\C)]_{\0}\pd  V_{\2}. \ee
Similarly, we can define the  derived intermediate subalgebras,
\be\label{derived3grade k }
\mathfrak{k}_\sz(\f{g}):=[\mathfrak{g}_{\0}, \mathfrak{g}_{\0}] \pd V.
\ee
The  positively graded  component $V$ is an ideal, which  we shall refer to as a  nilpotent algebra $N(V)$,
\be
a b=a+b,\quad a,b \in N(V).
\ee
For a Lie algebra $\f{g}$ with complex representation $V$ of dimension $n$ we will use the notation $\f{g}\ltimes N_{n}$ to denote $\f{g}$ semi-direct sum the  nilpotent algebra $N_{n}=N(V)$. Note, $V$ with a graded involution $\tau(\f{g}_i)=\f{g}_{-i}$ yields a Jordan triple system with triple product $(abc)=[[a,\tau(b)],c]$. We can quotient such an ideal, obtaining:
 \be
 \widetilde{\mathfrak{k}}(\f{g}):=\mathfrak{k}(\f{g})/V_{\2}=[{\mathfrak{h}}\oplus\mathfrak{gl}_1(\C)]_{\0},\qquad \widetilde{\mathfrak{k}}_{\sz}(\f{g}):=\mathfrak{k}_\sz(\f{g})/V=\mathfrak{h},
 \ee
which are reductive Lie algebras; $V$ is a nilpotent radical ideal.

Although we have presented the complexified case here, we shall be principally concerned with real forms determined by composition algebras over the reals, as discussed in the subsequent sections.
\subsection{Composition algebras}

A \emph{quadratic norm} on a vector space $V$ over a field $\F$ is a map $\bn:V\to\F$ such that:
\begin{enumerate}
\item  $\bn(\lambda a)=\lambda^2\bn(a),~~  \lambda\in\F, a\in V$
\item The form
$
\langle a, b\rangle:=\bn(a+b)-\bn(a)-\bn(b)
$
is bilinear.
\end{enumerate}

A  \emph{composition} algebra $\mathds{A}$  over  $\R$ has an identity element $e_0$ and a non-degenerate quadratic norm satisfying
\begin{equation}\label{comp}
 \qquad \bn(ab)=\bn(a)\bn(b),\quad \forall~~ a,b \in\alg,
\end{equation}
where we denote the multiplicative product of the algebra by juxtaposition; see \cite{Springer:2000} for a comprehensive review. Regarding ${\R}\subset\alg$ as the scalar multiples of the identity $ \R e_0$, we may decompose $\mathds{A}$ into its ``real'' and ``imaginary'' parts $\alg={\R}\oplus \alg'$, where $\alg'\subset\alg$ is the subspace orthogonal to $\R$. An arbitrary element $a\in\alg$ may thus be written $a=\text{Re}(a) +\text{Im}(a)$. Here  $\text{Re}(a)\in\R e_0$, $\text{Im}(a)\in \alg'$ and
\be
\text{Re}(a):=\frac{1}{2}(a+\overline{a}), \qquad \text{Im}(a):=\frac{1}{2}(a-\overline{a}),
\ee
where we have defined conjugation  using the bilinear form,
\be
\overline{a}:=\blf{a}{e_0}e_0-a.
\ee

A composition algebra $\mathds{A}$  is said to be \emph{division} if it contains no zero divisors,
\begin{equation*}
ab=0\quad \Rightarrow\quad a=0\quad\text{or}\quad b=0,
\end{equation*}
which is the  case if and only if $\bn$ is positive semi-definite. $\alg$ is then referred to as a normed division algebra.
Hurwitz's celebrated theorem states that there are exactly four normed division algebras \cite{Hurwitz:1898}: the reals, complexes, quaternions and octonions, denoted respectively by $\R, \C, \Q$ and $\Oct$. We will also consider their split forms $\C_s, \Q_s$ and $\Oct_s$ which have indefinite norms of signature $(1,1), (2,2)$ and $(4,4)$, respectively, and are therefore not division.

We can build each algebra starting from the reals and using the Cayley-Dickson doubling procedure. From an algebra $\alg_n$ of dimension $n$, define the algebra $\alg_{2n}:=\alg_n\pd\alg_n$ of dimension $2n$ with  conjugation, norm and multiplication given by
\be
\overline{(a, b)}:=(\overline{a}, -b),\qquad \bn(a,b):=\bn(a)+\varepsilon\bn(b)
\ee
and
\be
(a_1, b_1)\cdot(a_2, b_2): = (a_1a_2- \varepsilon b_2 \overline{b}_1, \overline{a}_1b_2+ a_2b_1),
\ee
respectively, where $\varepsilon=\pm1$. If $\alg_n$ is division, then $\alg_{2n}$ is division or split according as  $\varepsilon$ is $+1$ or $-1$, respectively. If $\alg_n$ is split, then $\alg_{2n}$ is split for either choice of   $\varepsilon$.

\subsection{Symmetries on algebras}
We will make use of three symmetry algebras defined on composition algebras:  orthogonal, derivation and triality. The resulting Lie algebras are summarised in \autoref{tab:syms}.
\paragraph{Orthogonal algebra:} The algebra of orthogonal transformations on  a vector space $V$ over a field $\F$ with non-degenerate inner-product $\langle , \rangle$ is defined by,
\be
\mathfrak{so}(V):=\{\alpha\in \Hom_\F(V)\; |\; \langle\alpha(a), b\rangle+\langle a, \alpha(b)\rangle=0\},
\ee
with Lie bracket given by the commutator. See  \autoref{tab:syms} for $\f{so}(\alg)$.

\paragraph{Derivation algebra:} The algebra of derivations on  a unital algebra $\mathcal{A}$ over a field $\F$ is defined as the set of $\F$-linear homomorphisms satisfying the Leibniz rule,
\be
\mathfrak{der}(\mathcal{A}):=\{\alpha\in \Hom_\F(\mathcal{A})\; |\; \alpha(ab)=\alpha(a)b+a\alpha(b)\},
\ee
with Lie bracket given by the commutator. See  \autoref{tab:syms} for $\f{der}(\alg)$.

\paragraph{Triality algebra:} The algebra of trialities on  a unital algebra $\mathcal{A}$ over a field $\F$ is defined as the set of triples of orthogonal transformations satisfying the triality property, which generalises the derivation property,
\be
\mathfrak{tri}(\mathcal{A}):=\{(\alpha_1, \alpha_2, \alpha_3) \in 3\f{so}(\mathcal{A})\; |\; \alpha_1(ab)=\alpha_2(a)b+a\alpha_3(b)\},
\ee
with Lie bracket given by the commutator. See  \autoref{tab:syms} for $\f{tri}(\alg)$.

  \begin{table}[ht]
 \begin{center}
 \footnotesize
\begin{tabular}{|l|ccccc|}
\hline
\hline
   &$\C$ &  $\T$  & $\Q$& $\s$ & $\Oct$   \\
 \hline
 &&&&&\\
   $\f{so}(\alg)$ & $\mathfrak{so}(2)$   &- &$\mathfrak{so}(3)\oplus\mathfrak{so}(3)$ & - & $\f{so}(8)$ \\[10pt]

     $\f{so}(\alg_s)$ &  $\mathfrak{so}(1,1)$   &$\text{Im}(\T_s)\oplus  \text{Im}(\T_s)$ & $\mathfrak{sl}_2\oplus \mathfrak{sl}_2$ & $\f{tri}(\Q_s)\oplus\f{so}(1,1)\pd\Q_s\otimes\Q_{s}^{+}$  & $\f{so}(4,4)$ \\[10pt]

   \hline
    &&&&&\\

   $\f{der}(\alg)$ & $\varnothing$   &- &$\mathfrak{so}(3)$ & - & $\f{g}_{2(-14)}$ \\[10pt]

     $\f{der}(\alg_s)$ &  $\varnothing$   &$\text{Im}(\T_s)$ & $\mathfrak{sl}_2$ & $\f{der}(\Q_s)\pd \Q_s$  & $\f{g}_{2(2)}$ \\[10pt]

     \hline
      &&&&&\\

        $\f{tri}(\alg)$ & $\mathfrak{so}(2)\oplus\mathfrak{so}(2)$   &- &$\mathfrak{so}(3)\oplus\mathfrak{so}(3)\oplus\mathfrak{so}(3)$ & - & $\f{so}(8)$ \\[10pt]

     $\f{tri}(\alg_s)$ &  $\mathfrak{so}(1,1)\oplus\mathfrak{so}(1,1)$   &$\text{Im}(\T_s)\oplus\text{Im}(\T_s)\oplus\text{Im}(\T_s)$ & $\mathfrak{sl}_2\oplus \mathfrak{sl}_2\oplus \mathfrak{sl}_2$ & $\mathfrak{tri}(\Q_s)\oplus\f{so}(1,1) \pd \Q_s\otimes\Q_{s}^{+}$  & $\f{so}(4,4)$ \\[10pt]
   \hline
   \hline
\end{tabular}
\caption{Lie algebras on the relevant composition algebras over $\R$. Here, $\f{sl}_2$ is understood to mean $\f{sl}_2(\R)$. We omit $\R$ column since it  is trivial in all cases. The $\T_s$ and $\s_s$ entries are explained in \autoref{splits}. \label{tab:syms}}
 \end{center}
\end{table}

\subsection{Split-null extensions and their symmetries}\label{splits}

\subsubsection{The ternions}\label{tris}
The split-complexes $\C_s$ have a one-dimensional bimodule $M$. The split-ternions $\mathds{T}_s$ are defined as the split null extension of  $\C_s$ by $M$: $\mathds{T}_s=\C_s\pd M$ with multiplication, norm and conjugation, respectively given by
\be
(A, a)\cdot (B, b): = (AB, \overline{A}b+{B}a),\qquad
\mathbf{n}(A, a) := \mathbf{n}(A),\qquad
\overline{(A, a)}:=(\overline{A}, -a),
\ee
where $A,B\in\Cs$ and $a, b\in M$. Alternatively, the ternions are isomorphic to  three dimensional  subalgebras of the split-quaternions $\Q_s$, for example the span of $\{e_0, e_2, (e_1+e_3)\}$.  This space is clearly closed under  addition and multiplication, and since $\mathbf{n}((e_1+e_3))=0$ it is isomorphic to $\Ts$. Equivalently, this algebra may be realised as  upper triangular $2\times 2$ matrices over the reals.

This possibility follows from the fact that, unlike the quaternions,  the split-quaternions  admit a 3-grading,
\be\label{Hgrading}
\Q_s=\R^- \pd\C_{s}^{\0}\pd\R^+.
\ee
If one regards $\Q_s$ as the Cayley-Dickson double $\Q_s=\C_s\pd\C_s$ and defines the grade zero piece $\C_{s}^{\0}$ as the image of the inclusion $\C_s \hookrightarrow \Q_s: A \mapsto (A, 0)$, then the subspace of elements of the form  $(0, B)$ is a left module,
\be
(A , 0)\cdot(0, B) = (0, \overline{A}B),
\ee
which may be decomposed under $(i,0)\in\C_{s}^{0}$ into  grade 1 and $-1$ subspaces as follows. Since $i^2=1$, $\C_s$ decomposes into  positive/negative eigenspaces under the left-multiplication operator $L_i(a)=ia$. Explicitly, consider the basis for $\C_s$ given by
\be
i_+:=\half (1+i),\quad i_-:=\half (1-i)
\ee
so that
\be
ii_\pm=\pm i_\pm,\qquad
i_\pm i_\pm=i_\pm, \qquad
i_+i_-=0.
\ee
 The $\pm 1$ graded components of $\Q_s$ (\ref{Hgrading}) are  given by,
\be
\R^{+} \cong \{(0, b^+i_+) | b^+\in\R \}, \quad \R^{-} \cong \{(0, b^-i_- ) | b^-\in\R \}.
\ee
Writing $A=A^+i_++A^-i_-$, the product of two elements of $\Q_s$ (\ref{Hgrading}) is given by
\be
(a^-, A, a^+)\cdot(b^-, B, b^+)=(A^+b^-+B^-a^-, AB-b^+a_-i_++b^-a^+i_-, A^-b^++B^+a^+).
\ee
Hence,  the ternions can be defined as the intermediate algebra of the split-quaternions,
\be\label{pain-0}
\T_s\cong \f{int}(\Q_s),
\ee
 since they are isomorphic to the positive grade component $\C_{s}^{\0}\pd\R^+$ in $\Q_s$, embedded by the obvious inclusion $(A, a^+)\mapsto(0, A, a^+)$. Through the standard bijection  $\Q_s\cong \R[2]$,
\be\label{Hmatrices}
\begin{array}{llllllllll}
(1,0)\mapsto e_0\mapsto& \begin{pmatrix}1&0\\0&1\end{pmatrix},&\quad (0,1)\mapsto e_1\mapsto& \begin{pmatrix}0&1\\-1&0\end{pmatrix},\\[12pt]
(i,0)\mapsto e_2\mapsto &\begin{pmatrix}-1&0\\0&1\end{pmatrix},&\quad
(0,i)\mapsto e_3\mapsto &\begin{pmatrix}0&1\\1&0\end{pmatrix},
\end{array}
\ee
under which
\be\label{Hmatricesnorm}
\overline{a}=\epsilon a^t\epsilon^t,\qquad \bn(a)=\det(a),
\ee
from the span $\left\{ e_{0},e_{2},e_{1}+e_{3}\right\} $, one obtains a third description of the ternions as upper-triangular  $2\times 2$ matrices over the reals, as pointed out above.


\paragraph{Orthogonal algebra:} The orthogonal algebras are defined  in terms of the norm and product of $\alg$ only, which both carry over to $\T_s$ allowing us to define analogously $\f{so}(\T_s)$. However, we shall first consider the ternion preserving subalgebra  in $\f{so}(\Q_s)$, denoted $\f{so}_{\T_s}(\Q_s)$.

As reported in  \autoref{tab:syms}, the inner product $\langle a, b\rangle$ on $\Q_s$ is invariant under two commuting $\SL_2(\R)$ factors,
\be
a\mapsto pa\overline{q}, \qquad p,q \in \Q_s\quad\text{s.t.} \quad \bn(p)=\bn(q)=1.
\ee The corresponding $\f{sl}_2(\R)\oplus\f{sl}_2(\R)$ action is given by,
\be
a\mapsto ma-an, \qquad m, n\in \text{Im}(\Q_s).
\ee
In terms of the matrix representation given by \eqref{Hmatrices} and \eqref{Hmatricesnorm}, the $\SL_2(\R)\times \SL_2(\R)$ action is given by left  and right multiplication by independent determinant 1 matrices. Similarly, the $\f{sl}_2(\R)\oplus \f{sl}_2(\R)$ action is given by left  and right multiplication by independent traceless matrices. Hence, the ternion preserving subalgebra $\f{so}_{\T_s}(\Q_s)\subset\f{so}(\Q_s)$   is given by the subset of upper-triangular elements in $\f{sl}_2(\R)\oplus \f{sl}_2(\R)$, yielding the non-reductive Lie algebra,
\be
\f{so}_{\T_s}(\Q_s)\cong [\f{so}(1,1)\pd\R]\oplus[\f{so}(1,1)\pd\R]\cong \text{Im}(\T_s)\oplus  \text{Im}(\T_s).
\ee
The restriction homomorphism $\f{so}_{\T_s}(\Q_s)\rightarrow\f{so}(\T_s)$ is an isomorphism (in contrast to the sextonionic case), and thus:
\be
\f{so}(\T_s)\cong\text{Im}(\T_s)\oplus  \text{Im}(\T_s).
\ee

Under the 3-grading \eqref{Hgrading} each $\f{sl}_2(\R)$ decomposes as,
\be
\f{sl}_2(\R)=\R_{\mii} \pd \f{so}(1,1)_{\0}\pd\R_{\2},
\ee
which should be regarded as a 5-grading, with trivial grade one component, induced by a trivial (in the sense that it spans the full algebra) extremal triple. Consequently,
\be
\f{int}(\f{sl}_2)\cong \f{so}(1,1)_{\0}\pd\R_{\2}\cong \text{Im}(\T_s)
\ee
and,  in this sense, $\f{so}(\T_s)$ should be regarded as an intermediate algebra,
\be
\f{so}(\T_s)\cong\f{int}(\f{sl}_2)\oplus\f{int}(\f{sl}_2).
\ee

\paragraph{Derivation algebra:} The algebra of derivations on $\Q_s$ is given by identifying the two $\f{sl}_2(\R)$ factors in $\f{der}(\Q_s)$,
\be
a\mapsto ma-am, \qquad m\in \text{Im}(\Q_s).
\ee
Clearly, the identity in $\Q_s$ is annihilated by derivations; $\text{Re}(\Q_s)$ and $\text{Im}(\Q_s)$ are one- and three- dimensional irreducible representations of $\f{der}(\Q_s)\cong \f{sl}_2(\R)$, respectively.

The ternion preserving subalgebra $\f{der}_{\T_s}(\Q_s)\subset\f{der}(\Q_s)$   is given by,
\be
\f{int}(\f{sl}_2)\cong \f{so}(1,1)_{\0}\pd\R_{\2}\cong \text{Im}(\T_s),\ee
which is isomorphic under  \eqref{Hmatrices} to $2\times 2$ traceless upper-triangular  real matrices.  Again, the restriction homomorphism $\f{der}_{\T_s}(\Q_s)\rightarrow\f{der}(\T_s)$ is an isomorphism, and therefore:
\be
\f{der}(\T_s)\cong \f{so}(1,1)\pd\R\cong \text{Im}(\T_s).
\ee

\paragraph{Triality algebra:} A triple $(p_i, q_i)\in\text{Im}(\Q_s)\oplus\text{Im}(\Q_s)$, $i=1,2,3$ of elements in $\f{so}(\Q_s)$ belong to $\f{tri}(\Q_s)$ if,
\be
p_1ab-abq_1=(p_2a-aq_2)b+a(p_3b-bq_3).
\ee
Setting $a=1$ and $b=1$ separately, it then follows that a generic element of $\f{tri}(\Q_s)$ can be written as the triple of related $\f{so}(\Q_s)$ elements $(p_1, q_1), (p_2, q_1), (p_1, -p_2)$, and therefore $\f{tri}(\Q_s)\cong\f{sl}_2(\R)\oplus\f{sl}_2(\R)\oplus\f{sl}_2(\R)$. The  ternion-preserving subalgebra $\f{tri}_{\T_s}(\Q_s)\subset\f{tri}(\Q_s)$ is therefore given by
\be
\f{tri}_{\T_s}(\Q_s)\cong\text{Im}(\T_s)\oplus\text{Im}(\T_s)\oplus\text{Im}(\T_s).
\ee
Again, it follows from $\f{der}_{\T_s}(\Q_s)\cong\f{der}(\T_s)$ that $\f{tri}_{\T_s}(\Q_s)\cong\f{tri}(\T_s)$.
\subsubsection{The sextonions}\label{sexts}
The sextonions, first constructed in \cite{kleinfeld1968extensions}, have been treated in some detail in \cite{westbury2006sextonions, landsberg2006sextonions, Marrani:2015nta} and we shall therefore be rather less  explicit in this section. Instead, we will use the representation theoretic conventions of the high energy physics community to help bridge any gaps in the notation.

The split-quaternions  admit a unique two-dimensional non-associative bimodule $M$ \cite{jacobson1954}. The sextonions $\mathds{S}_s$ can be defined as the split null extension of  $\Q_s$ by $M$: $\mathds{S}_s=\Q_s\pd M$, with multiplication, norm and conjugation respectively given by
\be
(A, a)\cdot (B, b) = (AB, \overline{A}b+{B}a),\qquad
\mathbf{n}(A, a) = \mathbf{n}(A),\qquad
\overline{(A, a)}=(\overline{A}, -a),
\ee
where $A,B\in\Q_s$ and $a, b\in M$. Alternatively, they are isomorphic to  six-dimensional maximal subalgebras of the split-octonions \cite{racine1974maximal}. Using $\Q_s\cong\R[2]$ and the Cayley-Dickson construction $\Oct_s\cong\Q_s\pd\Q_s$, a sextonionic subalgebra is given by all elements of the form $(A, b)$ for $A\in\R[2]$ and $b\in \R[2]_{+}$, where we denote by  $\R[2]_{+}$ ($\R[2]_-$) the subalgebra in $\R[2]$ of  matrices with arbitrary left (right) column   and  zero right (left) column. We will also denote $\R[2]_{\pm}$ by $\Q^{\pm}_{s}$. A description of the sextonions in terms of Zorn matrices is given in  \cite{Marrani:2015nta}.

Regarding the split octonions as the eight-dimensional vector (equivalently, spinor or conjugate spinor) representation of $\f{so}(4,4)\cong\f{so}(\Oct_s)$, denoted $\rep{8}_v$, we can decompose with respect to two split-quaternion subalgebras $\f{so}(\Q_s)\oplus\f{so}(\Q_s)\cong\f{so}(2,2)\oplus\f{so}(2,2)\cong\f{sl}_2(\R)\oplus\f{sl}_2(\R)\oplus\f{sl}_2(\R)\oplus\f{sl}_2(\R)$:
 \be
 \begin{array}{cccccccccccccccccccc}
\f{so}(4,4) &=& \f{sl}_2(\R)\oplus\f{sl}_2(\R)\oplus\f{sl}_2(\R)\oplus\f{sl}_2(\R)&\pd  &\Q_s\otimes\Q_s\\[5pt]
 \rep{28}&= &\rep{(3,1,1,1}\oplus\rep{(1, 3,1,1)}\oplus\rep{(1, 1,3,1)}\oplus\rep{(1, 1, 1, 3)}&\pd  &\rep{(2,2,2,2)}
 \end{array}
 \ee
under which,
 \be
 \begin{array}{cccccccccccccccccccc}
\Oct_s &=& \Q_s&\pd  &\Q_s\\\
 \rep{8}_v&= &\rep{(2, 2, 1, 1)}& \pd  &\rep{(1, 1, 2, 2)}.
 \end{array}
 \ee
 Further decomposing one of the four $\f{sl}_2(\R)$'s (say, the last one) with respect to $\f{so}(1,1)$, we obtain the 5-grading
  \be\label{so8split}
 \begin{array}{cccccccccccccccccccc}
 \R_{\mii}&\pd  & [\Q_s\otimes\Q_{s}^{-}]_{\mi} &\pd  & [\f{sl}_2(\R)\oplus\f{sl}_2(\R)\oplus\f{sl}_2(\R)\oplus\f{so}(1,1)]_{\0}&\pd  &[\Q_s\otimes\Q_{s}^{+}]_{\1}&\pd  &\R_{\2}\\[5pt]
 \rep{1}_{\mii}&\pd  &\rep{(2,2,2)}_{\mi} &\pd  & [\rep{(3,1,1)}\oplus\rep{(1, 3,1)}\oplus\rep{(1, 1,3)}\oplus\rep{(1, 1, 1)}]_{\0}&\pd  &\rep{(2,2,2)}_{\1}&\pd  &\rep{1}_{\2}
 \end{array}
 \ee
and 3-grading
 \be\label{octsplit}
 \begin{array}{cccccccccccccccccccc}
\Oct_s &=\Q_{s\mi}^{+}&\pd & \Q_{s\0}&\pd  &\Q_{s\1}^{+}\\[5pt]
 \rep{8}_v&= \rep{(1, 1, 2)}_{\mi}&\pd & \rep{(2, 2, 1)}_{\0}& \pd  &\rep{(1, 1, 2)}_{\1}.
 \end{array}
 \ee
 In this language, the sextonions are the positive grade component of \eqref{octsplit},
 \be\label{sext}
 \s_s\equiv \rep{(2, 2, 1)}_{\0} \pd  \rep{(1, 1, 2)}_{\1}.
 \ee
In this regard, $\s_s$ is isomorphic to an intermediate algebra $\f{int}(\Oct_s)$ of   Kantor-Koecher-Tits form.

\paragraph{Orthogonal algebra:} The sextonion-preserving subalgebra $\f{so}_{\s_s}(\Oct_s)\subset\f{so}(\Oct_s)$   is given by the positive grade component of \eqref{so8split}, yielding the non-reductive Lie algebra,
  \be\label{soso}
 \begin{array}{cccccccccccccccccccc}
\f{so}_{\s_s}(\Oct_s)\cong  & [\f{sl}_2(\R)\oplus\f{sl}_2(\R)\oplus\f{sl}_2(\R)\oplus\f{so}(1,1)]_{\0}&\pd  &[\Q_s\otimes\Q_{s}^{+}]_{\1}&\pd  &\R_{\2}\\[5pt]
 & [\rep{(3,1,1)}\oplus\rep{(1, 3,1)}\oplus\rep{(1, 1,3)}\oplus\rep{(1, 1, 1)}]_{\0}&\pd  &\rep{(2,2,2)}_{\1}&\pd  &\rep{1}_{\2}
 \end{array}
 \ee
which is the extremal intermediate algebra $\f{ext}(\f{so}(\Oct_s))$, as described in \eqref{com}. It follows immediately from the gradings that the sextonion subalgebra \eqref{sext} is annihilated by, and only by, the grade two element in \eqref{soso}. Hence, the restriction homomorphism $\f{so}_{\s_s}(\Oct_s)\rightarrow\f{so}(\s_s)$ is surjective with 1-dimensional kernel. Then, $\f{so}(\s_s)$ is given by quotient map $\f{ext}(\f{g})\rightarrow\widetilde{\mathfrak{ext}}(\f{g})$ described in \eqref{com} for $\f{g}=\f{so}(\Oct_s)$:
\be
\f{so}(\s_s)\cong\widetilde{\f{ext}}(\f{so}(\Oct_s))\cong\f{ext}(\f{so}(\Oct_s))/\R_{\2}\cong [\f{sl}_2(\R)\oplus\f{sl}_2(\R)\oplus\f{sl}_2(\R)\oplus\f{so}(1,1)]_{\0}\pd[\Q_s\otimes\Q_{s}^{+}]_{\1}.
\ee
Note, although as a vector space $\f{so}(\s_s)$ can be identified with the grade zero and one components of $\f{so}_{\s_s}(\Oct_s)$, their Lie algebra structures are different due to the quotient by the grade two ideal $\R_{\2}$.

\paragraph{Derivation algebra:} The derivation algebra of $\s_s$ was described in detail in \cite{westbury2006sextonions, Marrani:2015nta}. We briefly recall these results here. For $\f{der}(\Oct_s)\cong \f{g}_{2(2)}$ we have,
 \be
 \begin{array}{cccccccccccccccccccc}
 \f{g}_{2(2)} &=& \R_{\mii}&\pd  & \R^{4}_{\mi} &\pd  & [\f{gl}_1(\R)\oplus\f{sl}_2(\R)]_{\0}&\pd  &\R^{4}_{\1}&\pd  &\R_{\2}\\[5pt]
 \rep{14}&=&  \rep{1}_{\mii}&\pd  & \rep{4}_{\mi} &\pd  & [\rep{1}\oplus\rep{3}]_{\0}&\pd  &\rep{4}_{\1}&\pd  &\rep{1}_{\2}
 \end{array}
 \ee
 while, as a $ \f{g}_{2(2)}$ representation, the split-octonions
 \be
  \begin{array}{cccccccccccccccccccc}
\Oct_s &=& \text{Re}(\Oct_s)&\pd  &\text{Im}(\Oct_s)\\[5pt]
 \rep{8}&=&  \rep{1}&\pd  & \rep{7}
 \end{array}
 \ee
branch under $\f{gl}_1(\R)\oplus\f{sl}_2(\R) \subset \f{sl}_2(\R)\oplus\f{sl}_2(\R)\subset \f{g}_{2(2)}$ as
\be
 \rep{1}  \pd \rep{7}\rightarrow \rep{2}_{\mi}\pd \underbrace{\rep{1}_{\0}\pd \rep{3}_{\0} \pd   \rep{2}_{\1}}_{\s_s}
\ee
 so that $\f{der}_{\s_s}(\Oct_s)\subset\f{der}(\Oct_s)$ is given by
 \be
 \f{der}_{\s_s}(\Oct_s) \cong     \f{ext}(\f{g}_{2(2)})=[\f{gl}_1(\R)\oplus\f{sl}_2(\R)]_{\0}\pd  \R^{4}_{\1}\pd  \R_{\2}
 \ee
 and $\f{der}(\s_s)$ is obtained by taking the quotient  of the grade two 1-dimensional ideal annihilating $\s_s$,
  \be
 \f{der}(\s_s) \cong   \f{ext}(\f{g}_{2(2)})/ \R_{\2}= \widetilde{\f{ext}}(\f{g}_{2(2)})=[\f{gl}_1(\R)\oplus\f{sl}_2(\R)]_{\0}\pd  \R^{4}_{\1}.
 \ee

\paragraph{Triality algebra:} For three $\alpha_i\in\f{so}(\Oct_s)$ satisfying the triality condition
 \be
 \alpha_1(ab)=\alpha_2(a)b+a\alpha_3(b)
 \ee
$\alpha_j$ and $\alpha_k$ are uniquely determined by $\alpha_i$, $i\not=j\not=k$. Hence, $\f{tri}(\Oct_s)\cong\f{so}(\Oct_s)$. This property remains true when restricting to any sextonionic subalgebra and, therefore, we have,
\be
\f{tri}_{\s_s}(\Oct_s)\cong\f{ext}(\f{tri}(\Oct_s))\cong [\f{sl}_2(\R)\oplus\f{sl}_2(\R)\oplus\f{sl}_2(\R)\oplus\f{so}(1,1)]_{\0}\pd[\Q_s\otimes\Q_{s}^{+}]_{\1}\pd\R_{\2}
\ee
and
\be
\f{tri}(\s_s)\cong\f{ext}(\f{tri}(\Oct_s))/\R_{\2}=\widetilde{\f{ext}}(\f{tri}(\Oct_s))= [\f{sl}_2(\R)\oplus\f{sl}_2(\R)\oplus\f{sl}_2(\R)\oplus\f{so}(1,1)]_{\0}\pd[\Q_s\otimes\Q_{s}^{+}]_{\1}.
\ee
In \cite{landsberg2006sextonions}, the extended magic square is constructed using the intermediate algebra
\be
\f{tri}_{\s_s}(\Oct_s)_\sz:=\f{ext}_\sz(\f{tri}(\Oct_s))\cong \f{sl}_2(\R)\oplus\f{sl}_2(\R)\oplus\f{sl}_2(\R)\pd\Q_s\otimes\Q_{s}^{+}\pd\R,
\ee
and $\f{tri}_{\s_s}(\Oct_s)_\sz$ was denoted $\f{tri}^\star(\s_s)$ therein. Note, although as vector spaces $\f{tri}_{\s_s}(\Oct_s)_\sz$ and $\f{tri}(\s_s)$ are the same, they have distinct Lie algebra structures. In particular, $\f{tri}_{\s_s}(\Oct_s)_\sz$ is a subalgebra of $\f{tri}(\Oct_s)$, while there is no natural inclusion of $\f{tri}(\s_s)$ into $\f{tri}_{\s_s}(\Oct_s)$, and thus into $\f{tri}(\Oct_s)$.

\section{The extended magic square}\label{msquare}

The standard magic square can be defined equivalently \cite{Barton:2003} using the Barton-Sudbery \cite{Barton:2003} (also cfr. \cite{Evans:2009ed}), Vinberg \cite{Vinberg:1966} and Tits \cite{Tits:1966} constructions,
\be\label{msconstructions}
\mathfrak{m}(\alg_1, \alg_2)=\left\{\begin{array}{l} \f{tri}(\alg_1)\oplus\f{tri}(\alg_2) \pd   3 (\alg_1\otimes\alg_2),\\[5pt] \f{der}(\alg_1)\oplus\f{der}(\alg_2) \pd   \f{sa}(3, \alg_1\otimes \alg_2),\\ [5pt]
\f{der}(\alg_1)\oplus\f{der}(\J_{3}(\alg_2) ) \pd   \text{Im} \alg_1\otimes \J^{'}_{3}(\alg_2),\end{array}\right.
\ee
where $\alg_1, \alg_2 = \R, \C,  \Q, \Oct$. Here, $\J_3(\alg)$ denotes the Jordan algebra of $3\times3$ Hermitian matrices over $\alg$ and $\J'_{3}(\alg)$ its subspace of traceless elements.  The space of anti-Hermitian traceless $n\times n$ matrices over $\alg_1\otimes \alg_2$ is denoted $\f{sa}(n, \alg_1\otimes \alg_2)$. Details of the commutators and the isomorphisms between these Lie algebras can be found in \cite{Barton:2003}. The choice of division or split $\alg_1, \alg_2$ yields different  real forms. For the Tits construction, further possibilities are given by allowing the Jordan algebra to be Lorentzian, denoted $\J_{2,1}$, as described in \cite{Cacciatori:2012cb}; equivalently, one can introduce overall signs in the definition of the commutators between  distinct components in the Vinberg or Barton-Sudbery constructions, as described in \cite{Anastasiou:2013hba}. In \autoref{complexms} below, we will begin by discussing the complexified algebras, before  presenting the various reals forms in \autoref{realms}.

\subsection{The extended magic square: complex}\label{complexms}
In this section we will use $\alg$ to denote  the complexified composition algebras. The extended magic square can be obtained by including $\T$ and $\s$ in any one of the three constructions \eqref{msconstructions}.  We will discuse each in turn, first for the sextonions, which are slightly more subtle, and then for the ternions.

\paragraph{Sextonions} We start with the Barton-Sudbery triality construction given in \cite{landsberg2006sextonions}. Requiring  the entries $\mathfrak{m}(\s, \alg)$ to be  Lie subalgebras of $\mathfrak{m}(\Oct, \alg)$ there are two possible definitions  of the sextonionic column:
 \be\label{tri-pre}
  \begin{array}{lllllllllllll}
\mathfrak{m}(\s, \alg)&:=& \f{tri}_\s(\Oct)\oplus\f{tri}(\alg) &\pd &  3 (\s\otimes\alg),\\[5pt]
\mathfrak{m}(\s, \alg)_\sz&:=&   \f{tri}_\s(\Oct)_\sz \oplus\f{tri}(\alg) &\pd &  3 (\s\otimes\alg).
 \end{array}
 \ee
Using the obvious definition, the sextonionic row is given by $\mathfrak{m}(\alg, \s)_{(\sz)}$, which is isomorphic to $\mathfrak{m}(\s, \alg)_{(\sz)}$.  Note,  while one could in principle use either $\mathfrak{m}(\s, \alg)$ or  $\mathfrak{m}(\s, \alg)_\sz$ to define the sextonionic entries of the magic square, both being subalgebras in $\mathfrak{m}(\Oct, \alg)$,  it is the latter which satisfies the Deligne dimension formulae \cite{landsberg2006sextonions}. In fact, there are two further possibilities,
 \be\label{tri}
  \begin{array}{lllllllllllll}
\widetilde{\mathfrak{m}}(\s, \alg)&:=& \f{tri}(\s)\oplus\f{tri}(\alg) &\pd &  3 (\s\otimes\alg),\\[5pt]
\widetilde{\mathfrak{m}}(\s, \alg)_\sz&:=&   \f{tri}(\s)_\sz \oplus\f{tri}(\alg) &\pd &  3 (\s\otimes\alg),
 \end{array}
 \ee
  which are not  subalgebras of $\mathfrak{m}(\Oct, \alg)$, but are  relevant to supergravity, as will be discussed in \autoref{sugra}.

Similarly, there are two possible definitions  of the sextonionic row in terms of the Vinberg construction, which yield the same algebras:
  \be\label{vin}
  \begin{array}{lllllllllllll}
\mathfrak{m}(\s, \alg)&\cong&\f{der}_\s(\Oct) \oplus\f{der}(\alg) &\pd &  \f{sa}(3, \s\otimes\alg),\\[5pt]
\mathfrak{m}(\s, \alg)_\sz&\cong&  \f{der}_\s(\Oct)_\sz\oplus\f{der}(\alg) &\pd &  \f{sa}(3, \s\otimes\alg),
 \end{array}
 \ee
where we have defined $\f{der}_\s(\Oct)_{(\sz)}:=\f{ext}_{(\sz)}(\f{der}(\Oct))$ in the sense of \eqref{com}. Again, using the obvious definition, the sextonionic column is given by $\mathfrak{m}(\alg, \s)_{(\sz)}$, which is isomorphic to $\mathfrak{m}(\s, \alg)_{(\sz)}$.

Finally, the Tits construction was shown in \cite{westbury2006sextonions} to yield $\mathfrak{m}(\s, \alg)$ as given by  \eqref{tri} and \eqref{vin}, using:
\be\label{tits}
  \begin{array}{lllllllllllll}
\mathfrak{m}(\s, \alg)&\cong& \f{der}_\s(\Oct)\oplus\f{der}(\J_{3}(\alg) )& \pd &  \text{Im} \s\otimes \J'_{3}(\alg),\\[5pt]
\mathfrak{m}(\s, \alg)_{\sz}&\cong& \f{der}_\s(\Oct)_\sz\oplus\f{der}(\J_{3}(\alg) )& \pd &  \text{Im} \s\otimes \J'_{3}(\alg).
 \end{array}
\ee
Unlike the Barton-Sudbery and Vinberg formulae, the Tits construction is not manifestly symmetric under the interchange of $\alg_1$ and $\alg_2$ and, hence, it is not \emph{a priori} true that $\mathfrak{m}(\alg, \s)\cong\mathfrak{m}(\s, \alg)$ when using \eqref{tits}.  The interchange symmetry does  however  hold
so long as we use, instead of $\f{der}(\J_{3}(\s))$, the $\J_{3}(\s)$-preserving subalgebra $\f{der}_{\J_{3}(\s)}(\J_{3}(\Oct))\subset\f{der}(\J_{3}(\Oct) )$ in the definition of $\mathfrak{m}(\alg, \s)$:
\be\label{suff1}
\mathfrak{m}(\alg, \s)_{\sz}:= \f{der}(\alg)\oplus\f{der}_{\J_{3}(\s)}(\J_{3}(\Oct) ) \pd   \text{Im} \alg\otimes \J'_{3}(\s).
\ee  This is most readily checked by decomposing $\f{der}(\J_3(\Oct))\cong \f{f}_{4}$ with respect to a extremal triple with centralizer  $\f{der}(\J_{3}(\Q))\cong \f{sp}_6$, which gives  the 5-grading:
\be\label{derref}
\begin{array}{ccccccccccccccccccc}
\f{f}_4&=&\C_{\mii} &\pd  & V_{\mi}&\pd  &[\f{gl}_1(\C)\oplus\f{sp}_6]_{\0}&\pd  & V_{\1}&\pd  &\C_{\2}\\[5pt]
\rep{52}&\rightarrow&\rep{1}_{\mii}&\pd  &\rep{14'}_{\mi}&\pd  &[\rep{1}\oplus  \rep{21}]_{\0}&\pd  &\rep{14'}_{\1}&\pd  &\rep{1}_{\2}.
\end{array}
\ee
The Jordan algebra $\J_3(\Oct)$ transforms reducibly under $\f{der}(\J_3(\Oct))\cong \f{f}_4$ as a $\rep{1}\pd\rep{26}$, which branches with respect to the same extremal triple as,
\be
\begin{array}{ccccccccccccccccccc}
\J_3(\Oct)&=&\C\mathds{1}&\pd& \J'_3(\Oct)&=&\overline{\J}_{3}(\Q^-) &\pd    &\C\mathds{1}&\pd  & \J'_3(\Q)&\pd  &\J_{3}(\Q^+)\\[5pt]
\rep{27}&=&\rep{1}&\pd&\rep{26}&\rightarrow&\rep{6}_{\mi}& \pd  &\rep{1}_{\0} &\pd  &\rep{14}_{\0}&\pd  &\rep{6}_{\1},
\end{array}
\ee
where $\J_{3}(\Q^\pm)$ is the space of $3\times3$ Hermitian matrices over the 2-dimensional pure imaginary left (right) $\Q$-module, $\Q^\pm$, whereas $\rep{14}$ and $\rep{14'}$ respectively denote the rank-2 and rank-3 skew-symmetric irreprs. of $\f{sp}_6$. A maximal ($3+3 \dim \s =21$)-dimensional subalgebra $\J_3(\s)$ is given by the positive grade components
$\rep{1}_{\0} \pd  \rep{14}_{\0} \pd \rep{6}_{\1}$, while $\J'_3(\s)$ can be identified with $\rep{14}_{\0} \pd \rep{6}_{\1}$. In this regard, $\J_3(\s)$ is isomorphic to an intermediate algebra $\f{int}(\J_3(\Oct))$ of   Kantor-Koecher-Tits form, just as $\s$ is to  $\Oct$: $\s \cong \f{int}(\Oct)$. Hence, $\f{der}_{\J_3(\s)}(\J_3(\Oct))$ is given by the positive grade components of \eqref{derref}, and is an extremal intermediate algebra in the sense of \eqref{com},
\be
\f{der}_{\J_3(\s)}(\J_3(\Oct))\cong \f{ext}(\f{der}(\J_3(\Oct)))\cong [\f{gl}_1(\C)\oplus\f{sp}_6]_{\0}\pd   V_{\1}\pd  \C_{\2}.
\ee
It also follows that
\be
\f{der}(\J_3(\s))\cong \widetilde{\f{ext}}(\J_3(\Oct))\cong  \f{ext}(\f{der}(\J_3(\Oct)))/\C\cong [\f{gl}_1(\C)\oplus\f{sp}_6]_{\0}\pd  V_{\1}
\ee
and we can define,
\be
\f{der}_{\J_3(\s)}(\J_3(\Oct))_\sz:=\f{ext}_\sz(\f{der}(\J_3(\Oct)))\cong \f{sp}_6 \pd   V \pd  \C.
\ee
We then have that $\mathfrak{m}(\alg, \s)_{(\sz)}$ given in \eqref{suff1} is isomorphic to $\mathfrak{m}(\s, \alg)_{(\sz)}$ given by \eqref{tits}. Hence, the four possibilities, ${\mathfrak{m}}(\alg_1, \alg_2)_{(\sz)}$  and $\widetilde{\mathfrak{m}}(\alg_1, \alg_2)_{(\sz)}$, can be constructed using any one of three formulae \eqref{msconstructions}. All three  yield the same non-reductive Lie algebras which can be identified  as  extremal  intermediate algebras of the octonionic row/column of the magic square:
\be\label{four-S}
  \begin{array}{lllllllllllll}
\mathfrak{m}(\alg, \s)&\cong &\mathfrak{m}(\s, \alg)&\cong& \f{ext}(\mathfrak{m}(\alg, \Oct)),\\[5pt]
\mathfrak{m}(\alg, \s)_\sz&\cong &\mathfrak{m}(\s, \alg)_\sz&\cong& \f{ext}_\sz(\mathfrak{m}(\alg, \Oct)),\\[5pt]
\widetilde{\mathfrak{m}}(\alg, \s)&\cong &\widetilde{\mathfrak{m}}(\s, \alg)&\cong& \widetilde{\mathfrak{ext}}(\mathfrak{m}(\alg, \Oct)),\\[5pt]
\widetilde{\mathfrak{m}}(\alg, \s)_\sz&\cong &\widetilde{\mathfrak{m}}(\s, \alg)&\cong& \widetilde{\mathfrak{ext}}_\sz(\mathfrak{m}(\alg, \Oct)).
 \end{array}
 \ee

The relevant 5-graded decompositions and resulting entries in the magic square are summarised below. First, we have the decompositions,
\be
\begin{array}{ccccccccccccccccccc}
\f{f}_4&=&\C_{\mii} &\pd  & V_{\mi}&\pd  &[\f{gl}_1\oplus\f{sp}_6]_{\0}&\pd  & V_{\1}&\pd  &\C_{\2}\\[5pt]
\rep{52}&\rightarrow&\rep{1}_{\mii}&\pd  &\rep{14'}_{\mi}&\pd  &[\rep{1}\oplus  \rep{21}]_{\0}&\pd  &\rep{14'}_{\1}&\pd  &\rep{1}_{\2}\\
\\
\f{e}_6&=&\C_{\mii} &\pd  & V_{\mi}&\pd  &[\f{gl}_1\oplus\f{sl}_6]_{\0}&\pd  & V_{\1}&\pd  &\C_{\2}\\[5pt]
\rep{78}&\rightarrow&\rep{1}_{\mii}&\pd  &\rep{20}_{\mi}&\pd  &[\rep{1}\oplus  \rep{35}]_{\0}&\pd  &\rep{20}_{\1}&\pd  &\rep{1}_{\2}\\
\\
\f{e}_7&=&\C_{\mii} &\pd  & V_{\mi}&\pd  &[\f{gl}_1\oplus\f{so}_{12}]_{\0}&\pd  & V_{\1}&\pd  &\C_{\2}\\[5pt]
\rep{133}&\rightarrow&\rep{1}_{\mii}&\pd  &\rep{32}'_{\mi}&\pd  &[\rep{1}\oplus  \rep{66}]_{\0}&\pd  &\rep{32}'_{\1}&\pd  &\rep{1}_{\2}\\
\\
\f{e}_8&=&\C_{\mii} &\pd  & V_{\mi}&\pd  &[\f{gl}_1\oplus\f{e}_7]_{\0}&\pd  & V_{\1}&\pd  &\C_{\2}\\[5pt]
\rep{248}&\rightarrow&\rep{1}_{\mii}&\pd  &\rep{56}_{\mi}&\pd  &[\rep{1}\oplus  \rep{133}]_{\0}&\pd  &\rep{56}_{\1}&\pd  &\rep{1}_{\2}\\
\end{array}
\ee
For the $(\s, \s)$ entry, we take a second extremal $\f{sl}_2$ triple inside the 0-grade $\f{e}_{7\0}\in \f{e}_8$, which commutes with the first yielding the bigrading,
\be
\begin{array}{ccccccccccccccccccc}
\f{e}_8&=&\begin{array}{ccccccccccccccccccc}
& &\C_{(0,2)}& &\\
& V_{(-1,1)}&S'_{(0,1)}& V_{(1,1)}&\\
 \C_{(-2,0)} & S_{(-1,0)}&[\f{gl}_1\oplus\f{gl}_1\oplus\f{so}_{12}]_{(0,0)}& S_{(1,0)}&\C_{(2,0)}\\
 & V_{(-1,-1)}&S'_{(0,-1)}& V_{(1,-1)}&\\
 & &\C_{(0,-2)}& &\\
 \end{array}
 \\[8pt]
 \\
\rep{248}&\rightarrow&\begin{array}{ccccccccccccccccccc}
& &\rep{1}_{(0,2)}& &\\
& \rep{12}_{(-1,1)}&\rep{32'}_{(0,1)}& \rep{12}_{(1,1)}&\\
 \rep{1}_{(-2,0)} & \rep{32}_{(-1,0)}&[\rep{1}\oplus\rep{1}\oplus\rep{66}]_{(0,0)}& \rep{32}_{(1,0)}&\rep{1}_{(2,0)}\\
 &\rep{12}_{(-1,-1)}&\rep{32'}_{(0,-1)}& \rep{12}_{(1,-1)}&\\
 & &\rep{1}_{(0,-2)}& &\\
 \end{array}\\
\end{array}
\ee
Note that, unlike the other $\alg, \s$ cases, there are non-trivial  commutators amongst  the strictly positive grade $(1,0)$ and $(0,1)$  components which are not given by a symplectic invariant: $[S'_{10}, S_{01}]\subseteq V_{(11)}$.

For $\f{m}(\s, \alg)\cong \f{m}(\alg, \s)$ we obtain the extremal intermediate algebras $\f{ext}(\f{m}(\Oct, \alg))$:
\be\label{ppain-1}
\begin{array}{cllllllllllllllll}
\alg&\f{m}(\s, \alg)\\[8pt]
\R&\f{sp}_{6\frac{1}{2}} &:=&  [\f{gl}_1\oplus\f{sp}_{6}]_{\0} \pd   \rep{14}'_{\1}\pd   \rep{1}_\2 \\
&&=&  [\f{gl}_1\oplus\f{sp}_{6}] \ltimes H_{14}\\[10pt]
\C&\f{sl}_{6\frac{1}{2}} &:= &[\f{gl}_1\oplus\f{sl}_{6}]_{\0}  \pd   \rep{20}_{\1}\pd   \rep{1}_\2 \\
&&=& [\f{gl}_1\oplus\f{sl}_{6}] \ltimes H_{20}\\[10pt]
\Q&\f{so}_{12\frac{1}{2}} &:=&[\f{gl}_1\oplus\f{so}_{12}]_{\0}\pd   \rep{32'}_{\1}\pd   \rep{1}_\2\\
&&=&  [\f{gl}_1\oplus\f{so}_{12}] \ltimes H_{32}\\[10pt]
\s&\f{so}_{12(\frac{1}{2}+  \frac{1}{2})}&:=& [\f{gl}_1\oplus\f{gl}_1\oplus\f{so}_{12}]_{(0,0)}\pd  \rep{32} _{(1,0)}\pd  \rep{32}' _{(0,1)}\pd \rep{12} _{(1,1)} \pd \rep{1}_{(0,2)} \pd \rep{1}_{(2,0)} \\
&&=&  [\f{gl}_1\oplus\f{gl}_1\oplus\f{so}_{12} ]\ltimes H_{32}\ltimes H_{32+  12}\\[10pt]
\Oct&\f{e}_{7\frac{1}{2}} &:=&  [\f{gl}_1\oplus\f{e}_{7}]_{\0} \pd   \rep{56}_{\1}\pd   \rep{1}_\2 \\
&&=& [\f{gl}_1\oplus\f{e}_{7}] \ltimes H_{56},
\end{array}
\ee
which constitute the sextonionic row/column of the extended magic square given in  \autoref{tab:ems1}.

For $\f{m}(\s, \alg)_\sz \cong \f{m}(\alg, \s)_\sz$, we obtain the derived extremal intermediate algebras $\f{ext}_\sz(\f{m}(\Oct, \alg))$:
\be
\begin{array}{cllllllllllllllll}
\alg&\f{m}(\s, \alg)_\sz\\[8pt]
\R&\f{sp}^{\sz}_{6\frac{1}{2}} &=& \f{sp}_{6} \pd   \rep{14}'\pd   \rep{1} &=& \f{sp}_{6} \ltimes H_{14}\\[5pt]
\C&\f{sl}^{\sz}_{6\frac{1}{2}} &= &\f{sl}_{6}  \pd   \rep{20}\pd   \rep{1} &=&\f{sl}_{6} \ltimes H_{20}\\[5pt]
\Q&\f{so}^{\sz}_{12\frac{1}{2}} &=&\f{so}_{12}\pd   \rep{32'}\pd   \rep{1}&=& \f{so}_{12} \ltimes H_{32}\\[5pt]
\s&\f{so}^{\sz}_{12(\frac{1}{2}+  \frac{1}{2})}&=&\f{so}_{12}\pd  [\rep{32'} \pd  \rep{1}] \pd   [\rep{32} \pd  \rep{12}
\pd  \rep{1}]&=& \f{so}_{12} \ltimes H_{32}\ltimes H_{32+  12}\\[5pt]
\Oct&\f{e}^{\sz}_{7\frac{1}{2}} &=&  \f{e}_{7} \pd   \rep{56}\pd   \rep{1} &=& \f{e}_{7} \ltimes H_{56}
\end{array}
\ee

\paragraph{Ternions} The ternion row/column is given in much the same way. Again, there are multiple (in this case, two) possible definitions for each of the three magic square formulae \eqref{msconstructions} (for the sextonionic row/column, there are four possibilities \eqref{four-S}). The principal difference with respect to the sextonionic case is that we can use the ternion algebras $\f{der}(\T)$ and $\f{tri}(\T)$ directly in the definition of $\mathfrak{m}(\T, \alg)$ since they are subalgebras of $\f{der}(\Q)$ and $\f{tri}(\Q)$, respectively. Recall, $\f{der}(\T)$ and $\f{tri}(\T)$ are the Kantor-Koecher-Tits intermediate  algebras \eqref{3grade k }  of $\f{der}(\Q)$ and $\f{tri}(\Q)$, respectively. The second possibility $\mathfrak{m}(\T, \alg)_\sz$ is given by using the derived subalgebras, in the sense of \eqref{derived3grade k }, denoted $\f{der}(\T)_\sz$ and $\f{tri}(\T)_\sz$. It is the former that we use in the definition of the extended magic square given in \autoref{tab:ems1}.

Using the Barton-Sudbery formula we have the columns:
 \be\label{pain-1}
  \begin{array}{lllllllllllll}
\mathfrak{m}(\T, \alg)&=& \f{tri}(\T)\oplus \f{tri}(\alg)&\pd &  3 (\T\otimes\alg),\\[5pt]
\mathfrak{m}(\T, \alg)_{\sz}&=&   \f{tri}(\T)_\sz\oplus \f{tri}(\alg)&\pd &  3 (\T\otimes\alg).
 \end{array}
 \ee
Similarly,  the Vinberg construction gives:
  \be\label{pain-2}
  \begin{array}{lllllllllllll}
\mathfrak{m}(\T, \alg)&=& \f{der}(\T)\oplus \f{der}(\alg)&\pd &  \f{sa}(3, \T\otimes\alg),\\[5pt]
\mathfrak{m}(\T, \alg)_{\sz}&=&  \f{der}(\T)_\sz  \oplus\f{der}(\alg)&\pd &  \f{sa}(3, \T\otimes\alg).
 \end{array}
 \ee
The manifest symmetry in both cases leads to the obvious definitions of the rows $\mathfrak{m}(\alg, \T)_{(\sz)}$. The equivalence of the two constructions can be established following the same arguments used in \cite{Barton:2003} for the standard magic square.

Finally, the Tits construction for $\mathfrak{m}(\T, \alg)_{(\sz)}$ yields the columns (isomorphic to \eqref{pain-1} and \eqref{pain-2})
  \be\label{pain-3}
  \begin{array}{lllllllllllll}
\mathfrak{m}(\T, \alg)&=& \f{der}(\T)\oplus \f{der}(\J_3(\alg))&\pd &  \text{Im}(\T)\otimes\J'_3(\alg),\\[5pt]
\mathfrak{m}(\T, \alg)_{\sz}&=&  \f{der}(\T)_\sz  \oplus\f{der}(\J_3(\alg))&\pd &   \text{Im}(\T)\otimes\J'_3(\alg).
 \end{array}
 \ee
Again, to define the rows we must examine $\J_{3}(\T)$. Recall, as a $\f{der}(\J_3(\Q))\cong\f{sp}_6$ module, $\J_3(\Q)$ transforms reducibly as a $\rep{1}\pd\rep{14}$ and decomposes under $\f{der}(\J_3(\C))\oplus\f{gl}_1(\C)\subset\f{der}(\J_3(\Q))$ as ($\f{der}(\J_3(\C))\cong \f{sl}_3)$
\be\label{pain-3}
\underbrace{\rep{1}\pd\rep{14}}_{\J_3(\Q)}\rightarrow \overline{\rep{3}}_{\mii}\pd\underbrace{\rep{1}_{\0}\pd\underbrace{\rep{8}_{\0}\pd\rep{3}_{\2}}_{\J'_3(\T)}}_{\J_3(\T)}
\ee
so that a maximal ($3+3\dim\T=12$)-dimensional $\J_3(\T)$ subalgebra can be identified with the positive grade components in \eqref{pain-3}. Hence, from the treatment of Sec. 2.1, $\J_3(\T)$ is isomorphic to an intermediate algebra $\f{int}(\J_3(\Q))$ of the  Kantor-Koecher-Tits type, just as $\T$ is of $\Q$ (cfr. \eqref{pain-0}).
 Decomposing $\f{der}(\J_3(\Q))\cong \f{sp}_6$ with respect to $\f{der}(\J_3(\C))\cong \f{sl}_3$ according as the Kantor-Koecher-Tits  construction we obtain,
\be\label{derref}
\begin{array}{ccccccccccccccccccc}
\f{sp}_6&=&\overline{V}_{\mii} &\pd  &[\f{gl}_1(\C)\oplus\f{sl}_3(\C)]_{\0}&\pd    &V_{\2}\\[5pt]
\rep{21}&\rightarrow&\overline{\rep{6}}_{\mii}&\pd    &[\rep{1}\oplus  \rep{8}]_{\0}&\pd   &\rep{6}_{\2},
\end{array}
\ee
and we conclude that (recall \eqref{3grade k }):
\be
\f{der}_{\J_3(\T)}(\J_3(\Q))\cong \f{ k }(\f{der}(\J_3(\Q))\cong [\f{gl}_1\oplus\f{sl}_3]_{\0}\pd   V_{\2}\cong \f{der}(\J_3(\T)),
\ee
where $V$ is given by $\text{Sym}^{2}(\C^3)$. As before,  we define (recall \eqref{derived3grade k }):
\be
 \f{der}(\J_3(\T))_\sz:=\f{ k }_\sz(\f{der}(\J_3(\Q))\cong  \f{sl}_3 \pd   V.
\ee
We  then have
  \be
  \begin{array}{lllllllllllll}
\mathfrak{m}(\alg, \T)&=& \f{der}(\alg)\oplus \f{der}(\J_3(\T))&\pd &  \text{Im}(\alg)\otimes\J'_3(\T),\\[5pt]
\mathfrak{m}(\alg, \T)_\sz&=&  \f{der}(\alg)  \oplus\f{der}(\J_3(\T))_\sz&\pd &   \text{Im}(\alg)\otimes\J'_3(\T).
 \end{array}
 \ee
so that both $\mathfrak{m}(\alg_1, \alg_2)$ and $\mathfrak{m}(\alg_1, \alg_2)_\sz$ can be constructed using the Tits formula for either $\alg_1$ or $\alg_2$ equal to $\T$. Thus, all three constructions \eqref{msconstructions} yield the same non-reductive Lie algebras which can be identified as the Kantor-Koecher-Tits intermediate algebras of the quaternionic row/column of the magic square:
\be
  \begin{array}{lllllllllllll}
\mathfrak{m}(\alg, \T)&\cong &\mathfrak{m}(\T, \alg)&\cong& \f{ k }(\mathfrak{m}(\alg, \Q))\\[5pt]
\mathfrak{m}(\alg, \T)_\sz&\cong &\mathfrak{m}(\T, \alg)_\sz&\cong& \f{ k }_\sz(\mathfrak{m}(\alg, \Q)).
 \end{array}
 \ee
The relevant Kantor-Koecher-Tits 3-graded decompositions and resulting entries in the magic square are summarised below. First, we have the decompositions:
\be
\begin{array}{ccccccccccccccccccc}
\f{sp}_6&=& V_{\mii}&\pd  &[\f{gl}_1\oplus\f{sl}_3]_{\0}&\pd  & V_{\2}\\[5pt]
\rep{21}&\rightarrow&\overline{\rep{6}}_{\mii}&\pd  &[\rep{1}\oplus  \rep{8}]_{\0}&\pd  &\rep{6}_{\2}\\
\\
\f{sl}_6&=& V_{\mii}&\pd  &[\f{gl}_1\oplus\f{sl}_3\oplus\f{sl}_3]_{\0}&\pd  & V_{\2}\\[5pt]
\rep{35}&\rightarrow&{\rep{(\overline{3}, 3)}}_{\mii}&\pd  &[\rep{(1,1)}\oplus  \rep{(8,1)}\oplus  \rep{(1,8)}]_{\0}&\pd  &\rep{(3, \overline{3})}_{\2}\\
\\
\f{so}_{12}&=& V_{\mii}&\pd  &[\f{gl}_1\oplus\f{sl}_6]_{\0}&\pd  & V_{\2}\\[5pt]
\rep{66}&\rightarrow&\overline{\rep{15}}_{\mii}&\pd  &[\rep{1}\oplus  \rep{35}]_{\0} &\pd  &\rep{15}_{\2}\\
\\
\f{e}_7&=& V_{\mii}&\pd  &[\f{gl}_1\oplus\f{e}_6]_{\0}&\pd  & V_{\2}\\[5pt]
\rep{133}&\rightarrow&\overline{\rep{27}}_{\mii}&\pd  &[\rep{1}\oplus  \rep{78}]_{\0}&\pd  &\rep{27}_{\2}\\
\end{array}
\ee
For the $(\T, \T)$ entry, we take a second Jordan pair inside the 0-grade $\f{sl}_{6(0)}\in \f{so}_{12}$ yielding the bigrading,
\be
\begin{array}{ccccccccccccccccccc}
\f{so}_{12}&=&\begin{array}{ccccccccccccccccccc}
 V_{(-2,2)}&S'_{(0,2)}& V_{(2,2)}&\\
 \overline{S}_{(-2,0)}&[\f{gl}_1\oplus\f{gl}_1\oplus\f{sl}_3\oplus\f{sl}_3]_{(0,0)}& S_{(2,0)}\\
  V_{(-2,-2)}&\overline{S}'_{(0,-2)}& V_{(2,-2)}&\\
 \end{array}
 \\[8pt]
 \\
\rep{66}&\rightarrow&\begin{array}{ccccccccccccccccccc}
 \rep{(1, {3})}_{(-2,2)}&\rep{(3,\overline{3})}_{(0,2)}& \rep{(\overline{3}, 1)}_{(2,2)}\\
 \rep{(\overline{3}, \overline{3})}_{(-2,0)}&[\rep{(1,1)}\oplus\rep{(1,1)}\oplus\rep{(8, 2)}\oplus\rep{(2, 8)}]_{(0,0)}& \rep{(3,3)}_{(2,0)}\\
 \rep{({3}, 1)}_{(-2,-2)}&\rep{(\overline{3}, 3)}_{(0,-2)}& \rep{(1, \overline{3})}_{(2,-2)}\\
 \end{array}\\
\end{array}
\ee
Note, unlike the other $\alg, \T$ cases, there are  non-trivial commutators in the strictly positive grade components $[S'_{20}, S_{02}]\subseteq V_{(22)}$.

For $\f{m}(\T, \alg)\cong \f{m}(\alg, \T)$, we obtain the Kantor-Koecher-Tits intermediate algebras $\f{ k }(\f{m}(\Q, \alg))$:
\be\label{pppain-1}
\begin{array}{ccllllllllllllllll}
\alg&\f{m}(\T, \alg)\\[8pt]
\R&\f{sl}_{3\frac{1}{4}} &:=& [\f{gl}_1\oplus\f{sl}_{3}]_{\0} \pd   \rep{6}_{\2} \\[8pt] 
\C&[\f{sl}_{3}\oplus\f{sl}_{3}]_{\frac{1}{4}} &:= & [\f{gl}_1\oplus\f{sl}_{3}\oplus\f{sl}_{3}]_{\0} \pd   \rep{(3, \overline{3})}_{\2}\\[8pt] 
\T&[\f{sl}_{3}\oplus\f{sl}_{3}]_{\frac{1}{4}+\frac{1}{4}} &:=&[\f{gl}_1\oplus\f{gl}_1\oplus\f{sl}_{3}\oplus\f{sl}_{3}]_{(0,0)} \pd   \rep{(3, \overline{3})}_{(0,2)}\pd   \rep{(3, 3)}_{(2,0)} \pd  (\overline{\rep{3}}, 1)_{(2,2)}\\[8pt] 
\Q&\f{sl}_{6\frac{1}{4}} &:=& [\f{gl}_1\oplus\f{sl}_{6}]_{\0}\pd  \rep{15}_{\2}\\[8pt] 
\Oct&\f{e}_{6\frac{1}{4}} &:=&  [\f{gl}_1\oplus \f{e}_{6}]_{\0} \pd   \rep{27}_{\2}, \\[8pt] 
\end{array}
\ee
which constitute the ternionic row/column of the extended magic square given in Table 1.

For $\f{m}(\T, \alg)_\sz\cong \f{m}(\alg, \T)_\sz$, we analogously obtain the derived Kantor-Koecher-Tits intermediate algebras $\f{ k }_\sz(\f{m}(\T, \alg))$:\be
\begin{array}{ccllllllllllllllll}
\alg&\f{m}(\T, \alg)_\sz\\[8pt]
\R&\f{sl}^{\sz}_{3\frac{1}{4}} &:=& \f{sl}_{3} \pd   \rep{6}\\[8pt] 
\C&[\f{sl}_{3}\oplus\f{sl}_{3}]^{\sz}_{\frac{1}{4}} &:= & [\f{sl}_{3}\oplus\f{sl}_{3}] \pd   \rep{(3, \overline{3})} \\[8pt] 
\T&[\f{sl}_{3}\oplus\f{sl}_{3}]^{\sz}_{\frac{1}{4}+\frac{1}{4}} &:=&[\f{sl}_{3}\oplus\f{sl}_{3}]\pd   \rep{(3, \overline{3})}\pd   \rep{(3, 3)} \pd  \rep{(\overline{3}, 1)}\\[8pt] 
\Q&\f{sl}^{\sz}_{6\frac{1}{4}} &:=&\f{sl}_{6}\pd  \rep{15} \\[8pt] 
\Oct&\f{e}^{\sz}_{6\frac{1}{4}} &:=&  \f{e}_{6} \pd   \rep{27} \\[8pt] 
\end{array}
\ee

Finally, for the $(\T, \s)$ entry we can either consider a Jordan pair in the simple part of the 5-grading $\f{e}_7=\C_{\mii}\pd  V_{\mi}\pd  [\f{gl}_1\oplus\f{so}_{12}]_{\0}\pd   V_{\1}\pd  \C_{\2}$, or, equivalently ($\f{sl}_{6(\frac{1}{4}+\frac{1}{2})}\cong \f{sl}_{6(\frac{1}{2}+\frac{1}{4})}$), an extremal $\f{sl}_2$ triple in the simple part of the 3-grading $\f{e}_{7}= W_{\mi}\pd  [\f{gl}_1\oplus\f{e}_6]_{\0}\pd   W_{\1}$. In either case we obtain (up to Hermitian conjugation) the bigrading,
\be
\begin{array}{ccccccccccccccccccc}
\f{e}_7&=&\begin{array}{ccccccccccccccccccc}
& \overline{V}_{(2,-1)}&A_{(2,0)}&\overline{V}_{(2,1)}&\\
 \C_{(0,-2)} & S_{(0,-1)}&[\f{gl}_1\oplus\f{gl}_1\oplus\f{sl}_{6}]_{(0,0)}& S_{(0,1)}&\C_{(0,2)}\\
 &{V}_{(-2,-1)} &\overline{A}_{(-2,0)}& {V}_{(-2,1)}&\\
 \end{array}
 \\[8pt]
 \\
\rep{133}&\rightarrow&\begin{array}{ccccccccccccccccccc}
& \overline{\rep{6}}_{(2,-1)} &\rep{15}_{(2,0)}&\overline{\rep{6}}_{(2,1)} &\\
\rep{1}_{(0,-2)} & \rep{20}_{(0,-1)}&[\rep{1}\oplus\rep{1}\oplus\rep{35}]_{(0,0)}& \rep{20}_{(0,1)}&\rep{1}_{(0,2)}\\
 &{\rep{6}}_{(-2,-1)} &\overline{\rep{15}}_{(-2,0)}&{\rep{6}}_{(-2,1)} &\\
 \end{array}\\
\end{array}
\ee
which yields the three possible intermediate algebras:
\be
\begin{array}{lllllllll}
\f{m}(\T, \s)&=&\f{sl}_{6(\frac{1}{2}+\frac{1}{4})} &:=&[\f{gl}_1\oplus\f{gl}_1\oplus\f{sl}_{6}]_{(0,0)}\pd  \rep{20}_{(0,1)}\pd  \rep{15}_{(2,0)}\pd  \rep{1}_{(0,2)} \pd  \overline{\rep{6}}_{(2,1)},\\[8pt]
\f{m}(\T, \s)^{+}_{\sz}&=&\f{sl}^{\sz}_{+6(\frac{1}{2}+\frac{1}{4})} &:=&[\f{gl}_1\oplus\f{sl}_{6}]_{(0)}\pd  \rep{20}_{\0}\pd  \rep{15}_{\2}\pd  \rep{1}_{\0} \pd  \overline{\rep{6}}_{\2},\\[8pt]
\f{m}(\T, \s)_{\sz}&=&\f{sl}^{\sz}_{6(\frac{1}{2}+\frac{1}{4})} &:=&\f{sl}_{6}\pd  \rep{20}\pd  \rep{15}\pd  \rep{1}\pd  \overline{\rep{6}}.
\end{array}
\ee
The extended magic square $\f{m}(\alg_1, \alg_2)$, where $\alg_1, \alg_2=\R, \C, \T,  \Q, \s, \Oct$ is presented in \autoref{tab:ems1}.

\FloatBarrier

\subsection{The extended magic square: real forms}\label{realms}

Over the reals, the ternions and sextonions exist only in split form: $\T_s\subset\Q_s$ and $\s_s\subset\Oct_s$. Hence, to include both the $\T_s, \s_s$ row and column we must use split $\alg_1, \alg_2$. Using the standard definition of the commutators in \eqref{msconstructions}, the resulting real forms are given in \autoref{tab:emsreal1}.
  \begin{table}[ht]
 \begin{center}
\begin{tabular}{|c|cccccccc|}
\hline
\hline
  && $\R$ &$\C_s$ & $\T_s$  & $\Q_s$ & $\s_s$ & $\Oct_s$ & \\
 \hline
 &&&&&&&&\\
   $\R$ && $\mathfrak{so}(3)$ & $\mathfrak{sl}_3(\R)$   &$\mathfrak{sl}_{3\frac{1}{4}}(\R)$ &$\mathfrak{sp}_6(\R)$ & $\mathfrak{sp}_{6\frac{1}{2}}(\R)$ & $\mathfrak{f}_{4(4)}$  & \\[12pt]
  $\C_s$ && $\mathfrak{sl}_3(\R)$ & $\mathfrak{sl}_3(\R)\oplus \mathfrak{sl}_3(\R)$   &$[\mathfrak{sl}_{3(\R)}\oplus \mathfrak{sl}_{3}(\R)]_\frac{1}{4}$& $\mathfrak{sl}_6(\R)$ & $\mathfrak{sl}_{6\frac{1}{2}}(\R)$ & $\mathfrak{e}_{6(6)}$   &\\[12pt]
    $\T_s$ && $\mathfrak{sl}_{3\frac{1}{4}}(\R)$ & $[\mathfrak{sl}_{3}(\R)\oplus \mathfrak{sl}_{3}(\R)]_\frac{1}{4}$ & $[\mathfrak{sl}_{3}(\R)\oplus \mathfrak{sl}_{3}(\R)]_{\frac{1}{2}}$  & $\mathfrak{sl}_{6\frac{1}{4}}(\R)$ & $\mathfrak{sl}_{6\frac{3}{4}}(\R)$ & $\mathfrak{e}_{6(6)\frac{1}{4}}$   &\\[12pt]
  $\Q_s$ && $\mathfrak{sp}_6(\R)$ & $\mathfrak{sl}_6(\R)$  & $\mathfrak{sl}_{6\frac{1}{4}}(\R)$& $\mathfrak{so}_{6,6}$ &  $\mathfrak{so}_{6,6\frac{1}{2}}$ & $\mathfrak{e}_{7(7)}$  & \\[12pt]
  $\s_s$ && $\mathfrak{sp}_{6\frac{1}{2}}(\R)$ & $\mathfrak{sl}_{6\frac{1}{2}}(\R)$ & $\mathfrak{sl}_{6\frac{3}{4}}(\R)$ & $\mathfrak{so}_{6,6\frac{1}{2}}$ & $\mathfrak{so}_{6,6\frac{1}{2}+\frac{1}{2}}$ & $\mathfrak{e}_{7(7)\frac{1}{2}}$  & \\[12pt]
   $\Oct_s$ && $\mathfrak{f}_{4(4)}$ & $\mathfrak{e}_{6(6)}$ & $\mathfrak{e}_{6(6)\frac{1}{4}}$ & $\mathfrak{e}_{7(7)}$ & $\mathfrak{e}_{7(7)\frac{1}{2}}$ & $\mathfrak{e}_{8(8)}$&   \\[12pt]
   \hline
   \hline
\end{tabular}
\caption{The  real form of the extended magic square with real split algebras $\alg_1, \alg_2$. Here we have used the Tits construction with Euclidean Jordan algebras. Using instead Lorentzian Jordan algebras, leave all but the $(\R, \R)$ slot invariant: $\mathfrak{so}(3)\rightarrow\f{sl}_2(\R)$. \label{tab:emsreal1}}
 \end{center}
\end{table}
Using a Lorentzian Jordan algebra in the Tits construction, $\J_{2,1}(\alg)$, we obtain the same set of real forms for all but the $(\R,\R)$ entry, which is given by $\mathfrak{sl}_2(\R)$, as can be checked by comparison with \cite{Cacciatori:2012cb}. The same table can also be given by the Barton-Sudbery and Vinberg formulae, by adjusting the definition of the commutators as described in \cite{Anastasiou:2013hba}. With respect to the global symmetries of supergravity, it is also useful to consider the semi-extended magic squares with  rows including $\T_s, \s_s$, but not the columns.  The possible sets of real forms for the conventional magic square construction  are presented in \autoref{tab:emsreal3}. The possible sets of real forms for the conventional magic square with Lorentzian Jordan algebras are presented in \autoref{tab:emsreal5}.
  \begin{table}[ht]
   \begin{subtable}[t]{\linewidth}
 \begin{center}
\begin{tabular}{|c|cccccc|}
\hline
\hline
  && $\R$ &$\C$ &  $\Q$  & $\Oct$ & \\
 \hline
 &&&&&&\\
   $\R$ && $\mathfrak{so}_3$ & $\mathfrak{su}_3$   &$\mathfrak{usp}_6$ &  $\mathfrak{f}_{4(-52)}$  & \\[12pt]
  $\C_s$ && $\mathfrak{sl}_3(\R)$ & $\mathfrak{sl}_3(\C)$   & $\mathfrak{su}^{\star}_6$ & $\mathfrak{e}_{6(-26)}$   &\\[12pt]
    $\T_s$ && $\mathfrak{sl}_{3\frac{1}{4} }(\R)$ & $ \mathfrak{sl}_{3\frac{1}{4} }(\C)$   & $\mathfrak{su}^{\star}_{6\frac{1}{4} }$  & $\mathfrak{e}_{6(-26)\frac{1}{4} }$   &\\[12pt]
  $\Q_s$ && $\mathfrak{sp}_6(\R)$ & $\mathfrak{su}_{3,3}$  & $\mathfrak{so}^{\star}_{12}$  & $\mathfrak{e}_{7(-25)}$  & \\[12pt]
  $\s_s$ && $\mathfrak{sp}_{6\frac{1}{2}}(\R)$ & $\mathfrak{su}_{3,3\frac{1}{2}}$ & $\mathfrak{so}^{\star}_{12\frac{1}{2}}$ & $\mathfrak{e}_{7(-25)\frac{1}{2}}$  & \\[12pt]
   $\Oct_s$ && $\mathfrak{f}_{4(4)}$ & $\mathfrak{e}_{6\2}$  & $\mathfrak{e}_{7(-5)}$ & $\mathfrak{e}_{8(-24)}$&   \\[12pt]
   \hline
   \hline
\end{tabular}
\caption{Real form of the semi-extended magic square with  $\alg_1=\R, \C_s, \T_s,  \Q_s, \s_s, \Oct_s$ and  $\alg_2= \R, \C, \Q, \Oct$, using a Euclidean Jordan algebra. \label{tab:emsreal3a}}
\hspace{0.1in}
\begin{tabular}{|c|cccccc|}
\hline
\hline
  && $\R$ &$\C$ &  $\Q$  & $\Oct_s$ & \\
 \hline
 &&&&&&\\
   $\R$ && $\mathfrak{so}_3$ & $\mathfrak{su}_3$   &$\mathfrak{usp}_6$ &  $\mathfrak{f}_{4(4)}$  & \\[12pt]
  $\C_s$ && $\mathfrak{sl}_3(\R)$ & $\mathfrak{sl}_3(\C)$   & $\mathfrak{su}^{\star}_6$ & $\mathfrak{e}_{6(6)}$   &\\[12pt]
    $\T_s$ && $\mathfrak{sl}_{3\frac{1}{4} }(\R)$ & $ \mathfrak{sl}_{3\frac{1}{4} }(\C)$   & $\mathfrak{su}^{\star}_{6\frac{1}{4} }$  & $\mathfrak{e}_{6(6)\frac{1}{4} }$   &\\[12pt]
  $\Q_s$ && $\mathfrak{sp}_6(\R)$ & $\mathfrak{su}_{3,3}$  & $\mathfrak{so}^{\star}_{12}$  & $\mathfrak{e}_{7(7)}$  & \\[12pt]
  $\s_s$ && $\mathfrak{sp}_{6\frac{1}{2}}(\R)$ & $\mathfrak{su}_{3,3\frac{1}{2}}$ & $\mathfrak{so}^{\star}_{12\frac{1}{2}}$ & $\mathfrak{e}_{7(7)\frac{1}{2}}$  & \\[12pt]
   $\Oct_s$ && $\mathfrak{f}_{4(4)}$ & $\mathfrak{e}_{6\2}$  & $\mathfrak{e}_{7(-5)}$ & $\mathfrak{e}_{8(8)}$&   \\[12pt]
   \hline
   \hline
\end{tabular}
\caption{Real form of the semi-extended magic square with  $\alg_1=\R, \C_s, \T_s,  \Q_s, \s_s, \Oct_s$ and  $\alg_2= \R, \C, \Q, \Oct_s$, using a Euclidean Jordan algebra. \label{tab:emsreal3b}}

\vspace{0.1in}
\begin{tabular}{|c|cccccc|}
\hline
\hline
  && $\R$ &$\C$ &  $\Q_s$  & $\Oct_s$ & \\
 \hline
 &&&&&&\\
   $\R$ && $\mathfrak{so}_3$ & $\mathfrak{su}_3$   &$\mathfrak{sp}_6(\R)$ &  $\mathfrak{f}_{4(4)}$  & \\[12pt]
  $\C_s$ && $\mathfrak{sl}_3(\R)$ & $\mathfrak{sl}_3(\C)$   & $\mathfrak{sl}_6(\R)$ & $\mathfrak{e}_{6(6)}$   &\\[12pt]
    $\T_s$ && $\mathfrak{sl}_{3\frac{1}{4} }(\R)$ & $ \mathfrak{sl}_{3\frac{1}{4} }(\C)$   & $\mathfrak{sl}_{6\frac{1}{4} }(\R)$  & $\mathfrak{e}_{6(6)\frac{1}{4} }$   &\\[12pt]
  $\Q_s$ && $\mathfrak{sp}_6(\R)$ & $\mathfrak{su}_{3,3}$  & $\mathfrak{so}_{6,6}$  & $\mathfrak{e}_{7(7)}$  & \\[12pt]
  $\s_s$ && $\mathfrak{sp}_{6\frac{1}{2}}(\R)$ & $\mathfrak{su}_{3,3\frac{1}{2}}$ & $\mathfrak{so}_{6,6\frac{1}{2}}$ & $\mathfrak{e}_{7(7)\frac{1}{2}}$  & \\[12pt]
   $\Oct_s$ && $\mathfrak{f}_{4(4)}$ & $\mathfrak{e}_{6\2}$  & $\mathfrak{e}_{7(7)}$ & $\mathfrak{e}_{8(8)}$&   \\[12pt]
   \hline
   \hline
\end{tabular}
\caption{Real form of the semi-extended magic squares with  $\alg_1=\R, \C_s, \T_s,  \Q_s, \s_s, \Oct_s$ and  $\alg_2= \R, \C, \Q_s, \Oct_s$, using a Euclidean Jordan algebra.\label{tab:emsreal3c}}
 \end{center}
 \end{subtable}
 \caption{Real forms of the semi-extended magic squares with  composition and split  $\alg_1, \alg_2$, using a Euclidean Jordan algebra. \label{tab:emsreal3}}
\end{table}

  \begin{table}[ht]
     \begin{subtable}[t]{\linewidth}
 \begin{center}
\begin{tabular}{|c|cccccc|}
\hline
\hline
  && $\R$ &$\C$ &  $\Q$  & $\Oct$ & \\
 \hline
 &&&&&&\\
   $\R$ && $\mathfrak{sl}_2(\R)$ & $\mathfrak{su}_{2,1}$   &$\mathfrak{usp}_{4,2}$ &  $\mathfrak{f}_{4(-20)}$  & \\[12pt]
  $\C_s$ && $\mathfrak{sl}_3(\R)$ & $\mathfrak{sl}_3(\C)$   & $\mathfrak{su}^{\star}_6$ & $\mathfrak{e}_{6(-26)}$   &\\[12pt]
    $\T_s$ && $\mathfrak{sl}_{3\frac{1}{4} }(\R)$ & $ \mathfrak{sl}_{3\frac{1}{4} }(\C)$   & $\mathfrak{su}^{\star}_{6\frac{1}{4} }$  & $\mathfrak{e}_{6(-26)\frac{1}{4} }$   &\\[12pt]
  $\Q_s$ && $\mathfrak{sp}_6(\R)$ & $\mathfrak{su}_{3,3}$  & $\mathfrak{so}^{\star}_{12}$  & $\mathfrak{e}_{7(-25)}$  & \\[12pt]
  $\s_s$ && $\mathfrak{sp}_{6\frac{1}{2}}(\R)$ & $\mathfrak{su}_{3,3\frac{1}{2}}$ & $\mathfrak{so}^{\star}_{12\frac{1}{2}}$ & $\mathfrak{e}_{7(-25)\frac{1}{2}}$  & \\[12pt]
   $\Oct_s$ && $\mathfrak{f}_{4(4)}$ & $\mathfrak{e}_{6\2}$  & $\mathfrak{e}_{7(-5)}$ & $\mathfrak{e}_{8(-24)}$&   \\[12pt]
   \hline
   \hline
\end{tabular}
\caption{Real form of the semi-extended magic squares with  $\alg_1=\R, \C_s, \T_s,  \Q_s, \s_s, \Oct_s$ and  $\alg_2= \R, \C, \Q, \Oct$, using a Lorentzian  Jordan algebra.\label{tab:emsreal5a}}
\vspace{0.1in}

\begin{tabular}{|c|cccccc|}
\hline
\hline
  && $\R$ &$\C$ &  $\Q$  & $\Oct_s$ & \\
 \hline
 &&&&&&\\
   $\R$ && $\mathfrak{sl}_2(\R)$ & $\mathfrak{su}_{2,1}$   &$\mathfrak{usp}_{4,2}$ &  $\mathfrak{f}_{4(4)}$  & \\[12pt]
  $\C_s$ && $\mathfrak{sl}_3(\R)$ & $\mathfrak{sl}_3(\C)$   & $\mathfrak{su}^{\star}_6$ & $\mathfrak{e}_{6(6)}$   &\\[12pt]
    $\T_s$ && $\mathfrak{sl}_{3\frac{1}{4} }(\R)$ & $ \mathfrak{sl}_{3\frac{1}{4} }(\C)$   & $\mathfrak{su}^{\star}_{6\frac{1}{4} }$  & $\mathfrak{e}_{6(6)\frac{1}{4} }$   &\\[12pt]
  $\Q_s$ && $\mathfrak{sp}_6(\R)$ & $\mathfrak{su}_{3,3}$  & $\mathfrak{so}^{\star}_{12}$  & $\mathfrak{e}_{7(7)}$  & \\[12pt]
  $\s_s$ && $\mathfrak{sp}_{6\frac{1}{2}}(\R)$ & $\mathfrak{su}_{3,3\frac{1}{2}}$ & $\mathfrak{so}^{\star}_{12\frac{1}{2}}$ & $\mathfrak{e}_{7(7)\frac{1}{2}}$  & \\[12pt]
   $\Oct_s$ && $\mathfrak{f}_{4(4)}$ & $\mathfrak{e}_{6\2}$  & $\mathfrak{e}_{7(-5)}$ & $\mathfrak{e}_{8(8)}$&   \\[12pt]
   \hline
   \hline
\end{tabular}
\caption{Real form of the semi-extended magic square with  $\alg_1=\R, \C_s, \T_s,  \Q_s, \s_s, \Oct_s$ and  $\alg_2= \R, \C, \Q, \Oct_s$, using a Lorentzian  Jordan algebra.\label{tab:emsreal5b}}
\vspace{0.1in}

\begin{tabular}{|c|cccccc|}
\hline
\hline
  && $\R$ &$\C$ &  $\Q_s$  & $\Oct_s$ & \\
 \hline
 &&&&&&\\
   $\R$ && $\mathfrak{sl}_2(\R)$ & $\mathfrak{su}_{2,1}$   &$\mathfrak{sp}_{6}(\R)$ &  $\mathfrak{f}_{4(4)}$  & \\[12pt]
  $\C_s$ && $\mathfrak{sl}_3(\R)$ & $\mathfrak{sl}_3(\C)$   & $\mathfrak{sl}_6(\R)$ & $\mathfrak{e}_{6(6)}$   &\\[12pt]
    $\T_s$ && $\mathfrak{sl}_{3\frac{1}{4} }(\R)$ & $ \mathfrak{sl}_{3\frac{1}{4} }(\C)$   & $\mathfrak{sl}_{6\frac{1}{4} }(\R)$  & $\mathfrak{e}_{6(6)\frac{1}{4} }$   &\\[12pt]
  $\Q_s$ && $\mathfrak{sp}_6(\R)$ & $\mathfrak{su}_{3,3}$  & $\mathfrak{so}_{6,6}$  & $\mathfrak{e}_{7(7)}$  & \\[12pt]
  $\s_s$ && $\mathfrak{sp}_{6\frac{1}{2}}(\R)$ & $\mathfrak{su}_{3,3\frac{1}{2}}$ & $\mathfrak{so}_{6,6\frac{1}{2}}$ & $\mathfrak{e}_{7(7)\frac{1}{2}}$  & \\[12pt]
   $\Oct_s$ && $\mathfrak{f}_{4(4)}$ & $\mathfrak{e}_{6\2}$  & $\mathfrak{e}_{7(7)}$ & $\mathfrak{e}_{8(8)}$&   \\[12pt]
   \hline
   \hline
\end{tabular}
\caption{Real form of the semi-extended magic squares with  $\alg_1=\R, \C_s, \T_s,  \Q_s, \s_s, \Oct_s$ and  $\alg_2= \R, \C, \Q_s, \Oct_s$, using a Lorentzian  Jordan algebra.\label{tab:emsreal5c}}
 \end{center}
 \end{subtable}
\caption{Real forms of the semi-extended magic squares with  composition and split   $\alg_1, \alg_2$, using a Lorentzian Jordan algebra. \label{tab:emsreal5}}
\end{table}

\FloatBarrier

\section{Supergravity and the extended magic square}\label{sugra}

\subsection{Maximally supersymmetric theories in $D=5,4,3$}\label{max}
An important  and well-supported conjecture is that  the U-dualities of M-theory compactified on an $n$-torus are given by  a sequence of  discrete exceptional  groups $E_{n(n)}(\Z)$ \cite{Hull:1994ys}.   The corresponding low energy effective field theories are given  by $D=11$ supergravity compactified on an $n$-torus and classically the U-dualities manifest themselves as non-compact global electric-magnetic duality symmetries, $\mathcal{G}=E_{n(n)}(\R)$ \cite{Cremmer:1979up}. The global symmetry group acts non-linearly on the scalar sector, which parametrises a symmetric space $\mathcal{G/H}$, where $\mathcal{H}$ is the maximal compact subgroup of $\mathcal{G}$. The U-dualities and corresponding global symmetries for M-theory and $D=11$ supergravity compactified on a $n$-torus are summarised in \autoref{uduality}. \begin{table}[ht]
\begin{tabular*}{\textwidth}{@{\extracolsep{\fill}}cM{c}M{c}M{c}c}
\hline
\hline
 $n$-torus   & \text{U-duality}  & \mathcal{G}                                         & \mathcal{H}                                        & \\
\midrule
 1   & \SO(1,1,\mathds{Z})              & \SO(1,1,\mathds{R})                        & -                                        & \\
 2     & \SL(2,\mathds{Z})\times \SO(1,1,\mathds{Z})          & \SL(2,\mathds{R})\times \SO(1,1,\mathds{R}) & \SO(2,\mathds{R})                         & \\
 3     & \SL(2,\mathds{Z})\times \SL(3,\mathds{Z})           & \SL(2,\mathds{R})\times \SL(3,\mathds{R})   & \SO(2,\mathds{R})\times \SO(3,\mathds{R})  & \\
 4     & \SL(5,\mathds{Z})          & \SL(5,\mathds{R})                          & \SO(5,\mathds{R})                         & \\
 5     & \SO(5,5,\mathds{Z})           & \SO(5,5,\mathds{R})                        & \SO(5,\mathds{R})\times \SO(5,\mathds{R})  & \\
 6     & E_{6(6)}(\mathds{Z})         & E_{6(6)}(\mathds{R})                      & \USp(8)                                   & \\
 7     & E_{7(7)}(\mathds{Z})             & E_{7(7)}(\mathds{R})                      & \SU(8)                                    & \\
 8     & E_{8(8)}(\mathds{Z})         & E_{8(8)}(\mathds{R})                      & \SO(16,\mathds{R})                        & \\
\hline
\hline
\end{tabular*}
\caption{U-dualities  (global symmetries) of M-theory ($D=11$, $\N=1$ supergravity) compactified on an $n$-torus. Possible discrete factors in $\mathcal{H}$ are omitted.}\label{uduality}
\end{table}

However, upon compactification the exceptional global symmetries $E_{n(n)}$ are  initially hidden;    a judicious choice of field dualisations is required to reveal them \cite{Cremmer:1979up}. If one instead  leaves all fields as they come the global symmetries are given by \cite{Cremmer:1997ct},
\be\label{eve-1}
\SO(1,1)\times\SL(n, \R)\ltimes\R^{n(n-1)(n-2)/6}.
\ee
Note, for $n\leq2$ this coincides with $E_{n(n)}$. Furthermore, for $n< 6$  the \emph{scalar sector} of the Lagrangian already enjoys the full $E_{n(n)}$,  but to realise the symmetry on  higher  $p$-form fields some dualisation is required. For $n\geq6$ the $(D-2)$-form potentials must be dualised to scalars in order to exhibit $E_{n(n)}$.

Of course, one is not committed to dualising all or nothing. For example, dualising only the Ramond-Ramond sector yields non-reductive extensions of T-duality for $n\geq5$,
\be
\Spin(n-1,n-1)\ltimes S_n,
\ee
where $S_n$ denotes the  spinor representation of $\Spin(n-1,n-1)$ \cite{Cremmer:1997ct}. Another option is to fully dualise the $(D+1)$-dimensional theory and then consider the possible dualisations of  the resulting $D$-dimensional fields. It is this final possibility for which the ternionic and sextonionic entries  of the magic square matter.

\subsubsection{Reduction of $D=11$ supergravity on $T^n$}

In  the subsequent  subsections, we follow closely the excellent account  given in \cite{Cremmer:1997ct}. The bosonic sector of $D=11$  supergravity is given by \cite{Cremmer:1978km},
\be\label{Lagg}
\mathcal{L}=R\star 1-\frac{1}{2} \star F_\4 \wedge F_\4 -\frac{1}{6}F_\4 \wedge F_\4 \wedge A_\3.
\ee
In addition to general coordinate transformations it has a 2-form gauge invariance,
\be
\delta A_\3=d\lambda_\2,
\ee
and the action scales homogeneously under the global transformation
\be\label{scale}
g\rightarrow \Omega^2 g,\qquad A_\3\rightarrow \Omega^3 A_\3.
\ee

The bosonic sector of  $D=11$  supergravity compactified on $T^n$ (we restrict our attention here to $11-n>2$) prior to dualization, may be found in \cite{Cremmer:1997ct,Cremmer:1998px}.     We use the  standard metric Ansatz
\be
ds_{11}^2 = e^{\ft{1}{3} \vec g\cdot\vec\phi} \, ds_{\sst D}^2 +
\sum_i e^{2\vec\gamma_i\cdot\vec\phi}\, (h^i)^2\ ,\label{met}
\ee
where
\be
h^i=dz^i + {\cal A}^i + {\cal A}^i{}_j\, dz^j\
\ee
and
\be
\vec \gamma_i=\ft{1}{6}\vec g -\ft{1}{2}\vec f_i.
\ee
We have split the  scalar fields deriving from the metric  into  the $n$-vector of dilatonic scalar fields coming from the diagonal components of the internal metric, denoted $\vec\phi$, and the rest, denoted ${\cal A}^i{}_j$, $i,j=1,2\ldots n$ for $j>i$. In $D$ dimensions the vectors $\vec g$ and $\vec f_i$ have $(11-D)=n$ components and are given by
\bea
\vec g &=&3 (s_1, s_2, \ldots, s_{n})\ ,\nonumber\\
\vec f_i &=& \Big(0,0,\ldots, 0, (10-i) s_i, s_{i+1},
s_{i+2}, \ldots, s_{n}\Big)\ ,\label{gfdef}
\eea
where $s_i = \sqrt{2/((10-i)(9-i))}$.

The original eleven-dimensional fields
$g_{\sst{MN}}$ and $A_{\sst{MNP}}$ will give then rise to the following
fields in $D$ dimensions,
\bea
g_{\sst{MN}} &\rightarrow & g_{\mu\nu} , \qquad \vec\phi,\qquad
{\cal A}_{\mu}^{i},\qquad {\cal A}^i{}_j  ,\nn\\
A_{MNP} &\rightarrow & A_{\mu\nu\rho} ,\qquad A_{\mu\nu k }, \qquad A_{\mu jk},
\qquad A_{ijk},
\label{dfields}
\eea
where the indices $i, j, k$ run over the $n$ internal toroidally-compactified dimensions. If we denote the rank $(p+1)$ field strengths of the rank $p$ potentials by a subscript $(p+1)$, the Lagrangian is
\bea\label{L}\nn
\frac{{\cal L}}{\sqrt{-g}} &=& R -\ft{1}{2} \, (\del\vec\phi)^2  -\ft{1}{4}\, \sum_i e^{\vec b_i\cdot \vec\phi}\, ({\cal F}_\2^i)^2
-\ft{1}{2}\, \sum_{i<j} e^{\vec b_{ij}\cdot \vec\phi}\,
({\cal F}_\1^i{}_j)^2\\
&&
-\ft{1}{48}\, e^{\vec a\cdot
\vec\phi}\, F_\4^2 -\ft{1}{12} \sum_i
e^{\vec a_i\cdot \vec\phi}\, (F_{\3 i})^2
-\ft{1}{4}\, \sum_{i<j} e^{\vec a_{ij}\cdot \vec\phi}\, (F_{\2ij})^2
-\ft{1}{2} \, \sum_{i<j<k} e^{\vec a_{ijk} \cdot\vec \phi}\,
(F_{\1ijk})^2 \label{dgenlag}\\\nn
&&
 + {\cal L}_{\sst{FFA}}\ ,\nn
\eea
where the ``dilaton vectors'' $\vec a$, $\vec a_i$, $\vec a_{ij}$,
$\vec a_{ijk}$,
$\vec b_i$, $\vec b_{ij}$ are constants that characterise the couplings of
the dilatonic scalars $\vec \phi$ to the various gauge fields\footnote{For some intriguing connections to octonions and `octavian integers', cfr. \cite{Anastasiou:2015ena}.}
\begin{subequations}
\bea
{F}_\4:&&\vec a = -\vec g\ \\
{F}_\3:&&\vec a_i = \vec f_i -\vec g \ \\
{F}_\2:&& \vec a_{ij} = \vec f_i + \vec f_j - \vec g\ \\
{F}_\1:&&\vec a_{ijk} = \vec f_i + \vec f_j + \vec f_k -\vec g\\
{\cal F}_\2:&&\vec b_i = -\vec f_i \ \\
{\cal F}_\1:&& \vec b_{ij} = -\vec f_i + \vec f_j\
\label{dilatonvec}
\eea
\end{subequations}
Note,  the field strengths appearing in the kinetic
terms are not simply the exterior derivatives of their associated
potentials, but also include non-linear Kaluza-Klein modifications.
On the other hand, the terms included in ${\cal L}_{\sst{FFA}}$, which
denotes the dimensional reduction of the $F_\4\wedge F_\4\wedge A_\3$
term in $D=11$ Lagrangian \eqref{Lagg}, are better expressed purely in terms of the potentials
and their exterior derivatives.

The complete details may be found in \cite{Cremmer:1997ct}, where it is shown that the symmetry of the Lagrangian before any dualization is the non-reductive Lie group,
\be\label{predual}
\SO(1,1)\times\SL(n, \R)\ltimes\R^{n(n-1)(n-2)/6}.
\ee
In brief, the $D=11$ general coordinate transformations $\delta x^M = -\xi^{M}(x^N)$ are restricted by the Kaluza-Klein Ansatz to
\be
\xi^{\mu}=\xi^{\mu}(x^\nu), \qquad \xi^{i}=\Lambda^{i}{}_{j} x^j +\xi^i(x^\nu)
\ee
where $\xi^{\mu}$, $\Lambda^{i}{}_{j} - \Lambda^{k}{}_{k}\delta^{i}{}_{j}/n$ and $\xi^i(x^\nu)$ are the $D=11-n$ general coordinate, global $\SL(n, \R)$ and Abelian 1-form gauge parameters, respectively. The trace part $\Lambda^{k}{}_{k}$ must be combined with the global scaling symmetry  \eqref{scale} to give a global $\SO(1,1)$ that leaves the volume of the internal space invariant.

The $D=11$ 3-form gauge transformation $\delta A_\3 = d\lambda_\2$ gives the $D=11-n$ gauge transforms for the corresponding  3-, 2-, and 1-forms. The axionic 0-forms $A_{(0)ijk}$, however, transform globally
\be
\delta A_{(0)ijk} = c_{ijk},
\ee
 since the preservation of the  Kaluza-Klein Ansatz requires $\lambda_{ij} = c_{ijk}x^k$. Clearly, the $c_{ijk}$ commute amongst themselves, but belong to  the $\wedge^3 \R^{n}\cong \R^{n(n-1)(n-2)/6}$ representation of $\SL(n, \R)$, since they must transform  as the $A_{(0)ijk}$ do. Consequently, prior to any dualisation, the total global symmetries are given by \eqref{predual}.

\subsubsection{$D=4$ to $D=3$ and $ \widetilde{\f{e}}_{7(7)\frac{1}{2}}$}\label{e71/2}

Here, we demonstrate that if one starts with the fully dualised theory in $D=4$, with scalars parametrising $E_{7(7)}(\R)/[\SU(8)/\Z_2]$, and dimensionally reduces on a circle, the resulting theory in $D=3$  has global symmetry algebra
$
\widetilde{\f{e}}_{7(7)\frac{1}{2}}.
$
In brief, as described in \cite{Cremmer:1997ct}, in addition to the 70 scalars parametrising  $E_{7(7)}/[\SU(8)/\Z_2]$ that are inherited directly from $D=4$, we have a further 56 scalars appearing from the reduction of the 28 $D=4$ 1-form potentials (once we dualise the resulting $D=4$ 1-forms), which have a shift symmetry that gives the non-reductive part of $\widetilde{\f{e}}_{7(7)\sst{\frac{1}{2}}}$. To explain this in full detail, let us first review the $D=4$  theory obtained by dimensional reduction and the dualisations required to reveal the $E_{7(7)}$  symmetry.

 \paragraph{$D=4$ dualisation and $E_{7(7)}$}

Let us begin with the undualised theory in $D=4$ described by the Lagrangian given in \eqref{L} and consider its symmetries following the discussion in \cite{Cremmer:1997ct}. Recall, the global symmetry of the scalar sector is given by ($n=7$ in \eqref{eve-1}) :
\be
\SO(1,1)\times\SL(7, \R)\ltimes\R^{35}.
\ee
The Abelian $\R^{35}$ corresponds to shifts of $A_{\0ijk}$ ($i,j,k=1,...,7$)
\be
\delta A_{\0ijk} = c_{ijk},
\ee
where the $c_{ijk}$ transform as the $\rep{35}\cong\wedge^3 \R^7$ of $\SL(7, \R)$.

In $D=4$, axionic 0-form potentials are dual to 2-form potentials. The terms in \eqref{L} for $D=4$ involving 2-form potentials are given by
\be
\mathcal{L}_3=-\sqrt{-g}\ft{1}{12}
e^{\vec a_i\cdot \vec\phi}\, (F_{\3 i})^2 -\ft{1}{72}A_{\0 ijk} dA_{\0 lmn}\wedge dA_{\2 p}\epsilon^{ijklmnp},
\ee
where the second term belongs to ${\cal L}_{\sst{FFA}}$ and
\be
F_{\3 i} = \gamma^{j}{}_{i} d A_{\2 j}+\gamma^{j}{}_{i} \gamma^{k}{}_{l}d A_{\1 jk}\wedge \mathcal{A}_{\1}^{l}+\frac{1}{2}\gamma^{j}{}_{i} \gamma^{k}{}_{m} \gamma^{l}{}_{n} d A_{\0 jkl}\wedge \mathcal{A}_{\1}^{m}\wedge \mathcal{A}_{\1}^{n},
\ee
with
\be
[\gamma^{-1}]^{i}{}_{j}=\delta^{i}{}_{j}+\mathcal{A}_{\0}^{i}{}_{j}.
\ee

The Bianchi identities for $F_{\3 i}$ modulo non-scalar terms (which do not affect the scalar Lagrangian)
\be
d(F_\3\gamma^{-1})=0,
\ee
can be imposed via a set of Lagrange multipliers $\chi^i$ in a first order Lagrangian
\be
\tilde{\mathcal{L}}_3 = -\ft{1}{2}
e^{\vec a_i\cdot \vec\phi}\, \star F_{\3 i} \wedge  F_{\3 i} -\ft{1}{72}A_{\0 ijk} dA_{\0 lmn}\wedge (F_{\3}\gamma^{-1})_{p}\epsilon^{ijklmnp}-\chi^i  d(F_\3\gamma^{-1})_i.
\ee
Varying with respect to $F_\3\gamma^{-1}$, one finds
\be
\underbrace{e^{\vec a_i\cdot \vec\phi}\, \star F_{\3 i}}_{\text{(no sum)}} =  [\gamma^{-1}]^{i}{}_{j}\left( d\chi^{j}-\ft{1}{72}A_{\0klm}dA_{\0npq}\epsilon^{klmnpqj}\right)\equiv \widetilde{F}_{\1}^{i},
\ee
which, being algebraic, can be substituted back into $\tilde{\mathcal{L}}_3$ to give the dualised  Lagrangian,
\be
\tilde{\mathcal{L}}_3  \rightarrow \widetilde{\mathcal{L}}_1=-\half\sqrt{-g}
e^{-\vec a_i\cdot \vec\phi}\, (\widetilde{F}_{\1 i})^2.
\ee

Note, the sign appearing in front of the  dilaton vector $\vec a_i$ has flipped. The total scalar kinetic terms in the Lagrangian are now given by
\be\label{D4scalarL}
{\cal L}_{\text{scalar}} = \half\sqrt{-g}\left(- (\del\vec\phi)^2
- \sum_{i<j} e^{\vec b_{ij}\cdot \vec\phi}\,
({\cal F}_\1^i{}_j)^2
- \sum_{i<j<k} e^{\vec a_{ijk} \cdot\vec \phi}\,
(F_{\1ijk})^2- \sum_{i}
e^{-\vec a_i\cdot \vec\phi}\, (\widetilde{F}_{\1 i})^2\right).
\ee
The $63=21+35+7$ dilaton vectors $\vec b_{ij}, \vec a_{ijk}, -\vec a_i$ constitute a set of positive $\f{e}_{7(7)}$ roots, so already one sees the $E_{7(7)}$ symmetry exposed \cite{Cremmer:1979up, Cremmer:1997ct}.

The crucial observation is that the newly introduced axion, $\chi^i$, multiplier terms have a global transformation rule involving not only the obvious constant shifts  by $k^i$, but also the $c_{ijk}$ of the $\R^{35}$ factor, as well as the scaling $\f{so}(1,1)$ with parameter $\mu$,
\be
\delta \chi^i = k^i-\ft{1}{72}c_{jkl} A_{\0mnp}\epsilon^{ijklmnp}-\ft{1}{72}\mu\chi^i.
\ee
Consequently, the  $\R^{35}$ transformations parametrised by $c_{ijk}$ no longer necessarily commute:
\be
[\delta_{c}, \delta_{c'}] = \delta_k,\qquad \text{where}\qquad  k^i=\ft{1}{36}c_{jkl}c'_{mnp}\epsilon^{ijklnmp}.
\ee
This is reflected by the dilaton vectors, which satisfy,
\be
\vec a_{ijk}+\vec a_{lmn}=-\vec a_p,
\ee
for $ijklmnp$ all different. Note that prior to dualisation the required $-\vec a_p$ were simply not available. A maximal subset of mutually commuting transformations is given by selecting a common direction, say:
\be
\Lambda^{\alpha}{}_{7}, \quad c_{\alpha\beta7},\quad k^{\alpha},\quad  \alpha, \beta=1,2,\ldots, 6
\ee
leaving $\R^{27}\cong\R^6\oplus\R^6\oplus\wedge^2\R^6$ as the Abelian subalgebra, which is precisely the maximal Abelian subalgebra of $\f{e}_{7(7)}$ (see e.g. \cite{burceb}, and Refs. therein).

The Lagrangian \eqref{D4scalarL} can be written in a manifestly $E_{7(7)}(\R)$ invariant form \cite{Cremmer:1997ct},
 \be
{\cal L}_{\text{scalar}}= \frac{1}{4}\sqrt{-g}\tr \left(\partial_\mu \mathcal{M}^{-1}\partial^\mu \mathcal{M}\right),\qquad
  \mathcal{M}= \mathcal{V}^\dagger \mathcal{V},
 \ee
 where $\mathcal{V}$ is the $E_{7(7)}(\R)/\SU(8)$ coset representative.
Under the global $E_{7(7)}(\R)$ group, ${\cal M}$ transforms as
 \be
 {\cal M} \rightarrow U^T{\cal M}U,
 \ee
where $U$ is in the \rep{56} representation of  $E_{7(7)}(\R)$.

To realise the $E_{7(7)}(\R)$ symmetry in the vector sector, one must further dualise the 21 pseudo-vectors $A_{\1ij}$ to vectors \cite{Cremmer:1979up}. The dual vectors $\tilde{A}_{\1ij}$ together with the 7 $\mathcal{A}_{\1}^{i}$ sit in the $\rep{28}$ of $\SL(8, \R)$ ($a,b=1,...,8$)
\be
A_{\1}^{ab} = (\mathcal{A}_{\1}^{i}, \tilde{A}_{\1ij}),
\ee
with field strength $F_{\2}^{ab}=dA_{\1}^{ab}$.
Defining the dual  field strengths in the $\rep{28}'$ of $\SL(8, \R)$ as
\be
\star G_{\2 ab} := -\frac{\delta \mathcal{L}}{\delta F_{\2}^{ab}}
\ee
 the complete bosonic Lagrangian can be written as
 \be\label{SL8Lagrangian}
 \mathcal{L}' = R\star 1 +\frac{1}{4}\tr \left(\star d\mathcal{M}^{-1}\wedge d\mathcal{M}\right) +\frac{1}{4}  F_{\2}^{ab} \wedge  G_{\2 ab},
 \ee
 which has manifest $\SL(8, \R)$ symmetry. The $E_{7(7)}(\R)$ cannot be made manifest (at least naively) at the Lagrangian level without breaking manifest general  coordinate covariance (cfr. \cite{Bossard:2010dq}). However, at the level of the equations of motion $F$ and $G$ can be placed in a doublet,
 \be
 H_\2 = \begin{pmatrix} F_{\2}\\ G_{\2}\end{pmatrix},
 \ee
 which transforms linearly in the $\rep{56}$ of $E_{7(7)}(\R)$, where $\rep{56}\rightarrow \rep{28}+\rep{28}'$ under $\SL(8, \R)\subset E_{7(7)}(\R)$.

For our purposes, it is useful to introduce a manifestly  $E_{7(7)}(\R)$-invariant Lagrangian  that must be supplemented by twisted-self-duality constraint, as described in \cite{Cremmer:1997ct, Borsten:2012pd}. Let us define the  ``doubled'' potential $ \mathbf{A}_{\1}=(A_{\1}^{ab}, B_{\1 ab})^T$ transforming as the $\rep{56}$ of $E_{7(7)}(\R)$, such that
  \be
 H_{\2} = \begin{pmatrix} F_{\2}\\ G_{\2}\end{pmatrix} = d \mathbf{A}_{\1}= d \begin{pmatrix} A_{\1}^{ab}\\ B_{\1 ab}\end{pmatrix},
 \ee
 and introduce the manifestly  $E_{7(7)}(\R)$-invariant Lagrangian,
  \be\label{doubleL}
 \mathcal{L}' = R\star 1 +\frac{1}{4}\tr \left(\star d \mathcal{M}^{-1}\wedge d\mathcal{M}\right) -\frac{1}{4}  \star H_{\2}^{T}  \wedge  \mathcal{M}  H_{\2},
 \ee
 with constraint \cite{Cremmer:1979up},
 \be\label{constraint}
 H_{\2}=\star \Omega \mathcal{M}   H_{\2},\qquad \Omega = \begin{pmatrix} 0&\mathds{1}\\ -\mathds{1}&0 \end{pmatrix}.
 \ee
 Note, the constraint must be applied to the equations of motions and not to the Lagrangian, as \eqref{constraint} implies
 \be
 \star H_{\2}^{T}  \wedge \mathcal{M}  H_{\2} \rightarrow H_{\2}^{T}  \wedge \Omega H_{\2} =0.
 \ee
 The doubled Lagrangian \eqref{doubleL}, where the potential $\mathbf{A}_{\1}$ is treated as the independent variable, together with  the constraint \eqref{constraint} is on-shell equivalent to the standard dualised Lagrangian \eqref{SL8Lagrangian} \cite{Cremmer:1997ct}.

\paragraph{Dualised $D=4$, $\N=8$ supergravity on $S^1$} Having reviewed how dualisation takes us from $\SO(1,1)\times\SL(7, \R)\ltimes\R^{35}$ to $E_{7(7)}$, let us now reduce to $D=3$ on a circle.
 Applying the standard Kaluza-Klein reduction Ansatz\footnote{Where it is understood that the objects on left (right) side  are  $D+1$ ($D$) dimensional.},
 \be
 \begin{array}{cllllllll}
 d{s}^2 &=& e^{2\alpha \phi(x)} ds^2+e^{2\beta \phi(x)}(dz +\mathcal{A}_{\1}(x)),\\[8pt]
 {A}_{\sst{(n)}}(x, z)&=& A_{\sst{(n)}}(x) +A_{\sst{(n-1)}}(x)\wedge dz,
  \end{array}
 \ee
where $\alpha^{-2}=2(D-1)(D-2)$ and $\beta=-(D-2)\alpha$,
to the $E_{7(7)}$-invariant Lagrangian, \eqref{doubleL} we obtain the $D=3$ Lagrangian:
\be\label{D3L}
\begin{split}
\mathcal{L}_3=&R\star 1 - \frac{1}{2}\star d\phi \wedge d\phi - \frac{e^{-2\phi}}{2}\star \mathcal{F}_{\2} \wedge \mathcal{F}_{\2} +\frac{1}{4}\tr \left(\star d \mathcal{M}^{-1}\wedge d\mathcal{M}\right) -\frac{e^{-\phi}}{4}  \star  H _{\2}^{T}  \wedge  \mathcal{M}   H _{\2}\\
&-\frac{e^{ \phi}}{4}  \star  H _{\1}^{T}  \wedge  \mathcal{M}   H _{\1},
\end{split}
\ee
where
\be
 \mathcal{F}_{\2} = d\mathcal{A}_{\1}, \qquad  H _{\2} = d\mathbf{A}_{\1} -H_\1 \wedge \mathcal{A}_{\1},\qquad  H _{\1}=d\mathbf{A}_{\sst{(0)}}.
\ee
The twisted-self-duality constraint \eqref{constraint} reduces to
\be\label{D3C}
 H _{\2}+ H _{\1}\wedge (dz +\mathcal{A}_{\1})= e^{\phi}\Omega \mathcal{M}  \star  H _{\1} +e^{-\phi}\Omega \mathcal{M}  \star  H _{\2} \wedge (dz +\mathcal{A}_{\1}),
\ee
which implies
\be\label{D3Cred}
 H _{\2} = e^{\phi}\Omega \mathcal{M}  \star  H _{\1}\quad\Leftrightarrow\quad  H _{\1} = e^{-\phi}\Omega \mathcal{M}  \star  H _{\2}.
\ee
It is here worth stressing that the constraint \eqref{D3Cred} interchanges the equations of motion and the Bianchi identities of $ H _\1$ and $ H _\2$, respectively,
\be
\begin{split}
d(e^{-\phi}\star \mathcal{M}  H _{\2})=0&\longleftrightarrow d H _{\1}=0,\\
d( H _{\2}+ H _{\1} \wedge \mathcal{A}_{\1})=0&\longleftrightarrow d(e^{\phi}\Omega \mathcal{M}  \star  H _{\1} +e^{-\phi}\Omega \mathcal{M}  \star  H _{\2} \wedge \mathcal{A}_{\1})=0,
\end{split}
\ee
thus halving the number of degrees of freedom to 56, as required.

The Lagrangian  \eqref{D3L} and constraint \eqref{D3C} are manifestly invariant under the following set of global transformations:
\begin{subequations}
\begin{align}
 \mathcal{M}&\rightarrow U^T \mathcal{M}U\\
\phi &\rightarrow\phi+2c \\
\mathcal{A}_{\1} &\rightarrow e^{2c}\mathcal{A}_{\1}\\
 \mathbf{A}_{\0}&\rightarrow e^{-c}U^{-1}( \mathbf{A}_{\0}+ \mathbf{X})\\
 \mathbf{A}_{\1}&\rightarrow e^{c}U^{-1} \mathbf{A}_{\1}
\end{align}
\end{subequations}
where  $c\in \text{GL}(1, \R) , U\in E_{7(7)}$, and $\mathbf{X}\in \R^{56}$. We see that the translations $\mathbf{X}\in \R^{56}$ themselves transform as the $\rep{56}$ of $E_{7(7)}$, with weight one under the $\text{GL}(1, \R)$. Putting these together, infinitesimally the total Lie algebra under which \eqref{D3L}-\eqref{D3C} is invariant is given by
\be
\widetilde{\f{e}}_{7(7)\sst{\frac{1}{2}}} := \widetilde{\mathfrak{m}}(\s_s, \Oct_s)=\f{tri}(\s_s)\oplus\f{tri}(\Oct_s) \pd   3 (\s_s\otimes\Oct_s)=[\f{gl}_1(\R)\oplus\f{e}_{7(7)}]_{\0} \pd   \rep{56}_{\1},
\ee
where the split real form of \eqref{tri} (Barton-Sudbery construction) has been recalled, together with the last definition of \eqref{ppain-1}. The sextonions give precisely the correct global symmetry algebra using the very same magic square construction! 

Note that, using the Tits construction,
\be\label{ppain-2}
\widetilde{\mathfrak{m}}(\s_s, \Oct_s)= \f{der}(\Oct_s)\oplus\f{der}(\J_{3}(\s_s) ) \pd   \text{Im} \Oct_s\otimes \J^{'}_{3}(\s_s).
\ee
This algebra can be regarded as the quasi-conformal algebra\footnote{As we are necessarily using the  split composition algebra $\s_s$ in $\J_{3}(\s_s)$, \eqref{ppain-2} is a quasi-conformal algebra in a split-signature generalised spacetime, to use the terminology of \cite{Gunaydin:2005zz}.}, as introduced in \cite{Gunaydin:2000xr}, of the sextonionic Jordan algebra  $\J_{3}(\s)$ using the Freudenthal triple system. This approach will be treated in future work.

The vectors  $H_\2$ can be eliminated in terms of $H_\1$, by first dualising and then imposing the constraint \eqref{D3Cred}, to give a more familiar form of Lagrangian without any constraints (the symmetries are preserved by this process). To dualise, first note that the Bianchi identity for $H_\2$ is given by
\be
d(H_\2+ H_\1 \wedge \mathcal{A}_{\1})=0.
\ee
We can thus introduce a Lagrange multiplier, $\mathbf{C}_{\0}$, transforming in the $\rep{56}$ of $E_{7(7)}$, to write \eqref{D3L} in a first order form for $H_\2$ :
\be\label{D3Lfirst}
\begin{split}
\mathcal{L}_3'=&R\star 1 - \frac{1}{2}\star d\phi \wedge d\phi - \frac{e^{-2\phi}}{2}\star \mathcal{F}_{\2} \wedge \mathcal{F}_{\2} +\frac{1}{4}\tr \left(\star d \mathcal{M}^{-1}\wedge d\mathcal{M}\right)-\frac{e^{ \phi}}{4}  \star  H _{\1}^{T}  \wedge  \mathcal{M}   H _{\1}\\
 &-\frac{e^{-\phi}}{4}  \star  H _{\2}^{T}  \wedge  \mathcal{M}   H _{\2}
+\mathbf{C}_{\0}^T\Omega d(H_\2+ H_\1 \wedge \mathcal{A}_{\1}).
\end{split}
\ee
The relevant system of equations are then given by the Bianchi identity for $H_\2$,
\be
d(H_\2+ H_\1 \wedge \mathcal{A}_{\1})=0,
\ee
 and the equation of motion for $H_\2$,
\be\label{eomH2}
d\mathbf{C}_{\0} = -\frac{e^{-\phi}}{2} \Omega \mathcal{M} \star H_\2.
\ee
The latter, being algebraic,  may be substituted back into \eqref{D3Lfirst}, yielding to
\be\label{D3Lfirst2}
\begin{split}
\mathcal{L}_3'=&R\star 1 - \frac{1}{2}\star d\phi \wedge d\phi - \frac{e^{-2\phi}}{2}\star \mathcal{F}_{\2} \wedge \mathcal{F}_{\2} +\frac{1}{4}\tr \left(\star d \mathcal{M}^{-1}\wedge d\mathcal{M}\right)-\frac{e^{ \phi}}{4}  \star  H _{\1}^{T}  \wedge  \mathcal{M}   H _{\1}\\
 &-e^{\phi}  \star  d\mathbf{C}_{\0} ^{T}  \wedge  \mathcal{M}   d\mathbf{C}_{\0} - d\mathbf{C}_{\0}^T\wedge\Omega  H_\1 \wedge \mathcal{A}_{\1}.
\end{split}
\ee
Using \eqref{eomH2} and the constraint \eqref{D3Cred}, one finally reaches the expression
\be\label{D3Lfirst3}
\begin{split}
\mathcal{L}_3'=&R\star 1 - \frac{1}{2}\star d\phi \wedge d\phi - \frac{e^{-2\phi}}{2}\star \mathcal{F}_{\2} \wedge \mathcal{F}_{\2} +\frac{1}{4}\tr \left(\star d \mathcal{M}^{-1}\wedge d\mathcal{M}\right)-\frac{e^{ \phi}}{2}  \star  H _{\1}^{T}  \wedge  \mathcal{M}   H _{\1}\\
 &+\frac{1}{2} H_\1^T\wedge\Omega  H_\1 \wedge \mathcal{A}_{\1},
\end{split}
\ee
which is invariant under the following set of global transformations:
\begin{subequations}
\begin{align}
 \mathcal{M}&\rightarrow U^T \mathcal{M}U,\\
\phi &\rightarrow\phi+2c, \\
\mathcal{A}_{\1} &\rightarrow e^{2c}\mathcal{A}_{\1},\\
 \mathbf{A}_{\0}&\rightarrow e^{-c}U^{-1}( \mathbf{A}_{\0}+ \mathbf{X}),
\end{align}
\end{subequations}
where $c\in \text{GL}(1, \R) , U\in E_{7(7)},  \mathbf{X}\in \R^{56}$, which infinitesimally constitute the overall Lie algebra $\widetilde{\f{e}}_{7(7)\sst{\frac{1}{2}}}$.  We should also emphasise at this stage that  maximally supersymmetric $\mathcal{N}=16$ supergravity is unique and all the descriptions considered here are on-shell equivalent; at the level of equations of motion  the full $\f{e}_{8(8)}$ symmetry can be made manifest in all cases.

As explained in \autoref{complexms}, $\widetilde{\f{e}}_{7(7)\sst{\frac{1}{2}}}$ in not a subalgebra  of the $\f{e}_{8(8)}$  algebra obtained by first reducing to $D=3$ and then performing the full dualisation; namely:
\be
\widetilde{\mathfrak{m}}(\s_s, \Oct_s)=\f{tri}(\s_s)\oplus\f{tri}(\Oct_s) \pd   3 (\s_s\otimes\Oct_s)
\ee
  is not a subalgebra  of
  \be
  \mathfrak{m}(\Oct_s, \Oct_s)= \f{tri}(\Oct_s)\oplus\f{tri}(\Oct_s)\pd   3 (\Oct_s\otimes\Oct_s).
  \ee
Recall, the $\mathfrak{m}(\Oct_s, \Oct_s)$-subalgebra
\be
\f{e}_{7(7)\sst{\frac{1}{2}}} :=\mathfrak{m}(\s_s, \Oct_s)= \f{tri}_{\s_s}(\Oct_s)\oplus\f{tri}(\Oct_s) \pd   3 (\s\otimes\Oct_s)
\ee
 is obtained from $\widetilde{\mathfrak{m}}(\s_s, \Oct_s)$ by adding a central extension. In terms of \eqref{D3Lfirst3}, this corresponds to  dualising  the graviphoton $\mathcal{A}_{\1}$.  In this case  we recover the full $\mathfrak{e}_{8(8)}$ symmetry,  but with the Lagrangian written  in a manifestly $\mathfrak{e}_{7(7)\sst{\frac{1}{2}}}$ symmetric form, where the central extension effecting $\widetilde{\f{e}}_{7(7)\sst{\frac{1}{2}}} \longrightarrow \mathfrak{e}_{7(7)\sst{\frac{1}{2}}}$ corresponds precisely to the global shift symmetry  of the axion, $\chi$, dual to $\mathcal{A}_{\1}$. Explicitly, introducing the Lagrange multiplier term $\chi d \mathcal{F}_{\2}$ enforcing the Bianchi $d\mathcal{F}_{\2}=0$ and dualising we obtain, 
 \be\label{D3Lfirst4}
\begin{split}
\mathcal{L}_3''=&R\star 1 - \frac{1}{2}\star d\phi \wedge d\phi - \frac{e^{2\phi}}{2}\star G_\1 \wedge G_\1 +\frac{1}{4}\tr \left(\star d \mathcal{M}^{-1}\wedge d\mathcal{M}\right)-\frac{e^{ \phi}}{2}  \star  H _{\1}^{T}  \wedge  \mathcal{M}   H _{\1},
\end{split}
\ee
 where 
 \be
G_\1 = d\chi+\frac{1}{2} H_\1^T\wedge\Omega  \mathbf{A}_{\sst{(0)}}.
 \ee
 The fully dualised Lagrangian \eqref{D3Lfirst4} is manifestly invariant under  the infinitesimal transformations of $\chi$ given by,
 \be
\delta \chi  = -2c\chi - \frac{1}{2} \mathbf{A}_{\sst{(0)}}^T\wedge\Omega  \mathbf{X} +\alpha,
\ee
where $c, \mathbf{X}$ are parameters of the grade-0 $\mathfrak{gl}_1(\R)$ and the grade-1 $\R^{56}$, respectively,  and $\alpha$ belongs to a new $\R$ factor. Computing the commutators, in particular
\be
[\delta^{\mathfrak{gl}_1}_c, \delta^{\R}_\alpha]\chi  = \delta^{\R}_{2c \alpha}\chi,
\ee
one sees that  the new $\R$ factors carries +2 weight under the $\mathfrak{gl}_1(\R)$ and commutes with everything else, promoting the manifest symmetry algebra to  $\mathfrak{e}_{7(7)\frac{1}{2}}$.

\subsubsection{$D=5$ to $D=4$ and $\f{e}_{6(6)\frac{1}{4}}$}\label{D5max}
In \cite{Cremmer:1980gs} it was shown that the maximally supersymmetric $D=5$ supergravity theory has non-compact global symmetry $E_{6(6)}(\R)$.  When dimensionally reducing from $D=11$, in order to make manifest the $\f{e}_{6(6)}$ structure of the scalar sector in $D=5$, one must first dualise the 3-form potential  terms appearing in  \eqref{Lagg}, as described in detail in \cite{Cremmer:1997ct}. This gives a total of 42 scalars parametrising
$E_{6(6)}(\R)/\USp(8)$.


The fully dualised bosonic Lagrangian with manifest $E_{6(6)}(\R)$-invariance can be written as
  \be\label{D5L}
 \mathcal{L}_5 = R\star 1 +\frac{1}{4}\tr \left(\star d \mathcal{M}^{-1}\wedge d\mathcal{M}\right) -\frac{1}{2}  \star F_{\2}^{T}  \wedge  \mathcal{M}  F_{\2}-\frac{1}{6} N(F_{\2}, F_{\2}, A_{\1}).
 \ee
Here $\mathcal{M}$ is built from the 42 scalar fields parametrising the coset $E_{6(6)}(\R)/\USp(8)$. Under the global $E_{6(6)}(\R)$ group, ${\cal M}$ transforms linearly,
 $
 {\cal M} \rightarrow U^T{\cal M}U,
 $
where $U$ is in the \rep{27'} representation of  $E_{6(6)}$, and  is invariant under the local $\USp(8)$. The 1-form potentials transform linearly in the $\rep{27}$ of  $E_{6(6)}$. In addition to the singlet $\rep{1}\in\rep{27'}\times \rep{27}$, used to construct the 1-form kinetic term, there is a singlet in the totally symmetric 3-fold tensor product $\rep{1}\in\rep{27}\times \rep{27}\times \rep{27}$, which is used to construct the topological cubic term. The 1-forms can be consider as elements of the  cubic Jordan algebra $\J_{3}(\Oct_s)$ of $3\times 3$ Hermitian matrices
over $\Oct_s$ and $N: \J_{3}(\Oct_s)\times\J_{3}(\Oct_s)\times\J_{3}(\Oct_s)\rightarrow \R$ is the totally symmetric trilinear cubic norm. See \cite{Borsten:2010aa} for more details. A detailed treatment concerning \eqref{D5L} can be found in \cite{Cremmer:1980gs}.

Dimensionally reducing on a circle, one obtains the $D=4$ Lagrangian,
\be\label{D5to4L}
\begin{split}
\mathcal{L}_4=&R\star 1 - \frac{1}{2}\star d\phi \wedge d\phi - \frac{e^{-\sqrt{3}\phi}}{2}\star \mathcal{F}_{\2} \wedge \mathcal{F}_{\2} +\frac{1}{4}\tr \left(\star d \mathcal{M}^{-1}\wedge d\mathcal{M}\right) -\frac{e^{- \phi/\sqrt{3}} }{2}  \star  F _{\2}^{T}  \wedge  \mathcal{M}   F_{\2}\\
&-\frac{e^{ 2  \phi /\sqrt{3} }}{2}  \star  F _{\1}^{T}  \wedge  \mathcal{M}   F _{\1}+\frac{1}{2}N(dA_\1, A_\1, F_\1),
\end{split}
\ee
where
\be\label{D5to4L-2}
F_{\2} = dA_\1-dA_\0\wedge \mathcal{A}_\1,\quad F_{\1} = dA_\0,\quad  F_{\2}= d\mathcal{A}_\1.
\ee
The $D=4$ Lagrangian \eqref{D5to4L} is invariant under the following set of global transformations:
\begin{subequations}
\begin{align}
 \mathcal{M}&\rightarrow U^T \mathcal{M}U,\\
\phi &\rightarrow\phi+2\sqrt{3}c, \\
\mathcal{A}_{\1} &\rightarrow e^{3c}\mathcal{A}_{\1},\\
{A}_{\1} &\rightarrow e^{c}U^{-1}{A}_{\1},\\
 {A}_{\0}&\rightarrow e^{-2c}U^{-1}( {A}_{\0}+ X),
\end{align}
\end{subequations}
where $c\in \text{GL}(1, \R) , U\in E_{6(6)},  \mathbf{X}\in \R^{27}$. The overall Lie algebra under which \eqref{D5to4L}-\eqref{D5to4L-2} is invariant is thus given by  the ternionic slot of the extended magic square, $\f{m}(\T_s, \Oct_s)$ (cfr. the split form of the last of \eqref{pppain-1}):
\be
\f{e}_{6(6)\frac{1}{4}}:=\f{m}(\T_s, \Oct_s)\cong[\f{gl}_1(\R)\oplus\f{e}_{6(6)}]_\0\pd\rep{27}_\2.
\ee

By the symmetry of the extended magic square given in \autoref{tab:emsreal1}, and using the Tits construction,
\be\label{pppain-2}
\mathfrak{m}(\T_s, \Oct_s)= \f{der}(\Oct_s)\oplus\f{der}(\J_{3}(\T_s) ) \pd   \text{Im} \Oct_s\otimes \J^{'}_{3}(\T_s).
\ee
This algebra can be regarded as the quasi-conformal algebra\footnote{As we are necessarily using the  split composition algebra $\T_s$ in $\J_{3}(\T_s)$, \eqref{pppain-2} is a quasi-conformal algebra in a split-signature generalised spacetime, to use the terminology of \cite{Gunaydin:2005zz}.}, as introduced in \cite{Gunaydin:2000xr}, of the ternionic  Jordan algebra  $\J_{3}(\T_s)$ using the Freudenthal triple system. This approach will be treated in future work.

One might anticipate that by reducing the amount of supersymmetry more of extended magic square would appear. Indeed, it was shown in \cite{Julia:1980gr} that   the algebras of the  magic square do appear, but  with the ``wrong'' real forms in the sense that are not given by any magic square formula, as classified in \cite{Cacciatori:2012cb}, and therefore  cannot be included in the extended magic squares of   \autoref{realms}.  However, relaxing the requirement of supersymmetry, the full $6\times6$ $\alg_1, \alg_2= \R, \C_s, \T_s, \Q_s, \s_s, \Oct_s$   square of  \autoref{tab:emsreal1} appears, as we describe in the following section (apart from the difference between $\mathfrak{ext}$ and $\widetilde{\mathfrak{ext}}$ extremal intermediate algebras; cfr. Sec. 2.1).

\subsection{Maximally non-compact bosonic (\emph{aka} magic non-supersymmetric) theories in $D=5,4,3$}

As emphasised above, the real forms obtained by oxidising $D=3$ supergravity theories yield symmetries in $D=5,4$  with real forms that cannot be accommodated by any (at least conventional) magic square.  However, following \cite{Cremmer:1999du} and dispensing with the requirement of supersymmetry, the full maximally non-compact extended magic square given by Table 3 appears. Since the principles involved are much the same, we will be more telegraphic in this subsection.

In \cite{Cremmer:1999du} all $D=3$ gravity theories coupled to a $\mathcal{G/H}$ scalar sigma model, where $\mathcal{G}$ is maximally non-compact, were considered\footnote{These theories, also named 'magic non-supersymmetric' Maxwell-Einstein gravity theories, have been recently discussed in \cite{Marrani:2017aqc}.}. For each such theory the highest dimension $D_{\text{max}}$ from which it could be obtained via Kaluza-Klein reduction on a torus was determined. The non-compact global symmetries of each  theory in $3\leq D\leq D_{\text{max}}$ were determined starting from the Dynkin diagram of the seed $D=3, \mathcal{G}$ theory. For example, the $D=3, E_{8(8)}$ theory can be oxidised all the way to $D=11,\varnothing$, where it is given by the bosonic subsector of $D=11$, $\N=1$ supergravity. Each intermediate theory in $3<D<11$ has global symmetry $E_{n(n)}$, where $n=11-D$, and  corresponds to the bosonic subsector of the maximally supersymmetric theory in that dimension.  The $D=3, E_{7(7)}$ theory, however, can only be oxidised to $D=10$, where it is given by a consistent, but non-supersymmetric, truncation of type IIB supergravity.  Aside from these two examples, starting from $D=3$, the $E_{n(n)}$ theory generically has  $D_{\text{max}}=n+2$, the third exception being the $D=3, E_{4(4)}\cong\SL(5, \R)$ theory, which admits a further oxidisation to $D=7$.   Let the global symmetry of the $D$-dimensional theory, where $3\leq D\leq D_{\text{max}}$, obtained by oxidising the $D=3, E_{n(n)}$ theory be denoted $\mathcal{G}(D, n)$. The global symmetries of this set of theories display a remarkable symmetry under  $D \rightarrow 11-n, n\rightarrow 11-D$:
\be
\mathcal{G}(D, n)=\mathcal{G}(11-n,  11-D).
\ee
Hence, we obtain  a ``magic triangle'' of  groups  with a perfect symmetry under reflection about the diagonal $D=11-n$ \cite{Cremmer:1999du}; this symmetry has been recently investigated in detail, also in relation to Kac-Moody very extended algebras and Ehlers truncations, in \cite{Marrani:2017aqc}. It is  worth remarking here that, descending from $D=D_{\text{max}}$ via Kaluza-Klein reduction, the symmetries $\mathcal{G}(D, n)$ derived in \cite{Cremmer:1999du} only appear after the appropriate dualisations have been performed.

The Lie algebras of $\mathcal{G}(D, n)$ for the subset of theories with $D=3,4,5$ and $n=8,7,6$ are given by the $\alg_1, \alg_2=\C_s, \Q_s, \Oct_s$ portion of the conventional magic square (namely, a $3\times3$ subsquare of the doubly-split magic square \cite{Barton:2003}; also see \cite{Cacciatori:2012cb}). If rather than continuing the sequence  with $E_{5(5)}=SO(5,5)$, we  use the $D=3, F_{4(4)}$ theory of \cite{Cremmer:1999du} (i.e., $D=3$, $\N=4$ magic supergravity theory based on $\J_3(\R)$ \cite{Gunaydin:1983rk, Gunaydin:1983bi, Gunaydin:1984ak}), we obtain a forth column, giving a set of theories in $D=3,4,5$ with global symmetry algebras given by the $\alg_1=\C_s, \Q_s, \Oct_s$ and $\alg_2=\R, \C_s, \Q_s, \Oct_s$ portion of the Freudenthal-Rosenfeld-Tits magic square, once again included in the aforementioned doubly-split magic square \cite{Barton:2003}; this is given in \autoref{bosnicsq} \footnote{Note that, strictly speaking, \autoref{bosnicsq} cannot be considered as a sub-Table of Table 3, due to the difference between $\widetilde{\mathfrak{ext}}$ (characterising the sextonionic row of \autoref{bosnicsq}) and $\mathfrak{ext}$ (characterising the ternionic row of \autoref{bosnicsq}) extremal intermediate algebras, as discussed in Sec. 2.1.}.

 \begin{table}[ht]
 \begin{center}
\begin{tabular}{|c|cccccc|}
\hline
\hline
  && $\R$ &$\C_s$   & $\Q_s$ & $\Oct_s$ & \\
 \hline
 &&&&&&\\
  $D=5, \C_s$ && $\mathfrak{sl}_3(\R)$ & $\mathfrak{sl}_3(\R)\oplus \mathfrak{sl}_3(\R)$   & $\mathfrak{sl}_6(\R)$ & $\mathfrak{e}_{6(6)}$   &\\[12pt]
    $D=4', \T_s$ && $\mathfrak{sl}_{3 \frac{1}{4}}(\R)$ & $[\mathfrak{sl}_3(\R)\oplus \mathfrak{sl}_3(\R)]_{\frac{1}{4}}$   & $\mathfrak{sl}_{6 \frac{1}{4}}(\R)$ & $\mathfrak{e}_{6(6)\frac{1}{4}}$   &\\[12pt]

  $D=4, \Q_s$ && $\mathfrak{sp}_6(\R)$ & $\mathfrak{sl}_6(\R)$  & $\mathfrak{so}_{6,6}$ & $\mathfrak{e}_{7(7)}$  & \\[12pt]
  $D=3', \s_s$ && $\widetilde{\mathfrak{sp}}_{6 \frac{1}{2}}(\R)$ & $\widetilde{\mathfrak{sl}}_{6 \frac{1}{2}}(\R)$  & $\widetilde{\mathfrak{so}}_{6,6\frac{1}{2}}$ & $\widetilde{\mathfrak{e}}_{7(7)\frac{1}{2}}$  & \\[12pt]

      $D=3, \Oct_s$ && $\mathfrak{f}_{4(4)}$ & $\mathfrak{e}_{6(6)}$ & $\mathfrak{e}_{7(7)}$ & $\mathfrak{e}_{8(8)}$&   \\[12pt]
   \hline
   \hline
\end{tabular}
\caption{Global symmetry algebras obtained by oxidising the $D=3$ bosonic theories with $\mathcal{G}=F_{4(4)}, E_{6(6)}, E_{7(7)}, E_{8(8)}$ to $D=4,5$. The non-reductive Lie algebras in the $D=4', \T_s$ row correspond to the symmetries of the Lagrangian obtained by Kaluza-Klein reducing the fully dualised $D=5$ theory with manifest symmetry given by the $D=5, \C_s$ row. The non-reductive Lie algebras in the $D=3', \s_s$ row correspond to the symmetries of the Lagrangian obtained by Kaluza-Klein reducing the fully dualised $D=4$ theory with global symmetry algebra given by the $D=4, \Q_s$ row, followed by a dualisation of the resulting 1-form potentials, excluding the graviphoton. \label{bosnicsq}}
 \end{center}
\end{table}

Again, it is important to keep in mind that, when starting from  $D_{max}$  and descending via Kaluza-Klein reductions, the algebras of  \autoref{bosnicsq} only appear once all the required dualisations have been performed. We can therefore  consider the global symmetry algebras, as they appeared in the treatment of \autoref{max}. In \autoref{max}, we have already discussed the $\Oct_s$ column as it is obtained from the bosonic subsector of $D=11$, $\N=1$ supergravity. Let us now consider the $\Q_s$ column starting with $\f{sl}_6(\R)$ in $D=5$. As shown in \cite{Cremmer:1999du}, the $D=3, E_{7(7)}$ theory oxidises to a consistent non-supersymmetric truncation of $D=9$ maximal supergravity, which can be  obtained by reducing on $S^1$ type IIB supergravity truncated to the metric and self-dual 5-form only. The Lagrangian of the $D$-dimensional theory  given by Kaluza-Klein reduction on an $(9-D)$-torus is given by ($i,j,k=1,...,9-D$)
\bea\label{LE7}\nn
\frac{{\cal L}}{\sqrt{-g}} &=& R -\ft{1}{2} \, (\del\vec\phi)^2  -\ft{1}{4}\, \sum_i e^{\vec b_i\cdot \vec\phi}\, ({\cal F}_\2^i)^2
-\ft{1}{2}\, \sum_{i<j} e^{\vec b_{ij}\cdot \vec\phi}\,
({\cal F}_\1^i{}_j)^2\\\nn
&&
-\ft{1}{48}\, e^{\vec a\cdot
\vec\phi}\, F_\4^2 -\ft{1}{12} \sum_i
e^{\vec a_i\cdot \vec\phi}\, (F_{\3 i})^2\\
&&-\ft{1}{4}\, \sum_{i<j} e^{\vec a_{ij}\cdot \vec\phi}\, (F_{\2ij})^2-\ft{1}{4} e^{\vec c\cdot \vec\phi}\, (F_{\2})^2\\\nn
&&-\ft{1}{2} \, \sum_{i<j<k} e^{\vec a_{ijk} \cdot\vec \phi}\,
(F_{\1ijk})^2 -\ft{1}{2} \, \sum_{i} e^{\vec c_{i} \cdot\vec \phi}\,
(F_{\1i})^2 \label{dgenlag}\\\nn
&&
 + {\cal L}_{\sst{FFA}}\ ,\nn
\eea
  In $D=5$ (post dualisation of the 3-form potential), one obtains five dilatonic scalars and $15=1+4+4+6$ axionic scalars which couple through the dilaton vectors,
  \be\label{life-1}
  -\vec a, \vec c_i, \vec b_{ij}, \vec a_{ijk},\qquad i,j,k=1,2,3,4,
  \ee
  which constitute a set of  positive roots for $\f{sl}_6(\R)$; for details, we address the reader to \cite{Cremmer:1999du}. There are also 15 vectors, which couple through,
  \be\label{life-2}
 - \vec a_i, \vec c, \vec b_{i}, \vec a_{ij},
  \ee
  which form a set of   weight vectors for the $\rep{15}$ (rank-2 antisymmetric irrepr.) of $\f{sl}_6(\R)$. The Lagrangian can then be written in a manifestly $\SL(6, \R)$-invariant form with the 20 scalar parametrising $\SL(6, \R)/\SO(6)$. Reducing further to $D=4$, we  obtain a  dilaton and 15 further axions, giving a total of 36 scalar fields. These parametrise the coset $\SO(6,6)/[\SO(6)\times\SO(6)]$. We also obtain a single additional 1-form potential, the $D=4$ graviphoton. This gives a total of 16 vectors, which appear with 16 dilaton vectors that form a set of positive weight vectors for the $\rep{32}$ of $\f{so}(6,6)$. On-shell, we can unify the 16 2-form field strengths with their electromagnetic duals so that they transform as the $\rep{32}$ of $\f{so}(6,6)$. However, at the Lagrangian level this is generally broken (while retaining general coordinate invariance) and  the largest non-compact global symmetry algebra is given by
 \be\label{life-3}
 [\f{gl}_1(\R)\oplus\f{sl}_6(\R)]_\0\pd \rep{15}_\2
 \ee
 as can be read off from the dilaton vectors \eqref{life-1}-\eqref{life-2}. By recalling the treatment of Sec. 3.1, one can recognize \eqref{life-3} to be nothing but the $(\T_s, \Q_s)$ entry of the real form of the extended magic square given by Table 3, named $\mathfrak{sl}_{6 \frac{1}{4}}(\R)$. From the symmetry of \autoref{tab:emsreal1}, and using the Tits construction, one can write
\be
\mathfrak{sl}_{6 \frac{1}{4}}(\R) \cong {\mathfrak{m}}(\T_s, \Q_s)= {\mathfrak{m}}(\Q_s, \T_s)=\f{der}(\Q_s)\oplus\f{der}(\J_{3}(\T_s) ) \pd   \text{Im}\Q_s\otimes \J^{'}_{3}(\T_s),
\ee
and thus recover this entry as the conformal algebra \cite{Gunaydin:1992zh} of the ternionic  Jordan algebra  $\J_{3}(\T_s)$. We will deal with these aspects more in detail in a future work.

In  $D=3$, \eqref{LE7} has  $41= 6+15+20$ axions, which couple to the six dilatons through
\be
\vec c_i, \vec b_{ij}, \vec a_{ijk},\qquad i,j,k=1,\ldots 6.
  \ee
There are $22=1+6+15$ vectors, which couple to the six dilatons through
\be
\vec c, \vec b_{i}, \vec a_{ij}.
  \ee
 Dualising the 22 vectors the associated dilaton vectors flip sign and we have in total $63$ axions, which couple to the seven dilatons through
\be
-\vec c, -\vec b_{i}, -\vec a_{ij}, \vec c_i, \vec b_{ij}, \vec a_{ijk},
  \ee
  which form a set of positive roots for $\mathfrak{e}_{7(7)}$ and the $70=7+63$ scalars parametrise $E_{7(7)}(\R)/\SU(8)$ \cite{Cremmer:1999du}.  If, however, we first stop in $D=4$ and dualise to make the $\SO(6,6)$ symmetry manifest (either going on-shell, or using the ``doubled'' 1-form and twisted-self-duality constraint, as described in \autoref{e71/2}) and then reduce to $D=3$, we are left with 36 scalars belonging to $\SO(6,6)/[\SO(6)\times\SO(6)]$, 16 scalars that correspond to the positive weight vectors of the  $\rep{32}$ of $\f{so}(6,6)$, a dilaton, 16 1-form potentials that correspond to the negative weight vectors of the  $\rep{32}$ of $\f{so}(6,6)$, and the  graviphoton 1-form. Dualising the 16 1-form potentials to scalars, one obtains in total 36 scalars belonging to $\SO(6,6)/[\SO(6)\times\SO(6)]$ and 32 scalars in the  $\rep{32}$ of $\SO(6,6)$. In direct analogy to the maximally supersymmetric case described in \autoref{e71/2}, the resulting overall global symmetry algebra is given by
  \be
  \widetilde{\mathfrak{so}}_{6,6\frac{1}{2}} \cong \widetilde{\mathfrak{m}}(\s_s, \Q_s) = \f{tri}(\s_s)\oplus \f{tri}(\Q_s) \pd 3 (\s_s\otimes\Q_s)
  \ee entry of the real form of the extended magic square given by Table 3. Equivalently, from the (split form specialization of the) discussion in Sec. 3.1, and using the Tits construction, one can characterize
  \be
  \widetilde{\mathfrak{so}}_{6,6\frac{1}{2}} \cong \widetilde{\mathfrak{m}}(\Q_s, \s_s) = \f{der}(\Q_s)\oplus\f{der}(\J_{3}(\s_s) ) \pd   \text{Im} \Q_s\otimes \J^{'}_{3}(\s_s)
  \ee
as the conformal algebra of $\J_{3}(\s_s)$. As mentioned above, a detailed investigation of these aspects will be performed elsewhere.

    Repeating this analysis for the $D=3, F_{4(4)}$ and $D=3, E_{6(6)}$ sequences, one obtains the split extended magic square of global symmetry algebras as given in \autoref{bosnicsq}, with the $\R$ and $\C_s$ columns given by automorphism  and reduced structure algebras,  respectively, of $\J_{3}(\alg_s)$. In particular, selecting $\alg_s=\T_s, \s_s$, one obtains :
    \be
    \begin{array}{lcccccccccc}
        &\text{Automorphism algebra of } \J_{3}(\alg_s)&\text{Reduced structure algebra of } \J_{3}(\alg_s)\\[8pt]
    D=4'&\mathfrak{sl}_{3 \frac{1}{4}}(\R)=\f{der}(\J_{3}(\T_s) )&[\mathfrak{sl}_3(\R)\oplus \mathfrak{sl}_3(\R)]_{\frac{1}{4}}=\f{der}(\J_{3}(\T_s) ) \pd   \J^{'}_{3}(\T_s)=\f{str}_0(\J_{3}(\T_s) )\\[8pt]
        D=3'&\widetilde{\mathfrak{sp}}_{6 \frac{1}{2}}(\R)=\f{der}(\J_{3}(\s_s) )&\widetilde{\mathfrak{sl}}_{6 \frac{1}{2}}(\R)=\f{der}(\J_{3}(\s_s) ) \pd   \J^{'}_{3}(\s_s)=\f{str}_0(\J_{3}(\s_s) )    \end{array}
    \ee
    For the meaning of the priming of $D$, cfr. the caption of Table 7. The symmetries of $\J_{3}(\T_s)$ and $\J_{3}(\s_s)$ will be investigated in a future work.

\subsection{Magic theories in $D=5,4,3$}

It has been well known since the seminal work of G\"{u}naydin,
Sierra and Townsend \cite{Gunaydin:1983bi,
Gunaydin:1983rk,Gunaydin:1984ak} that the geometry of 5-dimensional
$\mathcal{N}=2$ supergravity coupled to $n_V$ vector multiplets is
intimately related to the class of \emph{cubic} Jordan algebras. The $n_V$-dimensional space
$\mathcal{M}$ of scalar fields is given by a hypersurface with
vanishing fundamental second form in an $(n_V+1)$-dimensional
Riemannian space $\xi$. The hypersurface is defined by a homogeneous
cubic polynomial $N(X)=1$ in the coordinates $\{X\}$ of $\xi$. The
cubic norm $N$ may be used to construct a cubic (Euclidean \footnote{$\mathcal{N}=2$ Maxwell-Einstein supergravity theories in $D=4$ and $5$ based on cubic Lorentzian Jordan algebras $\J_{1,2}(\alg) \cong \J_{2,1}(\alg)$ have been considered, and their vector multiplets' scalar manifolds are non-homogeneous \cite{Gunaydin:2003yx,Gunaydin:2005df,Gunaydin:2005bf}.}) Jordan algebra
$\J_3$, which as a vector space is isomorphic to $\xi$. If   the
moduli space is assumed to be symmetric then
$\mathcal{M}=\Str_0(\J_3)/\Aut(\J_3)$, where $\Str_0(\J_3)$ and
$\Aut(\J_3)$ are the reduced structure (norm preserving) and the
automorphism  (Jordan product preserving) groups of $\J_3$,
respectively. $\Aut(\J_3)$ is the maximal compact subgroup of
$\Str_0(\J_3)$. In this case, for $\mathcal{N}=2$ supergravity coupled to $n_V$ vector multiplets with a \emph{non-factorisable}  cubic norm there are only four possibilities, $n_V=3\dim \alg+2$, where
$\alg=\R, \C, \Q, \Oct$ \cite{Jordan:1933vh}. These four theories are referred to as the ``magic''
supergravities, because their symmetries correspond to  a line of the single-split real form of the
Freudenthal-Rosenfeld-Tits magic square \cite{Gunaydin:1983bi,
Gunaydin:1983rk,Gunaydin:1984ak}.
Their Jordan algebras are given by $3\times 3$ Hermitian matrices
over $\alg=\R, \C, \Q, \Oct$, which we denote $\J_3(\alg)$:
\begin{equation}\label{eq:cubicnormJ33}
X=\begin{pmatrix}\alpha&c&\overline{b}\\\overline{c}&\beta&a\\
b&\overline{a}&\gamma\end{pmatrix}, \quad\text{where}\quad \alpha,
\beta, \gamma \in \R\quad\text{and}\quad a, b, c\in\alg.
\end{equation}
The cubic norm is defined as
\begin{equation}\label{eq:cubicnormexp}
N(X):=\alpha\beta\gamma-\alpha a\overline{a}-\beta b\overline{b}-\gamma c \overline{c} +(ab)c+\overline{c}(\overline{b}\overline{c}), \end{equation}
which for associative $\alg$ coincides with the matrix determinant. The Jordan product is given by
\begin{equation}X\circ Y=\frac{1}{2}(XY+YX),\qquad X, Y \in \mathfrak{J}^{\mathds{A}}_{3},
\end{equation}
where juxtaposition denotes the conventional matrix product.

The maximally supersymmetric $D=5, \mathcal{N}=8$ theory treated in \autoref{D5max} can  be defined in terms of the Jordan algebra $\J(\Oct_s)$ in a directly analogous fashion. Structurally, the $D=5$ magic supergravities and the maximally supersymmetric $D=5, \mathcal{N}=8$ theory are the same, but with the split composition algebra $\Oct_s$ replaced by one of the normed division algebras $\R, \C, \Q, \Oct$. Consequently, the analysis of \autoref{D5max} goes through without modification for all cases $\alg=\R, \C, \Q, \Oct$ except that the associated real forms are different. The resulting global symmetry algebras are presented in \autoref{magicsugra}. Note again that Tables 7 and 8 are not sub-tables of Tables 3 and 4a), due to the presence of tilded intermediate extremal algebras in their sextonionic row. In particular, while the sextonionic row of Tables 3,4,5 is a subalgebra of the octonionic row, the sectonionic row of Tables 7 and 8 is not.

 \begin{table}[ht]
 \begin{center}
\begin{tabular}{|c|cccccc|}
\hline
\hline
  && $\R$ &$\C$ &  $\Q$  & $\Oct$ & \\
 \hline
 &&&&&&\\
  $D=5, \C_s$ && $\mathfrak{sl}_3(\R)$ & $\mathfrak{sl}_3(\C)$   & $\mathfrak{su}^{\star}_6$ & $\mathfrak{e}_{6(-26)}$   &\\[12pt]
    $D=4', \T_s$ && $\mathfrak{sl}_{3\frac{1}{4} }(\R)$ & $ \mathfrak{sl}_{3\frac{1}{4} }(\C)$   & $\mathfrak{su}^{\star}_{6\frac{1}{4} }$  & $\mathfrak{e}_{6(-26)\frac{1}{4} }$   &\\[12pt]
  $D=4, \Q_s$ && $\mathfrak{sp}_6(\R)$ & $\mathfrak{su}_{3,3}$  & $\mathfrak{so}^{\star}_{12}$  & $\mathfrak{e}_{7(-25)}$  & \\[12pt]
  $D=3', \s_s$ && $\widetilde{\mathfrak{sp}}_{6\frac{1}{2}}(\R)$ & $\widetilde{\mathfrak{su}}_{3,3\frac{1}{2}}$ & $\widetilde{\mathfrak{so}}^{\star}_{12\frac{1}{2}}$ & $\widetilde{\mathfrak{e}}_{7(-25)\frac{1}{2}}$  & \\[12pt]
   $D=3, \Oct_s$ && $\mathfrak{f}_{4(4)}$ & $\mathfrak{e}_{6\2}$  & $\mathfrak{e}_{7(-5)}$ & $\mathfrak{e}_{8(-24)}$&   \\[12pt]
   \hline
   \hline
\end{tabular}
\caption{Global symmetry algebras of the magic supergravities in  $D=3, 3', 4, 4', 5$. The non-reductive Lie algebras in the $D=4', \T_s$ row correspond to the symmetries of the Lagrangian obtained by Kaluza-Klein reducing the fully dualised $D=5$ theory with manifest symmetry given by the $D=5, \C_s$ row. The non-reductive Lie algebras in the $D=3', \s_s$ row correspond to the symmetries of the Lagrangian obtained by Kaluza-Klein reducing the fully dualised $D=4$ theory with global symmetry algebra given by the $D=4, \Q_s$ row, followed by a dualisation of the resulting 1-form potentials, excluding the graviphoton.\label{magicsugra}}
\end{center}
\end{table}

\section{Conclusions}

We have   introduced an extended  $6\times 6$ magic square of not necessarily reductive Lie algebras, as given in   \autoref{tab:ems1}, generalising the original $4\times 4$ magic square
of Freudenthal-Rosenfeld-Tits  \cite{Freudenthal:1954,Tits:1955, Freudenthal:1959,Rosenfeld:1956, Tits:1966}. The additional  rows and columns are provided by two intermediate composition algebras:  $(i)$ the three-dimensional ternions $\T$, which sit in-between the complexes and quaternions, $\C\subset \T\subset \Q$ and $(ii)$ the six-dimensional sextonions $\s$, which sit in-between the  quaternions and octonions, $\Q\subset \s\subset \Oct$. Using the ternions and sextonions in the magic square formulae of Barton-Sudbery \cite{Barton:2003}, Vinberg \cite{Vinberg:1966} or Tits \cite{Tits:1966} \eqref{msconstructions}, we obtained the $6\times 6$ array of non-reductive Lie algebras described in detail in \autoref{complexms}. For the sextonions there is a subtlety : $\f{tri}(\s)$ is not a subalgebra of $\f{tri}(\Oct)$ and hence the $(\s, \alg)$-entry of \eqref{msconstructions} in which at least one of $\alg_1$ and $\alg_2$ is $\s$, denoted as $\widetilde{\mathfrak{m}}(\s, \alg) \cong \widetilde{\mathfrak{m}}(\alg, \s)$ (cfr. \eqref{tri}) is not a subalgebra of $\mathfrak{m}(\Oct, \alg)$. This can be remedied following \cite{landsberg2006sextonions}, namely by using the $\s$-preserving subalgebra $\f{tri}_\s(\Oct)\subset\f{tri}(\Oct)$ and defining $\mathfrak{m}(\s, \alg)=\f{tri}_\s(\Oct)\oplus\f{tri}(\alg) \pd   3 (\s\otimes\alg)$ (cfr. \eqref{tri-pre}), which is a Lie subalgebra of $\mathfrak{m}(\Oct, \alg)$. Then, as discussed in full generality in Sec. 2.1, $\widetilde{\mathfrak{m}}(\s, \alg)$ is related to $\mathfrak{m}(\s, \alg)$ by a quotient of a one-dimensional ideal.

With a view to their application to supergravity,  we classified all possible real forms of the extended magic square; see \autoref{tab:emsreal1}, \autoref{tab:emsreal3}  and \autoref{tab:emsreal5}. Since the ternions and sextonions over the reals only exist in a split form $\T_s, \s_s$, the full set of possibilities given in \cite{Cacciatori:2012cb} is restricted to these eight cases (with Table 3 counting twice; cfr. the end of its caption). It was then shown that two of these  real form (semi)-extended magic squares provide the global symmetry algebras in $D=5,4,3$ of maximally supersymmetric supergravity, of bosonic maximally non-compact coset (also named magic non-supersymmetric) theories \cite{Cremmer:1999du, Marrani:2017aqc}, and of the G\"unaydin-Sierra-Townsend $\mathcal{N}=2$ magic supergravities \cite{Gunaydin:1983rk, Gunaydin:1983bi, Gunaydin:1984ak} (cfr. Tables 7 and 8). The non-reductive Lie algebras occurring in the $\T_s$- and $\s_s$- rows appear either because the particular choice of dualisations of the various $p$-form potentials and/or because electromagnetic duality prevents the full global symmetry of the equations  of motions from manifesting itself at the Lagrangian level. An important feature  is that dualisation and Kaluza-Klein reduction do not commute with respect to the manifest symmetries of the Lagrangian. The example of $D=11$ nicely illustrates this point. If we reduce to $D=3$ and leave all 1-forms potentials as they come we have global symmetry algebra $\f{gl}_1(\R)\oplus\f{sl}(8, \R)\ltimes\R^{56}$.  If we then dualise all 1-form potentials to 0-form potentials we have global symmetry algebra $\f{e}_{8(8)}$ given by $\f{m}(\Oct_s, \Oct_s)$. If, however, we first stop in $D=4$ and dualise  all 1-form potentials to 0-form potentials (as well as dualising the pseudo-vectors to vectors) and only then continue down to $D=3$, we are left with the  non-reductive  algebra  $\tilde{\f{e}}
_{7\frac{1}{2}}$ given by the sextonionic entry $\tilde{\f{m}}(\s_s, \Oct_s)$, which is not a subalgebra of $\f{e}_{8(8)} \cong \f{m}(\Oct_s, \Oct_s)$.

There are a number of directions for future work. In particular, we intend to construct the rotation, reduced structure, conformal and quasi-conformal realisations, as introduced in \cite{Gunaydin:1992zh, Gunaydin:2000xr},  of the  ternionic and sextonionic entries of the extended magic square. Moreover, the sextonionic geometry  has many interesting features \cite{landsberg2006sextonions}, akin to yet distinct from that of the octonions,
and it would be of interest to explore how these might arise in, say, the duality orbits  of black hole solutions where we already know that  octonionic geometry  plays an important role.

An  obvious extension of the present analysis is to consider the Jordan algebra of $2\times 2$ Hermitian matrices $\J_2(\alg)$ for $\alg=\T,  \s$, which would yield the Kantor-Koecher-Tits and extremal intermediate algebras of $\f{so}(6, \C)$ and $\f{so}(10, \C)$ (or $\f{so}(3,3)$ and $\f{so}(5, 5)$ over the reals), respectively,  as their reduced structure algebras. Using $\J_2(\alg)$, for $\alg=\R, \C, \T, \s, \Oct$, in the Tits construction will yield a ``reduced'' (order two \cite{Barton:2003}) extended magic square, including the rotation, reduced structure, conform and quasi-conformal algebras of $\J_2(\T), \J_2(\s)$. This will developed in detail in future work. The $2\times 2$ Jordan algebras appear in the construction of $D=6$ maximal and magic supergravity theories \cite{Gunaydin:2010fi} and one would naturally  anticipate an application in this context, also extended to include the $D=6$ uplifts of magic non-supersymmetric bosonic theories.

One could also use $\J_3(\T)$, $\J_3(\s)$ to construct novel theories in $D=5$ dimensions in the spirit of the original treatment of the magic supergravites \cite{Gunaydin:1983rk, Gunaydin:1983bi, Gunaydin:1984ak}. These would not be supersymmetric, but would nonetheless share many common structural features. However, they would also have some rather novel aspects, principally due to the degenerate nature of the cubic norm in these cases. How exactly these theories would behave remains a question for future work.

We should also like to mention cubic Jordan algebras with Lorentzian norm, for which we presented the real forms (cfr. \autoref{tab:emsreal1} and \autoref{tab:emsreal3}) of the extended magic square (\autoref{tab:ems1}). After \cite{Gunaydin:2003yx, Gunaydin:2005df, Gunaydin:2005bf}, it is known that the Maxwell-Einstein supergravity theories based on such Jordan algebras have non-homogeneous vector multiplets' scalar manifolds. One could use $\J_{2,1}(\T)\cong\J_{1,2}(\T)$ and $\J_{2,1}(\s)\cong\J_{1,2}(\s)$ to construct theories in $D=4$ and $5$ dimensions, in the spirit of \cite{Gunaydin:2003yx, Gunaydin:2005df, Gunaydin:2005bf}. Again, these would not be supersymmetric, but would nonetheless share many features with magic supergravities. The detailed investigation of such theories is left to the future.

Finally,  a separate possible role that will explored in subsequent work is the relevance of the non-reductive algebras of the extended magic square to light-like Kaluza-Klein reductions\footnote{We thank Tristan McLoughlin for this suggesting this possibility.}.

\section*{Acknowledgements}

We are grateful to  Laurent~Manivel for stimulating discussions that initiated this project. The work of LB is supported by a Sch\"{o}dinger Fellowship. We also acknowledge  `Queen' for inspirational music.


\begin{thebibliography}{10}

\bibitem{Freudenthal:1954}
H.~Freudenthal, ``{Beziehungen der $E_7$ und $E_8$ zur oktavenebene I-II},''
  {\em Nederl. Akad. Wetensch. Proc. Ser.} {\bf 57} (1954)  218--230.

\bibitem{Tits:1955}
J.~Tits, ``{Interpr\'{e}tation g\'{e}om\'{e}triques de groupes de Lie simples
  compacts de la classe $E$},'' {\em M\'{e}m. Acad. Roy. Belg. Sci} {\bf 29}
  (1955)  3.

\bibitem{Freudenthal:1959}
H.~Freudenthal, ``{Beziehungen der $E_7$ und $E_8$ zur oktavenebene IX},'' {\em
  Nederl. Akad. Wetensch. Proc. Ser.} {\bf A62} (1959)  466--474.

\bibitem{Rosenfeld:1956}
B.~A. Rosenfeld, ``{Geometrical interpretation of the compact simple Lie groups
  of the class $E$}(russian),'' {\em Dokl. Akad. Nauk. SSSR} {\bf 106} (1956)
  600--603.

\bibitem{Tits:1966}
J.~Tits, ``Alg\'ebres alternatives, alg\'ebres de jordan et alg\'ebres de lie
  exceptionnelles,'' {\em Indag. Math.} {\bf 28} (1966)  223--237.

\bibitem{kleinfeld1968extensions}
E.~Kleinfeld, ``On extensions of quaternions,'' {\em Indian J. Math} {\bf 9}
  (1968) no.~443-446, 1967.

\bibitem{jeurissen}
R.~Jeurissen, ``{The automorphism groups of octave algebras},'' {\em Doctoral
  dissertation, University of Utrecht} (1970)  .

\bibitem{racine1974maximal}
M.~Racine, ``On maximal subalgebras,'' {\em Journal of Algebra} {\bf 30} (1974)
  no.~1, 155--180.

\bibitem{westbury2006sextonions}
B.~W. Westbury, ``Sextonions and the magic square,'' {\em Journal of the London
  Mathematical Society} {\bf 73} (2006) no.~2, 455--474.

\bibitem{landsberg2006sextonions}
J.~M. Landsberg and L.~Manivel, ``The sextonions and $e_{7 \frac{1}{2}}$,''
  {\em Advances in Mathematics} {\bf 201} (2006) no.~1, 143--179.

\bibitem{Marrani:2015nta}
A.~Marrani and P.~Truini,
  \href{http://dx.doi.org/10.1007/s11005-017-0966-7}{``Sextonions, zorn
  matrices, and $e_{7\frac{1}{2}}$,''{\em Letters in Mathematical Physics}
  (May, 2017)  }, \href{http://arxiv.org/abs/1506.04604}{{\tt arXiv:1506.04604
  [math.RA]}}.
\url{http://dx.doi.org/10.1007/s11005-017-0966-7}.

\bibitem{dray2015geometry}
T.~Dray and C.~A. Manogue, {\em The geometry of the octonions}.
\newblock World Scientific, 2015.

\bibitem{deligne1996exceptional}
P.~Deligne, ``The exceptional series of lie groups,'' {\em Comptes rendus de
  l'Acad{\'e}mie des sciences. S{\'e}rie 1, Math{\'e}matique} {\bf 322} (1996)
  no.~4, 321--326.

\bibitem{cohen1996computational}
A.~M. Cohen and R.~de~Man, ``Computational evidence for deligne's conjecture
  regarding exceptional lie groups,'' {\em Comptes rendus de l'Acad{\'e}mie des
  sciences. S{\'e}rie 1, Math{\'e}matique} {\bf 322} (1996) no.~5, 427--432.

\bibitem{Landsberg200259}
J.~Landsberg and L.~Manivel, ``Triality, exceptional lie algebras and deligne
  dimension formulas,''
  \href{http://dx.doi.org/http://dx.doi.org/10.1006/aima.2002.2071}{{\em
  Advances in Mathematics} {\bf 171} (2002) no.~1, 59 -- 85}.

\bibitem{Julia:1980gr}
B.~Julia, ``{Group disintegrations},'' in {\em Superspace and Supergravity},
  S.~Hawking and M.~Rocek, eds., vol.~C8006162 of {\em Nuffield Gravity
  Workshop}, pp.~331--350.
\newblock Cambridge University Press,
1980.
\newblock

\bibitem{Julia:1982gx}
B.~Julia, ``{Kac-Moody Symmetry Of Gravitation And Supergravity Theories},'' in
  {\em {American Mathematical Society summer seminar on Appication of Group
  Theory in Physics and Mathematical Physics Chicago, Illinois, July 6-16,
  1982}}, P.~S. M.~Flato and G.~Zuckerman, eds., vol.~21, p.~355.
\newblock Lectures in Applied Mathematics,
1982.
\newblock

\bibitem{Anastasiou:2015vba}
A.~Anastasiou, L.~Borsten, L.~J. Hughes, and S.~Nagy, ``{Global symmetries of
  Yang-Mills squared in various dimensions},''
{\em JHEP} {\bf 148} (2016)  1601.

\bibitem{Cremmer:1999du}
E.~Cremmer, B.~Julia, H.~Lu, and C.~N. Pope, ``{Higher dimensional origin of D
  = 3 coset symmetries},''
\href{http://arxiv.org/abs/hep-th/9909099}{{\tt arXiv:hep-th/9909099
  [hep-th]}}.

\bibitem{Marrani:2017aqc}
A.~Marrani, G.~Pradisi, F.~Riccioni, and L.~Romano, ``{Non-Supersymmetric Magic
  Theories and Ehlers Truncations},''
\href{http://arxiv.org/abs/1701.03031}{{\tt arXiv:1701.03031 [hep-th]}}.

\bibitem{Gunaydin:1983rk}
M.~G{\"u}naydin, G.~Sierra, and P.~K. Townsend, ``{Exceptional supergravity
  theories and the magic square},''
\href{http://dx.doi.org/10.1016/0370-2693(83)90108-9}{{\em Phys. Lett.} {\bf
  B133} (1983)  72}.

\bibitem{Gunaydin:1983bi}
M.~G{\"u}naydin, G.~Sierra, and P.~K. Townsend, ``{The geometry of $N=2$
  Maxwell-Einstein supergravity and Jordan algebras},''
\href{http://dx.doi.org/10.1016/0550-3213(84)90142-1}{{\em Nucl. Phys.} {\bf
  B242} (1984)  244}.

\bibitem{Gunaydin:1984ak}
M.~G{\"u}naydin, G.~Sierra, and P.~K. Townsend, ``{Gauging the $d = 5$
  Maxwell-Einstein supergravity theories: More on Jordan algebras},''
\href{http://dx.doi.org/10.1016/0550-3213(85)90547-4}{{\em Nucl. Phys.} {\bf
  B253} (1985)  573}.

\bibitem{Gunaydin:1992zh}
M.~Gunaydin, ``{Generalized conformal and superconformal group actions and
  Jordan algebras},''
\href{http://dx.doi.org/10.1142/S0217732393001124}{{\em Mod.Phys.Lett.} {\bf
  A8} (1993)  1407--1416}.

\bibitem{Gunaydin:2005zz}
M.~G{\"u}naydin and O.~Pavlyk, ``{Generalized spacetimes defined by cubic forms
  and the minimal unitary realizations of their quasiconformal groups},''
\href{http://dx.doi.org/10.1088/1126-6708/2005/08/101}{{\em JHEP} {\bf 08}
  (2005)  101}.

\bibitem{Gunaydin:2000xr}
M.~G{\"u}naydin, K.~Koepsell, and H.~Nicolai, ``{Conformal and quasiconformal
  realizations of exceptional Lie groups},''
\href{http://dx.doi.org/10.1007/PL00005574}{{\em Commun. Math. Phys.} {\bf 221}
  (2001)  57--76}.

\bibitem{Gunaydin:2007bg}
M.~Gunaydin, A.~Neitzke, B.~Pioline, and A.~Waldron, ``{Quantum Attractor
  Flows},'' \href{http://dx.doi.org/10.1088/1126-6708/2007/09/056}{{\em JHEP}
  {\bf 09} (2007)  056},
\href{http://arxiv.org/abs/0707.0267}{{\tt arXiv:0707.0267 [hep-th]}}.

\bibitem{Ferrara:1997uz}
S.~Ferrara and M.~G{\"u}naydin, ``{Orbits of exceptional groups, duality and
  BPS states in string theory},''
  \href{http://dx.doi.org/10.1142/S0217751X98000913}{{\em Int. J. Mod. Phys.}
  {\bf A13} (1998)  2075--2088},
\href{http://arxiv.org/abs/hep-th/9708025}{{\tt arXiv:hep-th/9708025}}.

\bibitem{Gunaydin:2005gd}
M.~G{\"u}naydin, ``{Unitary realizations of U-duality groups as conformal and
  quasiconformal groups and extremal black holes of supergravity theories},''
\href{http://dx.doi.org/10.1063/1.1923339}{{\em AIP Conf. Proc.} {\bf 767}
  (2005)  268--287}.

\bibitem{Gunaydin:2005mx}
M.~Gunaydin, A.~Neitzke, B.~Pioline, and A.~Waldron, ``{BPS black holes,
  quantum attractor flows and automorphic forms},''
\href{http://dx.doi.org/10.1103/PhysRevD.73.084019}{{\em Phys. Rev.} {\bf D73}
  (2006)  084019}.

\bibitem{Ferrara:2006xx}
S.~Ferrara and M.~Gunaydin, ``{Orbits and attractors for N = 2 Maxwell-Einstein
  supergravity theories in five dimensions},''
  \href{http://dx.doi.org/10.1016/j.nuclphysb.2006.09.016}{{\em Nucl. Phys.}
  {\bf B759} (2006)  1--19},
\href{http://arxiv.org/abs/hep-th/0606108}{{\tt arXiv:hep-th/0606108}}.

\bibitem{Bellucci:2006xz}
S.~Bellucci, S.~Ferrara, M.~G{\"u}naydin, and A.~Marrani, ``{Charge orbits of
  symmetric special geometries and attractors},''
  \href{http://dx.doi.org/10.1142/S0217751X06034355}{{\em Int. J. Mod. Phys.}
  {\bf A21} (2006)  5043--5098},
\href{http://arxiv.org/abs/hep-th/0606209}{{\tt arXiv:hep-th/0606209}}.

\bibitem{Borsten:2008wd}
L.~Borsten, D.~Dahanayake, M.~J. Duff, H.~Ebrahim, and W.~Rubens, ``{Black
  Holes, Qubits and Octonions},''
  \href{http://dx.doi.org/10.1016/j.physrep.2008.11.002}{{\em Phys. Rep.} {\bf
  471} (2009) no.~3--4, 113--219},
\href{http://arxiv.org/abs/0809.4685}{{\tt arXiv:0809.4685 [hep-th]}}.

\bibitem{Borsten:2009zy}
L.~Borsten, D.~Dahanayake, M.~J. Duff, and W.~Rubens, ``{Black holes admitting
  a Freudenthal dual},''
  \href{http://dx.doi.org/10.1103/PhysRevD.80.026003}{{\em Phys. Rev.} {\bf
  D80} (2009) no.~2, 026003},
\href{http://arxiv.org/abs/0903.5517}{{\tt arXiv:0903.5517 [hep-th]}}.

\bibitem{Borsten:2010aa}
L.~Borsten, D.~Dahanayake, M.~J. Duff, S.~Ferrara, A.~Marrani, {\em et al.},
  ``{Observations on Integral and Continuous U-duality Orbits in N=8
  Supergravity},'' \href{http://dx.doi.org/10.1088/0264-9381/27/18/185003}{{\em
  Class.Quant.Grav.} {\bf 27} (2010)  185003},
  \href{http://arxiv.org/abs/1002.4223}{{\tt arXiv:1002.4223 [hep-th]}}.

\bibitem{Borsten:2011ai}
L.~Borsten, M.~J. Duff, S.~Ferrara, A.~Marrani, and W.~Rubens, ``{Small
  Orbits},'' \href{http://dx.doi.org/10.1103/PhysRevD.85.086002}{{\em
  Phys.Rev.} {\bf D85} (2012)  086002},
\href{http://arxiv.org/abs/1108.0424}{{\tt arXiv:1108.0424 [hep-th]}}.

\bibitem{Marrani:2012uu}
A.~Marrani, C.-X. Qiu, S.-Y.~D. Shih, A.~Tagliaferro, and B.~Zumino,
  ``{Freudenthal Gauge Theory},''
  \href{http://dx.doi.org/10.1007/JHEP03(2013)132}{{\em JHEP} {\bf 03} (2013)
  132},
\href{http://arxiv.org/abs/1208.0013}{{\tt arXiv:1208.0013 [hep-th]}}.

\bibitem{Borsten:2008}
L.~Borsten, ``{$E_{7(7)}$ invariant measures of entanglement},''
  \href{http://dx.doi.org/10.1002/prop.200810542}{{\em Fortschr. Phys.} {\bf
  56} (2008) no.~7--9, 842--848}.

\bibitem{levay-2008}
P.~L\'{e}vay and P.~Vrana, ``Three fermions with six single-particle states can
  be entangled in two inequivalent ways,''
  \href{http://dx.doi.org/10.1103/PhysRevA.78.022329}{{\em Phys. Rev.} {\bf
  A78} (2008) no.~2, 022329}, \href{http://arxiv.org/abs/0806.4076}{{\tt
  arXiv:0806.4076 [quant-ph]}}.

\bibitem{Levay:2008mi}
P.~L\'{e}vay, M.~Saniga, and P.~Vrana, ``{Three-qubit operators, the split
  Cayley hexagon of order two and black holes},''
  \href{http://dx.doi.org/10.1103/PhysRevD.78.124022}{{\em Phys. Rev.} {\bf
  D78} (2008) no.~12, 124022},
\href{http://arxiv.org/abs/0808.3849}{{\tt arXiv:0808.3849 [quant-ph]}}.

\bibitem{Levay:2009}
P.~Vrana and P.~L{\'e}vay, ``Special entangled quantum systems and the
  freudenthal construction,'' {\em Journal of Physics A: Mathematical and
  Theoretical} {\bf 42} (2009) no.~28, 285303,
  \href{http://arxiv.org/abs/0902.2269}{{\tt arXiv:0902.2269 [quant-ph]}}.
  \url{http://stacks.iop.org/1751-8121/42/i=28/a=285303}.

\bibitem{LevayVrana}
G.~{Sierra}, ``{An application of the theories of Jordan algebras and
  Freudenthal triple systems to particles and strings},''
  \href{http://dx.doi.org/10.1088/0264-9381/4/2/006}{{\em Classical and Quantum
  Gravity} {\bf 4} (1987) no.~2, 227--236}.

\bibitem{Bars:1978yx}
I.~Bars and M.~Gunaydin, ``{Construction of Lie Algebras and Lie Superalgebras
  From Ternary Algebras},''
\href{http://dx.doi.org/10.1063/1.524309}{{\em J.Math.Phys.} {\bf 20} (1979)
  1977}.

\bibitem{Chiodaroli:2011pp}
M.~Chiodaroli, M.~Gunaydin, and R.~Roiban, ``{Superconformal symmetry and
  maximal supergravity in various dimensions},''
  \href{http://dx.doi.org/10.1007/JHEP03(2012)093}{{\em JHEP} {\bf 1203} (2012)
   093},
\href{http://arxiv.org/abs/1108.3085}{{\tt arXiv:1108.3085 [hep-th]}}.

\bibitem{Borsten:2013bp}
L.~Borsten, M.~J. Duff, L.~J. Hughes, and S.~Nagy, ``{A magic square from
  Yang-Mills squared},''
  \href{http://dx.doi.org/10.1103/PhysRevLett.112.131601}{{\em Phys.Rev.Lett.}
  {\bf 112} (2014)  131601},
\href{http://arxiv.org/abs/1301.4176}{{\tt arXiv:1301.4176 [hep-th]}}.

\bibitem{Anastasiou:2013hba}
A.~Anastasiou, L.~Borsten, M.~J. Duff, L.~J. Hughes, and S.~Nagy, ``{A magic
  pyramid of supergravities},''
  \href{http://dx.doi.org/10.1007/JHEP04(2014)178}{{\em JHEP} {\bf 1404} (2014)
   178},
\href{http://arxiv.org/abs/1312.6523}{{\tt arXiv:1312.6523 [hep-th]}}.

\bibitem{Chiodaroli:2014xia}
M.~Chiodaroli, M.~G{\"u}naydin, H.~Johansson, and R.~Roiban, ``{Scattering
  amplitudes in $ \mathcal{N}=2 $ Maxwell-Einstein and Yang-Mills/Einstein
  supergravity},'' \href{http://dx.doi.org/10.1007/JHEP01(2015)081}{{\em JHEP}
  {\bf 01} (2015)  081},
\href{http://arxiv.org/abs/1408.0764}{{\tt arXiv:1408.0764 [hep-th]}}.

\bibitem{Springer:2000}
T.~A. Springer and F.~D. Veldkamp, {\em Octonions, Jordan Algebras and
  Exceptional Groups}.
\newblock Springer-Verlag, Berlin, Heidelberg, New York, 2000.

\bibitem{Hurwitz:1898}
A.~Hurwitz, ``{Uber die komposition der quadratishen formen von beliebig vielen
  variabeln},'' {\em Nachr. Ges. Wiss. Gottingen} (1898)  309--316.

\bibitem{jacobson1954}
N.~Jacobson, ``Structure of alternative and jordan bimodules,'' {\em Osaka
  Math. J.} {\bf 6} (1954) no.~1, 1--71.

\bibitem{Barton:2003}
C.~H. Barton and A.~Sudbery, ``{Magic squares and matrix models of Lie
  algebras},'' \href{http://dx.doi.org/10.1016/S0001-8708(03)00015-X}{{\em Adv.
  in Math.} {\bf 180} (2003) no.~2, 596--647},
  \href{http://arxiv.org/abs/math/0203010}{{\tt arXiv:math/0203010}}.

\bibitem{Evans:2009ed}
J.~M. Evans, ``{Trialities and Exceptional Lie Algebras: Deconstructing the
  Magic Square},''
\href{http://arxiv.org/abs/0910.1828}{{\tt arXiv:0910.1828 [hep-th]}}.

\bibitem{Vinberg:1966}
E.~B. Vinberg, ``A construction of exceptional simple lie groups,'' {\em Tr.
  Semin. Vektorn. Tr. Semin. Vektorn. Tensorn. Anal.} {\bf 13} (1966) no.~7-9,
  .

\bibitem{Cacciatori:2012cb}
S.~L. Cacciatori, B.~L. Cerchiai, and A.~Marrani, ``{Squaring the Magic},''
  \href{http://dx.doi.org/10.4310/ATMP.2015.v19.n5.a1}{{\em Adv. Theor. Math.
  Phys.} {\bf 19} (2015)  923--954},
\href{http://arxiv.org/abs/1208.6153}{{\tt arXiv:1208.6153 [math-ph]}}.

\bibitem{Hull:1994ys}
C.~M. Hull and P.~K. Townsend, ``{Unity of superstring dualities},''
  \href{http://dx.doi.org/10.1016/0550-3213(94)00559-W}{{\em Nucl. Phys.} {\bf
  B438} (1995)  109--137},
\href{http://arxiv.org/abs/hep-th/9410167}{{\tt arXiv:hep-th/9410167}}.

\bibitem{Cremmer:1979up}
E.~Cremmer and B.~Julia, ``{The $SO(8)$ supergravity},''
\href{http://dx.doi.org/10.1016/0550-3213(79)90331-6}{{\em Nucl. Phys.} {\bf
  B159} (1979)  141}.

\bibitem{Cremmer:1997ct}
E.~Cremmer, B.~Julia, H.~Lu, and C.~Pope, ``{Dualization of dualities. 1.},''
  \href{http://dx.doi.org/10.1016/S0550-3213(98)00136-9}{{\em Nucl.Phys.} {\bf
  B523} (1998)  73--144},
\href{http://arxiv.org/abs/hep-th/9710119}{{\tt arXiv:hep-th/9710119
  [hep-th]}}.

\bibitem{Cremmer:1978km}
E.~Cremmer, B.~Julia, and J.~Scherk, ``{Supergravity theory in 11
  dimensions},''
\href{http://dx.doi.org/10.1016/0370-2693(78)90894-8}{{\em Phys. Lett.} {\bf
  B76} (1978)  409--412}.

\bibitem{Cremmer:1998px}
E.~Cremmer, B.~Julia, H.~Lu, and C.~Pope, ``{Dualization of dualities. 2.
  Twisted selfduality of doubled fields, and superdualities},''
  \href{http://dx.doi.org/10.1016/S0550-3213(98)00552-5}{{\em Nucl.Phys.} {\bf
  B535} (1998)  242--292},
\href{http://arxiv.org/abs/hep-th/9806106}{{\tt arXiv:hep-th/9806106
  [hep-th]}}.

\bibitem{Anastasiou:2015ena}
A.~Anastasiou and M.~J. Hughes, ``{Octonionic D=11 Supergravity and 'Octavian
  Integers' as Dilaton Vectors},''
\href{http://arxiv.org/abs/1502.02578}{{\tt arXiv:1502.02578 [hep-th]}}.

\bibitem{burceb}
D.~Burde and M.~Ceballos, ``{Abelian Ideals of Maximal Dimension for Solvable
  Lie Algebras},'' {\em J. of Lie Theory.} {\bf 22(3)} (2009)  ,
  \href{http://arxiv.org/abs/0911.2995}{{\tt arXiv:0911.2995 [math.RA]}}.

\bibitem{Bossard:2010dq}
G.~Bossard, C.~Hillmann, and H.~Nicolai, ``{E7(7) symmetry in perturbatively
  quantised N=8 supergravity},''
  \href{http://dx.doi.org/10.1007/JHEP12(2010)052}{{\em JHEP} {\bf 12} (2010)
  052},
\href{http://arxiv.org/abs/1007.5472}{{\tt arXiv:1007.5472 [hep-th]}}.

\bibitem{Borsten:2012pd}
L.~Borsten, M.~J. Duff, S.~Ferrara, and A.~Marrani, ``{Freudenthal Dual
  Lagrangians},'' \href{http://dx.doi.org/10.1088/0264-9381/30/23/235003}{{\em
  Class.Quant.Grav.} {\bf 30} (2013)  235003},
\href{http://arxiv.org/abs/1212.3254}{{\tt arXiv:1212.3254 [hep-th]}}.

\bibitem{Cremmer:1980gs}
E.~Cremmer, ``{Supergravities in 5 Dimensions},'' in {\em {In *Salam, A. (ed.),
  Sezgin, E. (ed.): Supergravities in diverse dimensions, vol. 1* 422-437. (In
  *Cambridge 1980, Proceedings, Superspace and supergravity* 267-282) and Paris
  Ec. Norm. Sup. - LPTENS 80-17 (80,rec.Sep.) 17 p. (see Book Index)}}.
\newblock
1980.
\newblock

\bibitem{Gunaydin:2003yx}
M.~Gunaydin and M.~Zagermann, ``{Unified Maxwell-Einstein and
  Yang-Mills-Einstein supergravity theories in five-dimensions},''
  \href{http://dx.doi.org/10.1088/1126-6708/2003/07/023}{{\em JHEP} {\bf 07}
  (2003)  023},
\href{http://arxiv.org/abs/hep-th/0304109}{{\tt arXiv:hep-th/0304109
  [hep-th]}}.

\bibitem{Gunaydin:2005df}
M.~Gunaydin, S.~McReynolds, and M.~Zagermann, ``{Unified N=2 Maxwell-Einstein
  and Yang-Mills-Einstein supergravity theories in four dimensions},''
  \href{http://dx.doi.org/10.1088/1126-6708/2005/09/026}{{\em JHEP} {\bf 09}
  (2005)  026},
\href{http://arxiv.org/abs/hep-th/0507227}{{\tt arXiv:hep-th/0507227
  [hep-th]}}.

\bibitem{Gunaydin:2005bf}
M.~Gunaydin, S.~McReynolds, and M.~Zagermann, ``{The R-map and the coupling of
  N=2 tensor multiplets in 5 and 4 dimensions},''
  \href{http://dx.doi.org/10.1088/1126-6708/2006/01/168}{{\em JHEP} {\bf 01}
  (2006)  168},
\href{http://arxiv.org/abs/hep-th/0511025}{{\tt arXiv:hep-th/0511025
  [hep-th]}}.

\bibitem{Jordan:1933vh}
P.~Jordan, J.~von Neumann, and E.~P. Wigner, ``{On an algebraic generalization
  of the quantum mechanical formalism}.''
\href{http://www.jstor.org/stable/pdfplus/1968117.pdf}{\textit{Ann. Math.}
  \textbf{35} (1934) no. 1, 29--64}.

\bibitem{Gunaydin:2010fi}
M.~Gunaydin, H.~Samtleben, and E.~Sezgin, ``{On the Magical Supergravities in
  Six Dimensions},''
  \href{http://dx.doi.org/10.1016/j.nuclphysb.2011.02.010}{{\em Nucl.Phys.}
  {\bf B848} (2011)  62--89},
\href{http://arxiv.org/abs/1012.1818}{{\tt arXiv:1012.1818 [hep-th]}}.

\end{thebibliography}

\providecommand{\href}[2]{#2}\begingroup\raggedright\endgroup

\end{document}